\documentclass[sigconf]{acmart}

\AtBeginDocument{%
  \providecommand\BibTeX{{%
    \normalfont B\kern-0.5em{\scshape i\kern-0.25em b}\kern-0.8em\TeX}}}
    
\usepackage{multirow}
\usepackage{enumitem}
\usepackage{dblfloatfix}
\usepackage{color,soul}
\usepackage[title]{appendix}
\usepackage{tabularx}
\usepackage{colortbl}
\usepackage{graphicx}
\usepackage{longtable}

\copyrightyear{2025}
\acmYear{2025}
\setcopyright{rightsretained}
\acmConference[CHI '25]{CHI Conference on Human Factors in Computing Systems}{April 26-May 1, 2025}{Yokohama, Japan}
\acmBooktitle{CHI Conference on Human Factors in Computing Systems (CHI '25), April 26-May 1, 2025, Yokohama, Japan}\acmDOI{10.1145/3706598.3713188}
\acmISBN{979-8-4007-1394-1/25/04}

\begin{document}

\title[Predicting Trust In Autonomous Vehicles]{Predicting Trust In Autonomous Vehicles: Modeling Young Adult Psychosocial Traits, Risk-Benefit Attitudes, And Driving Factors With Machine Learning}

\author{Robert Kaufman}
\email{rokaufma@ucsd.edu}
\orcid{0000-0003-1279-690X}
\affiliation{%
  \institution{University of California, San Diego}
  \streetaddress{9500 Gilman Dr}
  \city{La Jolla}
  \state{California}
  \country{USA}
  \postcode{92093}
}

\author{Emi Lee}
\email{eal003@ucsd.edu}
\orcid{0009-0005-4099-2045}
\affiliation{%
  \institution{University of California, San Diego}
  \city{La Jolla}
  \state{California}
  \country{USA}
}

\author{Manas Satish Bedmutha}
\email{mbedmutha@ucsd.edu}
\orcid{0000-0003-3427-2226}
\affiliation{%
  \institution{University of California, San Diego}
  \city{La Jolla}
  \state{California}
  \country{USA}
}

\author{David Kirsh}
\email{kirsh@ucsd.edu}
\orcid{0000-0002-6242-5851}
\affiliation{%
  \institution{University of California, San Diego}
  \city{La Jolla}
  \state{California}
  \country{USA}
}

\author{Nadir Weibel}
\email{weibel@ucsd.edu}
\orcid{0000-0002-3457-4227}
\affiliation{%
  \institution{University of California, San Diego}
  \city{La Jolla}
  \state{California}
  \country{USA}
}

\renewcommand{\shortauthors}{Kaufman et al.}

\begin{abstract}
Low trust remains a significant barrier to Autonomous Vehicle (AV) adoption. To design trustworthy AVs, we need to better understand the individual traits, attitudes, and experiences that impact people's trust judgements. We use machine learning to understand the most important factors that contribute to young adult trust based on a comprehensive set of personal factors gathered via survey (n = 1457). Factors ranged from psychosocial and cognitive attributes to driving style, experiences, and perceived AV risks and benefits. Using the explainable AI technique SHAP, we found that perceptions of AV risks and benefits, attitudes toward feasibility and usability, institutional trust, prior experience, and a person's mental model are the most important predictors. Surprisingly, psychosocial and many technology- and driving-specific factors were not strong predictors. Results highlight the importance of individual differences for designing trustworthy AVs for diverse groups and lead to key implications for future design and research.
\end{abstract}




\maketitle
\section{Introduction}
Potential benefits of Autonomous Vehicles (AVs) range from reduced traffic, less crashes, and lower environmental impact to improvements in efficiency and risk management~\cite{hewitt2019assessing}. However, widespread adoption of AVs is impeded by a lack of rider trust in AV decisions and the real-world consequences of adoption~\cite{kenesei2022trust}. AVs are a clear case of a more general lack of public trust in Artificial Intelligence (AI), spanning many domains~\cite{bedue2022can}. \color{black} Distrust in AI is particularly evident for opaque “black box” systems, whose decision-making processes are not human accessible \cite{gunning2019darpa, miller2019explanation}.

It has long been established that trust is a major determinant of adoption of AI-based technologies like AVs, where people's beliefs about a system directly inform their willingness to accept and rely on the system \cite{lee2004trust, muir1994trust}. The relationship between trust and adoption may be especially strong in high-stakes contexts like driving, where users must rely on the technology for personal safety \cite{hoff2015trust}. Highly influential models by \citet{choi2015investigating} and \citet{hoff2015trust} demonstrate how trust is a necessary pre-condition for intention to use autonomous vehicles: if a person does not trust an AV, they will not be willing to ride in it. \citet{liao2022designing} discuss how trust impacts users’ perception of risk and shapes their interactions with AI-based systems, meaning that even a perfectly safe AV may face adoption challenges if users do not trust it. \emph{In essence, this means that no matter how safe or high performing an AV is, without trust, adoption efforts will fail.}

Trust reflects the subjective, psychological confidence that users place in the technology, and does not automatically follow when a system is high-performing or safe \cite{muir1994trust}. In this way, trust serves as a bridge between vehicle performance and user acceptance -- without this bridge, AVs may not reach widespread use. Understanding user trust perceptions in AV research (as opposed to solely improving AV safety or performance) enables designers and researchers to address specific user perceptions and concerns that may hinder adoption, including those that move beyond driving performance \cite{kraus2021s}. By identifying which personal factors most influence trust, this study offers insights on the foundation of AV trust perceptions, informing the design of AVs that align with user needs beyond technical safety alone. \color{black}

There is substantial theoretical and experimental evidence supporting that a person's trust in AVs is influenced by their individual traits and experiences~\cite{hoff2015trust, kaufman2024developing, lee2004trust}. To develop more trustworthy AVs, it is therefore necessary to understand \emph{which} specific factors influence trust judgments, and by \emph{how much}. This can empower AV designers to address the unique needs of different individuals.

Much prior research has examined personal traits that may predict trust and adoption attitudes toward AVs~\cite{mosaferchi2023personality, choi2015investigating, liu2019public, ferronato2020examination, he2022modelling, li2020personality, bockle2021can}. However, many of these studies lack comprehensive coverage of relevant characteristics, limiting their utility. By focusing on only a narrow set of traits, potentially critical factors are left unexamined. For instance, several personal factors discussed by theoretical frameworks and trust research in other AI domains have not yet been explored in the context of AVs, including certain driving behaviors, cognitive traits, and institutional attitudes among others ~\cite{hoff2015trust, kaufman2024developing, lee2004trust}. Further, studying traits in isolation or in small sets prevents a clear understanding of the \emph{relative importance} of different factors in trust development. While various traits -- ranging from demographic characteristics to personality, risk assessment, and experience with AVs -- have been associated with trust, without a comprehensive approach that measures and models diverse factors together, our understanding of their relative significance remains incomplete. This makes it difficult for AV designers and researchers to prioritize factors of importance and develop personalized approaches to the human-AV trust problem. These two research gaps -- (1) unexplored factors and (2) lack of information on relative importance -- underscore the need for studies that account for a broader range of traits while using methods that can capture the interplay between them. In this work, we aim to address both gaps.

A concrete result of securing a more comprehensive understanding of how traits impact trust (in particular, which traits are more important) is designing AVs that can meet the specific needs of diverse individuals, as opposed to taking a one-size-fits-all approach to human-AV interaction. For example, explainable AI (XAI) explanations meant to increase trust via transparency for “black box” AI (and AV) decisions can provide \emph{tailored} communications to meet the needs of specific groups~\cite{gunning2019darpa, liao2021human, kaufman2024developing}. Tailoring communications to individual users has been shown to more effectively address informational needs, helping to calibrate system reliance to appropriate and optimal levels~\cite{schneider2019personalized, ma2023analysing, wang2019designing, kaufman2022cognitive}. Knowing \emph{who} has \emph{what} needs to address can guide the creation of more accessible and equitable AVs~\cite{hassija2024interpreting}, including tailored educational campaigns~\cite{pataranutaporn2023influencing} and inclusive policy guidelines\cite{jobin2019global}. Without meeting the needs of specific individuals or subpopulations, a future with ubiquitous AV adoption cannot be realized. 

In the present study, we use survey methods (n = 1457) to measure a broader range of variables than any previous research on this topic, bridging a diverse spectrum of driving and non-driving related research. We include variables ranging from cognitive and psychological traits to cultural values, driving style, and technology attitudes. We also include a wide range of specific risks and benefits in AVs, as these may be important contributors to trust perceptions~\cite{ayoub2021modeling}. Including such a wide range of variables allows us to build a more nuanced and comprehensive understanding of the factors that influence trust than has been previously possible.

\color{black} We are able to capture the complex interactions between variables and assess the \emph{relative} importance of the factors in our analysis by using machine learning (ML) to predict trust in AVs. The rationale for using ML as the basis of our inquiry stems from its ability to capture complex, nonlinear interactions among our chosen personal traits, attitudes, and trust that traditional regression methods may miss \cite{wang2020survey}. \citet{yarkoni2017choosing} argue that, unlike linear models which assume independent, additive effects, ML can reveal nuanced, multidimensional relationships and should be opted for in research where predictors have the potential to be interrelated, such as in the work presented here. ML models also handle high-dimensional data very effectively, addressing multicollinearity and retaining diverse features better than traditional regression, without overfitting \cite{breiman2001random}. By using ML, we can retain a comprehensive set of trust-related features without compromising model validity, offering a more holistic view of a wide range of factors potentially impacting trust. We apply SHapley Additive eXplanations (SHAP)~\cite{lundberg2020local} to “peek beneath the hood” of our models, adding interpretability to the predictive power of our ML models by showing precisely how important different factors were to predictions. This transparency makes our ML insights actionable, guiding AV design and policy toward strategies that directly address users’ trust-building needs. By utilizing these advanced analysis methods, we can build deeper insights into how a wide range of traits relate to trust judgements. 
\color{black}

Our research also highlights the importance of isolating subpopulations when predicting trust in AVs. Most studies in this area analyze data across general populations~\cite{ferronato2020examination, he2022modelling, li2020personality, bockle2021can}, which can obscure unique traits and concerns relevant to specific groups. For example, the common finding that age predicts a person's trust in AI may be more of a superficial correlation rather than a true driving factor~\cite{araujo2020ai, choung2023trust, abraham2017autonomous}. Though age may be helpful for predicting trust, age often correlates strongly with underlying variables, such as technical expertise, familiarity with AI systems, and openness to innovation~\cite{chen2011review, steinke2012trust}. These underlying variables are far \emph{more} informative of \emph{why} people of different ages may trust AI systems differently, giving insight into \emph{how} to design better systems for them. Subpopulation-focused approaches have proven effective in other areas of computing, such as online interventions for social media, where deeper cognitive traits outperformed broader indicators like age or political affiliation in predicting behavior~\cite{kaufman2022s, kaufman2024warning}. We apply similar techniques to this work by focusing on a specific subpopulation -- in this case, young adults. By holding broader factors constant, we can identify the more nuanced, informative factors that shape trust in this subpopulation.

We sought to prioritize early adopters and the next generation of transportation users, as these groups will likely play a key role in the adoption and shaping of AV design. Young adults, in particular, fit this profile due to their high comfort with new technologies~\cite{choung2023trust} and their potential to influence future transportation trends~\cite{berliner2019uncovering}. By studying young adults, we can better understand their unique perspectives and concerns -- a critical step towards aligning AV systems with the needs of these early adopters. Our findings not only inform AV design and communication strategies for this group but also offer a methodological template and approach for investigating other subpopulations, ultimately supporting the development of AV systems that support appropriate trust and acceptance across diverse user bases.

The results of this study show that young adult trust in autonomous vehicles (AVs) can be effectively predicted from personal factors, using machine learning. Our models achieved high accuracy (85.8\%) in classifying trust levels. By applying the explainable AI technique SHAP, we identified which factors most strongly predicted trust and how the value of each factor contributes to the model's predictions. Notably, perceptions of AV risks and benefits, attitudes toward AV feasibility, institutional trust, and prior experience were the most significant trust predictors, while personality, culture, cognitive preferences, driving style, and driving cognition were less impactful than expected, despite previous research suggesting their importance.

\vspace{1em}
\noindent
The work presented here builds new understanding on how personal factors can be used to predict trust in AVs, a crucial step towards designing future, more trustworthy human-AV interactions. In Summary, we contribute: 
\begin{itemize}[topsep=0pt, nolistsep]
    \item A comprehensive study on personal factors predicting young adult AV trust, including relative factor importance.
    \item A methodology and set of important factors that can be replicated for study in different subpopulations.
    \item A set of design implications based on this work, including high-priority areas for future research and design.
\end{itemize}

\section{Related Work}
\subsection{Personal Factors Impacting Trust in AVs}
\label{prior_personal}
\textbf{Individual Differences ---} Hoff and Bashir's~\cite{hoff2015trust} influential analysis of modern human-AI trust research posits that dispositional factors including culture and age, as well as situational and learned factors such as expertise and prior experience, impact how much a person will trust and rely on automated systems like AVs. According to Hoff and Bashir, these factors influence the system in conjunction with the context of use (e.g. driving from location A to location B) and the system's demonstrated performance in the moment. Similarly, \citet{kaufman2024developing} propose a framework for human-AV interaction based on situational awareness and joint action, emphasizing the role personal traits and contextual characteristics play in guiding interactions and achieving goals. Lee and See's~\cite{lee2004trust} review of trust foundations posits that trust success is shaped by a trustor's predispositions and environment. By all three accounts, personal traits clearly influence trust in the system. 

\textbf{Demographics ---} Prior work examining demographic factors like age have shown that younger individuals tend to adopt autonomous systems earlier than older individuals~\cite{hulse2018perceptions, abraham2017autonomous}, however, differences in age may be caused by underlying variables, including familiarity with AI, concern over risks, and self-efficacy~\cite{chen2011review, steinke2012trust, mosaferchi2023personality}. Gender differences commonly show that men are more likely to adopt AVs than women~\cite{hulse2018perceptions, mosaferchi2023personality}, however, these effects may also be moderated by underlying variables like risk preferences~\cite{mosaferchi2023personality} or vehicle anxiety~\cite{hohenberger2016and}. Education may also play a role: people with more education may be more willing to adopt autonomous vehicles~\cite{hudson2019people}. Recent work has even shown that people who lean politically liberal in the United States may be more willing to adopt AVs -- and see greater benefit to adoption -- than those who lean politically conservative, though this finding too may be motivated by underlying driving variables like cultural values and belief systems~\cite{mack2021politics, mosaferchi2023personality}. Socioeconomic status (SES) is understudied in the realm of AV trust prediction; SES may be important given its negative association with interpersonal trust in other domains~\cite{stamos2019investigating}. In the present study, we measure all of these demographic variables: age, gender, education, political orientation, and socioeconomic status to see how important they are for our trust models. Our assessment of trust in young individuals primarily associated with an undergraduate institution means we can hold age and education relatively constant, allowing us to examine deeper underlying factors which may drive trust in AVs.

\textbf{Personality and Culture ---} Personality and culture may impact a person's trust in automated vehicles, however, prior findings are conflicting. Using the Big Five Inventory (BFI)~\cite{rammstedt2007measuring}, \citet{ferronato2020examination} found that extroversion negatively correlated with trust, while openness, conscientiousness, and agreeableness showed positive associations. In contrast, \citet{bockle2021can} reported positive correlations for both extroversion and agreeableness with trust in automation, but none for openness. \citet{chien2016relation} found that agreeable and conscientious individuals were more trusting. They also measured Hofstede's Five Dimensions of Cultural Values via the Cultural Values Scale (CVSCALE)~\cite{yoo2011measuring}, which did not seem to have an impact on trust. \citet{chien2018effect}, however, found that Hofstede's Values such as power distance, uncertainty avoidance, and collectivism did impact automation trust. These often conflicting findings suggest that we should reevaluate the assumption that personality and culture are key drivers of trust, highlighting the need for clarification on their importance in relation to other factors. Cultural identity and race may also impact AI trust, one study attributing trust differences between black and white participants to prior experiences with institutional discrimination~\cite{lee2021included}. This highlights the important impact prior experiences and institutional trust may have on AI trust.

\textbf{Knowledge and Prior Experience with AVs ---} There is substantial evidence suggesting that a person's domain knowledge and past experiences with a system will impact their level of trust~\cite{hoff2015trust}. \citet{mosaferchi2023personality} find that individuals with higher AV knowledge tend to have higher trust in autonomous vehicles. Higher trust is not always a good thing; \emph{calibrating} trust to cases where trust is warranted is critical for sustained adoption. Preventing over-trust is an often neglected secondary benefit of knowledge, since it can inform when to rely on an AV and when to not~\cite{khastgir2018calibrating}. Prior experience interacting with AVs and AV technology may also be an important contributor to trust judgements: individuals with positive past experiences with AV technology -- even if they have never ridden in a fully autonomous AV itself -- may be more willing to trust and adopt AVs in the future~\cite{hoff2015trust}. By contrast, exposure to performance errors can deteriorate trust~\cite{luo2020trust, kaufman2024did}. In the present study, we examine how AV expertise and prior experience with AV and other autonomous vehicle technologies may impact trust judgements. We also examine prior experiences with driving collisions in general, as these may impact a person's judgements of driving safety and risk~\cite{butler1999post}. 

\textbf{Technology and Institutional Attitudes ---} Technology attitudes may also impact trust: those with a stronger technology affinity tend to have higher trust~\cite{mosaferchi2023personality, bennett2019attitudes, huangexploring}. We assess this using questions from the commonly used Affinity for Technology Interaction (ATI) scale~\cite{franke2019personal}. As previously noted, institutional trust may impact trust in AI-based systems~\cite{lee2021included}. We examine a person's trust in AV-specific institutions, including tech companies, automakers, and government regulators, as these may impact a person's trust in AVs~\cite{kaufman2022s}.

\textbf{Driving Behavior, Attitudes, Cognitions, and Abilities ---} The way a person drives on their own and how they think about themselves as a driver may impact how much they trust an AV. A person's own driving style -- such as whether they are an aggressive, passive, or distracted driver -- has demonstrated importance on rider satisfaction and willingness to rely on an AV's driving~\cite{zhang2024effects}. Specifically, defensive drivers showed better calibrated AV reliance than aggressive drivers. Similar factors have not been thoroughly examined for AV trust. The widely-used Multidimensional Driving Style Inventory (MDSI)~\cite{taubman2004multidimensional} is used in the present study to assess a person's driving style. \citet{kraus2020scared} study self-esteem, driving self-efficacy, and driving anxiety, showing a negative relationship between anxiety and trust in AVs. This effect was moderated by a person's confidence in their own ability. Similarly, \citet{hohenberger2016and} show driving anxiety as a major driver of trust. Anxiety specific to driving and driving concerns has been understudied in AV trust research. We leverage the Driving Cognitions Questionnaire (DCQ)~\cite{ehlers2007driving} to assess several dimensions of social and accident-related anxieties. The effects of self-esteem and self-efficacy have been contested in past AV trust research: some studies claim they are important~\cite{kraus2020scared, hegner2019automatic} while other studies claim they are not~\cite{lohaus2024automated}. We seek to clarify conflicting prior work by assessing driving anxiety, specific concerns and cognitions, self-esteem, self-efficacy, and style.

\textbf{Cognitive and Psychological Traits ---} Common AV adoption concerns involve perceptions of increased risk and worry over giving up driving control~\cite{howard2014public}. Traits such as a person's willingness to take risks~\cite{ajenaghughrure2020risk} and their desire for control~\cite{hegner2019automatic} have been shown to positively correlate with trust in AVs~\cite{choi2015investigating, liu2019public}. These are examined in the present study. Given that most interactions with fully autonomous vehicles require giving up agency to the vehicle, it is relevant to understand a person's tolerance towards decision ambiguity and rationalization. Preference for predictability and ambiguity can be measured via the Need for Closure~\cite{roets2011item} scale; these factors have been highlighted as potentially important for the design of autonomous vehicles~\cite{hamburger2022personality} and have clear implications for explainable AI~\cite{wang2019designing}. At present, they are understudied in AV trust research and are included in our analysis. Similarly, a person's preference for rational or intuitive decision-making may also impact their trust in AVs, as AVs operate using complex, logic-based algorithms for perception, recognition, and classification~\cite{levinson2011towards}. We study these cognitive and psychological traits and preferences to further understand their impact and importance for AV trust.

\textbf{Risks and Benefits of AVs ---} Perhaps the most obvious determinant of AV trust and adoption decisions are a driver's personal estimates of the risks and benefits of AV's~\cite{choi2015investigating, hoff2015trust}. In their model of trust in autonomous vehicles, \citet{ayoub2021modeling} found that perceptions of risks and benefits were the most important predictors of trust. Unfortunately, they failed to measure specifically which risks and benefits are important. As a whole, there are a plethora of potential benefits and risks that may impact a person's willingness to adopt an AV~\cite{fagnant2015preparing}. \citet{acheampong2019capturing} specify perceived benefits related to safety, self-image, and the ability to do other activities while the AV is driving, among others. The Autonomous Vehicle Acceptance Model (AVAM)~\cite{hewitt2019assessing} includes several risks and benefits in their questionnaire, including measures of performance expectancy, ease of use, realism of deployment, safety, and social influence. Likewise, the Checklist for Trust between People and Automation (CTPA)~\cite{jian2000foundations} includes automation concerns surrounding the potential harm, dependability, and reliability of the system, suggesting higher concerns will deteriorate trust. Additional concerns over cost, usefulness, and ease of use have also been highlighted~\cite{zhang2023human}. Several risks and benefits were included in the present study to delineate specifically which are the best predictors of trust.

\vspace{1em}
\noindent
\textbf{Summary ---} Despite the range of studies on trust, knowledge gaps remain on how certain untested variables impact \emph{AV} trust specifically, and how different variables impact AV trust \emph{relative} to each other. Most prior studies on AV trust utilize a small selection of variables and limited analysis techniques, impeding their ability to bring the community a comprehensive understanding of relative variable importance. Further, very little prior work is done on isolated populations, making it difficult to apply insights to specific populations of interest, such as young adults. The present study seeks to address these knowledge gaps.

\begin{figure*}[t]
    \centering    \includegraphics[width=.9\linewidth]{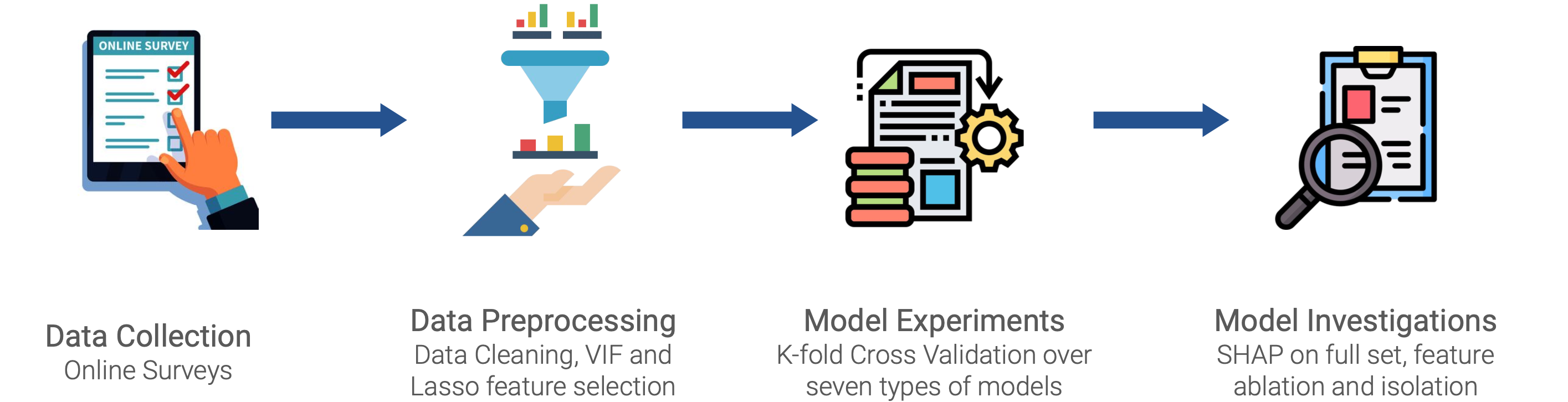}
    \vspace{-.5em}
    \caption{Overall process architecture, from data collection and processing to model development and explanations using SHAP.}
    \Description{Overall process architecture diagram for our study, moving from data collection and preprocessing to model development (experiments) and investigations (explanations using SHAP).}
    \label{fig:pipeline}
\end{figure*}

\subsection{Predicting Trust Using Machine Learning}
The application of machine learning (ML) techniques to predict trust in autonomous vehicles is a relatively new but promising research area. Using large datasets, advanced ML-based predictive models may provide more accurate predictions than traditional analysis and modeling approaches like ANOVA or linear regression when used for challenges like predicting trust~\cite{wang2020survey}. One of the main obstacles with using advanced ML techniques is a lack of explainability -- it is difficult to know how features of a dataset individually contribute to a model's conclusion. Without this information, we cannot know how important different variables are to predicting an outcome. Recent techniques developed for introducing transparency for AI-based systems -- such as SHapley Additive exPlanations (SHAP)~\cite{lundberg2020local} -- can be used to provide insight into a model's behavior, providing justification for decisions via coefficients for relative feature importance. In the case of predicting trust from personal traits, the use of advanced ML models combined with model explainability via SHAP allows us to understand \emph{if} specific personal factor variables are important for trust judgements as well as the relative importance of \emph{each} variable to the model's prediction. 

Traditional analyses like linear regression and ANOVA are by far the most common methods to understanding the personal factors which may predict trust. Some recent works, however, have used advanced machine learning techniques like Random Forest, XGBoost, Naive Bayes, decision trees, or SVM for such goals~\cite{wang2020survey}. For example, \citet{zolfaghar2012syntactical} used decision trees to successfully predict a person's trust in social web applications based on social trust factors like knowledge and personality. \citet{liao2022driver} use personality traits and driver style to categorize drivers into profiles. \citet{kraus2021s} build a hierarchical model to show the relationships between elemental traits (e.g. personality and self-esteem), situational traits (affinity for technology and interpersonal trust), and surface traits (e.g. trust and technology acceptance). \citet{liu2014generic} use decision trees to classify trust in large-scale systems like recommendation agents. Other studies have proposed frameworks for classifying trust using Support Vector Machines (SVMs)~\cite{lopez2015towards}.

For AVs specifically, a few recent studies have used ML techniques to predict trust from personal traits. \citet{ma2024understanding} predict reliance decisions in a simulated driving task using a random forest model. Unsurprisingly, they find that a person's prior situational trust and experience with AVs predicted their current reliance decision. \citet{bennett2019attitudes} model autonomous vehicle adoption attitudes for people with disabilities using demographics, anxiety, locus of control, and other attitudes towards AVs and technology as a whole. \citet{huangexploring} use Additive Bayesian Networks (ABNs) to explore factors related to trust in an Advanced Driver Assistance System (ADAS), finding prior system and technological experience were the two highest predictors. 

Outside of prediction from personal traits, some recent work has sought to assess trust from physiological signals using machine learning. \citet{yi2023measurement} use a multi-modal feature fusion network to predict trust from physiological signals like galvanic skin response and heart rate variability. Other studies have used XGBoost~\cite{ayoub2023real} and discriminant classification~\cite{akash2018classification} to accomplish similar ends. 

Using similar methods as the present study, \citet{ayoub2021modeling} used demographics, knowledge, feelings, and behavioral intentions to predict how a person responded to the question “how much would you trust an autonomous vehicle?” They included general ratings of the overall “risk” and “benefit” of AVs, however, they did not specify which risks or benefits, nor did they gather precise concerns. Nonetheless, they found that the risk and benefit scores were the highest contributors to their XGBoost model. We seek to build upon this effort by incorporating a much larger range of features, a more comprehensive measure of “trust” based on modern trust theory, and a more specific analysis focusing on a particular user group of interest (young adults). In addition, we include a wide range of \emph{specific} risks and benefits based on prior work, allowing us to understand \emph{which} risks and benefits may be most important.

In sum, the integration of machine learning and explainable AI techniques like SHAP for predicting trust in AVs provides a significant opportunity to advance our understanding of trust development in complex sociotechnical systems like autonomous driving~\cite{fraedrich2015transition}. These models not only offer high predictive accuracy but also provide valuable insights for datasets with large numbers of variables not available using other methods. As AV technology continues to evolve, the ability to predict and enhance user trust will be crucial for successful adoption and widespread acceptance of AVs. In the present work, we seek to build on prior research to advance the study of individual differences in trust, addressing knowledge gaps that will allow researchers and designers to build AVs that can meet the needs of diverse groups.

\section{Method}
In the present study, we use survey methods to assess a wide range of personal factors from a pool of young adults. Then, we build models predicting a person's trust in autonomous vehicles from those factors, using a variety of machine learning techniques. Following, we use SHAP~\cite{lundberg2020local} to make the top performing models explainable, giving us measures of feature importance. We add additional nuance to our study with ablation experiments knocking out specific feature groups from the models in order to derive greater insight into how factors contribute to AV trust. Figure~\ref{fig:pipeline} describes the overall modeling process. Similar to concurrent work on clustering~\cite{bedmutha2024exploring, saha2021person}, to look at trust differences between the “high” and “low” groups at the level of each factor, we also perform a more traditional descriptive analysis using Kruskal-Wallis tests as a means to provide comparison to our ML results.


\begin{table*}[t] 
  \caption{Distributions of participant age and highest completed education.}
  \Description{Table showing the distributions of participant age and highest completed education.}
  \label{demographic_table}
  \renewcommand{\arraystretch}{1.3}
  \vspace{-.5em}
  \begin{tabular}{p{0.1\linewidth}p{0.15\linewidth}p{0.15\linewidth}p{0.24\linewidth}p{0.24\linewidth}}
    \toprule
    \textbf{Age} & \textbf{18-19} & \textbf{20-21} & \textbf{22-23} & \textbf{24+}\\
    & 29.3\% & 48.2\% & 15.2\% & 7.3\% \\
    \midrule
    \textbf{Education} & \textbf{{\footnotesize < 2 years of college}} & \textbf{{\footnotesize 2 years of college}} & {\footnotesize \textbf{4-year degree (e.g. bachelor's)}} & {\footnotesize \textbf{Grad. Degree (e.g. master's, PhD)}}\\
    & 27.3\% & 59.3\% & 12.6\% & 0.8\% \\
    \bottomrule
\end{tabular}
\end{table*}

\subsection{Participants and Data Collection}
A total of 1457 participants completed the study and passed all quality control requirements. The survey was designed and deployed using Qualtrics~\cite{Qualtrics2024}, taking a median of 17.9 minutes to complete. Recruitment occurred via email lists and via SONA,\footnote{https://www.sona-systems.com} an undergraduate study pool where students are granted study credit for participation. As such, the vast majority of participants were enrolled at or affiliated with a large undergraduate university in the United States. The mean age of the study sample was 20.7 (SD = 2.4), with ages capped between 18 and 35 years. The sample was 73.9\% female. Table~\ref{demographic_table} shows distributions of age and education for our sample. Only participants who passed three attention checks were included. As an additional means of quality control, we screened out participants with low variability in survey responses (e.g. used the same pattern of answers) and those who had unrealistically low completion time. As a result of this quality control process, we are confident that the data used in our analysis is of high quality.

\begin{table*}[b] 
  \caption{Measures: Driving and Technology Behaviors, Attitudes, and Perceptions}
  \Description{Explanation of measures used in the study: Driving and Technology Behaviors, Attitudes, and Perceptions}
  \label{driving_factors}
  \renewcommand{\arraystretch}{1.3}
  \vspace{-.5em}
  \begin{tabular}{p{0.08\linewidth}>{\raggedright}p{0.13\linewidth}p{0.4\linewidth}p{0.31\linewidth}}
    \toprule
    \textbf{Category} & \textbf{Factor} & \textbf{Description} & \textbf{Scale and Source} \\
    \midrule
    \textbf{Driving Behavior} & Driving Style & 43-question Multidimensional Driving Style Inventory (MDSI) assessing driving behaviors and thoughts. & Strg. Disagree- Strg. Agree (1-5).~\cite{taubman2004multidimensional} \\
    & Driving Freq. & How often do you drive a car? & Daily - Never (1-7) * \\
    & Driving Collisions & How many driving collisions have you been a part of in the past 3 years?  & * \\
    \midrule
    \textbf{Driving Priors} & AV Knowledge and Expertise & 10 questions related to knowledge of autonomous vehicles and AI-based systems. & Strg. Disagree- Strg. Agree (1-5). Adapted from~\cite{kaufman2024effects} \\
    & AV Experience & Have you ever ridden in an autonomous vehicle? & Fully auto., Partially auto, No. * \\
    & Assistive Tech. & Have you ever used any of the following autonomous driving functionalities? List of 8 (e.g. Adaptive Cruise Control,  Lane Assist). & Never - Often (1-3), n/a.~\cite{abraham2017case} \\
    \midrule
    \textbf{Driving Attitudes} & Tech. Self-Efficacy & 5 questions similar to “I am good with technology.” & Strg. Disagree- Strg. Agree (1-5). Adapted from~\cite{teo2012understanding} \\
    & Affinity for Technology & 4 questions from the Affinity for Technology Interaction (ATI) scale measuring how much a person enjoys new technologies, e.g. “I am quick to incorporate new technologies into my life.” & Strg. Disagree- Strg. Agree (1-5).~\cite{franke2019personal} \\
    & AV Feasibility & The AV and the infrastructure necessary to use the AV are practically feasible. & Strg. Disagree- Strg. Agree (1-5).~\cite{hewitt2019assessing} \\
  \bottomrule
  \multicolumn{3}{l}{\textit{* For simple questions, the research team came up with question wording and scaling.}}
\end{tabular}
\end{table*}

\begin{table*}[h] 
  \caption{Measures: Demographic and Psychosocial}
  \Description{Explanation of measures used in the study: Demographic and Psychosocial}
  \label{demo_psych_factors}
  \renewcommand{\arraystretch}{1.3}
  \vspace{-.5em}
  \begin{tabular}{p{0.08\linewidth}>{\raggedright}p{0.14\linewidth}p{0.4\linewidth}p{0.31\linewidth}}
    \toprule
    \textbf{Category} & \textbf{Factor} & \textbf{Description} & \textbf{Scale and Source} \\
    \midrule
   \textbf{General} & Age & What is your age in years? & * \\
    & Education & Highest level of education completed. &  * \\
    & SES & What is your yearly household income (pre-tax)? &  * \\
    & Gender & What is your gender? & Adapted from~\cite{lindqvist2021gender} \\
    & Politics & Politically speaking, how would you best describe yourself? (US politics, N/A optional). & Strong Dem. - Strong Rep. (1-5).~\cite{kaufman2022s} \\
    & Trust in Institutions & Trust in technology companies, traditional automakers, and the government. & Strg. Disagree- Strg. Agree (1-5). Adapted from~\cite{funk2019trust} \\
    \midrule
    \textbf{Psycho-social} & Personality & Big Five Inventory (BFI): extroversion, agreeableness, conscientiousness, neuroticism, and openness. & Strg. Disagree- Strg. Agree (1-5).~\cite{rammstedt2007measuring}  \\
    & Culture & Hofstede's Dimensions of Cultural Values via the Cultural Values Scale (CVSCALE). 17 questions, 3 sub-scales: power distance, uncertainty avoidance, and collectivism. & Strg. Disagree- Strg. Agree (1-5).~\cite{yoo2011measuring} \\
    & Risk Willingness & How willing are you to take risks, in general? & Not at all - Very (1-10).~\cite{dohmen2005individual} \\
    & Need For Control & 5 questions from the Desirability Of Control Scale, e.g. “I prefer to be in control of situations involving my safety”. & Strg. Disagree- Strg. Agree (1-5).~\cite{burger1979desirability} \\
    & Decision Style & 3 questions on intuitive (gut feeling) vs rational (logic-based) decision-making style preferences. & Strg. Disagree- Strg. Agree (1-5).~\cite{scott1995decision} \\
    & Self-esteem & 5-question scale with questions similar to “On the whole, I am satisfied with myself.” & Strg. Disagree- Strg. Agree (1-4).~\cite{monteiro2022efficient} \\
    & Need For Closure & NFC is a 15-question scale assessing a person's preferences for ambiguity and unpredictability. Results are split into these two sub-categories. & Strg. Disagree- Strg. Agree (1-5).~\cite{roets2011item} \\
    \midrule
    \textbf{Driving-specific} & Driving Cognitions & We assess dimensions of social and accident-related thoughts that occur during driving via 13 questions of the Driving Cognitions Questionnaire (DCQ). & Strg. Disagree- Strg. Agree (1-5).~\cite{ehlers2007driving} \\
    & Driving Risk & How risky do you consider driving? & Not at all - Extremely (1-5) * \\
    & Passenger Anxiety & How anxious do you feel as a passenger when someone else is driving? & Not at all - Extremely (1-5) * \\
  \bottomrule
  \multicolumn{3}{l}{\textit{* For simple questions, the research team came up with question wording and scaling.}}
\end{tabular}
\end{table*}

\begin{table*}[h] 
  \caption{Measures: Risks, Benefits, and Trust}
  \Description{Explanation of measures used in the study: Risks, Benefits, and Trust}
  \label{risk_benefit_factors}
  \renewcommand{\arraystretch}{1.3}
  \vspace{-.5em}
  \begin{tabular}{p{0.08\linewidth}>{\raggedright}p{0.17\linewidth}p{0.6\linewidth}}
    \toprule
    \textbf{Category} & \textbf{Factor} & \textbf{Description} \\
    \midrule
    \textbf{Perceptions} & AV Feasibility & The AV and the infrastructure necessary to use the AV are practically feasible.  \\
    & Overall Risk-Benefit & How do you perceive the risk-benefit trade-off of using an autonomous vehicle? \newline \textit{[Scale: All Risk, No Benefit- No Risk, All Benefit (1-7)]}  \\
    & Mental Model & Autonomous vehicles make decisions like humans do. \\
    \midrule
    \textbf{Benefits} & Reduce Accidents & Autonomous vehicles can reduce traffic crashes. \\
    & Reduce Traffic & Autonomous vehicles can reduce traffic congestion. \\
    & Reduce Emissions & Autonomous vehicles can reduce vehicle emissions and pollution. \\
    & Improve Fuel Economy & Autonomous vehicles can improve fuel economy. \\
    & Reduce Transport Cost & Autonomous vehicles can reduce transport costs. \\
    & Improve Mobility & Autonomous vehicles can increase mobility for those who are unable to drive. \\
    & Free Time & Autonomous vehicles will give me more free time (e.g. to text, play games). \\
    & Increase Fun & An autonomous vehicle would make driving more interesting or fun. \\
    & Improve Efficiency & Using an autonomous vehicle would enable me to reach my destination quickly. \\
    \midrule
    \textbf{Risks} & System Failure & I am concerned about equipment and system failures in autonomous vehicles. \\
    & Legal Liability & I am concerned about the legal liability of drivers or owners of autonomous vehicles. \\
    & Hacking & I am concerned that the computer systems of autonomous vehicles may be hacked. \\
    & Data Privacy & I am concerned about sharing driving data with a company or government agency. \\
    & Performance (varied) & I am concerned that autonomous vehicles cannot handle varied weather and terrain. \\
    & Cost & I am concerned that autonomous vehicles will be too expensive. \\
    & Losing Control & I am concerned about giving up driving control to an autonomous vehicle. \\
    & Lacking Understanding & I am afraid that I would not understand an autonomous vehicle's decisions. \\
    & Ease of Use & I would find an autonomous vehicle easy to use. \\
    \midrule
    \textbf{Trust} & Dependability & Autonomous vehicles are more dependable than human drivers. \\
   \textbf{(Outcome)} & Adaptability & Autonomous vehicles will adapt well to new environments. \\
    & Goal Alignment & The goals for autonomous vehicles are aligned with my own. \\
    & Danger to Self & I believe that using an autonomous vehicle would be dangerous to me. \\
    & Danger to Others & I believe that autonomous vehicles would be dangerous to other drivers/pedestrians. \\
    & Safety & I would feel safe while using an autonomous vehicle. \\
    & General Trust & I would trust an autonomous vehicle. \\
    & Good Idea & Using an autonomous vehicle would be a good idea. \\
    & Positivity & Overall, I feel positive about autonomous vehicles. \\
    & Recommendation & I will recommend family members and friends to ride in autonomous vehicles. \\
  \bottomrule
  \multicolumn{3}{l}{\textit{All questions across categories were adapted from~\cite{hewitt2019assessing, fagnant2015preparing, acheampong2019capturing, jian2000foundations, zhang2023human}. Unless otherwise specified, scale: Strg. Disagree- Strg. Agree (1-5).}}
\end{tabular}
\end{table*}

\subsection{Survey Design \color{black} and Administration \color{black}}
The survey was designed to be as comprehensive as possible while considering limits on survey duration and participant attention. \color{black} Our questions were heavily based on validated questionnaires and survey measures to enhance the reliability and interpretability of responses; we relied on previously tested measures to ensure that our questions were well-aligned with established constructs and our own expectations. Given the use of validated questions, we deemed that a formal cognitive pretest was not necessary. However, we conducted a pretest with approximately 10 participants and qualitative feedback on the survey was solicited to ensure that questions were straightforward, clear, free of errors, and not too burdensome. The results from these pilot participants was not included in the final analysis. The pretest allowed us to refine any questions that appeared ambiguous or confusing to participants. The survey itself can be found in Supplementary Material.

Participants were provided with a brief description of autonomous vehicles at the beginning of the survey, specifying AVs as \textit{“vehicles that use technology, such as artificial intelligence and special cameras, to drive without the need for human intervention. They sense their environment, adjusting speed and direction to bring a passenger from one location to another safely.”} This was accompanied by a simple and clear cartoon visualization of an autonomous sedan in an everyday driving scenario (commuting amongst other vehicles). This introduction aimed to establish the context of interest as AVs focused on typical, day-to-day applications, as opposed to specialized uses like autonomous trucking, racing, or industrial AVs. We chose this commuting-oriented context to ensure that participants were evaluating AVs intended for personal, everyday driving tasks. By avoiding references to specific brands or technical specifications, we allowed participants to respond based on their own knowledge and attitudes toward AVs used in common, relatable scenarios. This approach was designed to capture perceptions of AVs in the specific context of everyday driving, reflecting a more realistic and relevant use case for general public adoption.
\color{black} 

\subsubsection{Survey Scales and Questionnaires}
Table \ref{driving_factors} shows factors related to driving-specific attitudes and behaviors, Table \ref{demo_psych_factors} shows demographic, psychosocial, and general attitude measures, and Table \ref{risk_benefit_factors} details risk and benefit measures as well as the trust measures used to form our composite trust outcome. Given the large number of variables, variable descriptions that are not relevant to the reported results are omitted from these tables; all variable names were designed to be descriptive and intuitive.

\subsubsection{Trust as an Outcome}
We predict a person's AV trust attitudes by creating a composite score of 10 individual trust questions designed to encompass several dimensions of trust, based on  current models of trust formation~\cite{jian2000foundations, lee2004trust, hoff2015trust}. These can be found in the “trust” section of Table \ref{risk_benefit_factors}; \texttt{Danger to Self} and \texttt{Danger to Others} are reverse-scored. Despite a large corpus of cognitive science and philosophy work demonstrating the multidimensional nature of trust formation~\cite{hoff2015trust, lee2004trust, kaufman2024developing, jian2000foundations}, most present studies use single-item outcomes (e.g. “do you trust AVs?”)~\cite{ayoub2021modeling}. People are notoriously bad at articulating complex implicit judgements like trust in systems~\cite{miller2019explanation}, and thus multidimensional measures provide a more valid approach to measurement. Scores above or below the middle value of scores were categorized as “high” and “low”, respectively. We note that these groups were of slightly different sizes: 44.5\% of participants were in the “high” condition. This was accounted for during model construction.

\subsection{Model Development}
\label{sec:model_development}
After deriving an exploratory understanding of the dataset and its features using descriptive statistics and Kruskal-Wallis tests (discussed in Section \ref{sec:descriptive}), we investigated the potential for machine learning systems to predict users trust in AVs, \color{black} as these advanced modeling methods provide a powerful and nuanced way to derive deep insights from complex data \cite{wang2020survey, yarkoni2017choosing}. \color{black} We discuss our approach to designing such a model below.

\textbf{Feature Selection ---}
Our survey data resulted in a set of features that could be used as predictors for our models predicting user trust. For survey questions that contributed to validated composite scores (like Big 5 personality or CVSCALE culture values), only the composites were kept in the dataset. For measures that provide meaning on their own (for example, each risk/benefit or specific driving behaviors), these were included individually as their own feature. 

Since this data comes from surveys, we took multiple steps to achieve a concise feature set and mitigate the possibility that some features could be correlated with each other. Following recent work around modeling personality traits~\cite{saha2021person}, we first filtered features using their Variance Inflation Factor (VIF) with respect to the raw trust scores, using a threshold of 10. This helped us detect and reduce multicollinearity. To further reduce multicollinearity, next we ran feature selection through Lasso regression~\cite{tibshirani1996regression} on the raw trust scores. \color{black} Lasso was opted for because it not only regularizes the model but also performs feature selection by driving less important coefficients to zero, making it more effective than ridge regression for identifying relevant predictors in high-dimensional datasets \cite{tibshirani1996regression}. \color{black} This resulted in a final set of 130 features, which was used for all our machine learning experiments and is henceforth referred to as the \textit{full feature set}.

\begin{table*}[t] 
  \caption{Descriptive Statistics and Kruskal–Wallis Tests For Low and High Trust Groups. This table shows the factors that had the largest individual differences between the high and low trust groups (top 15 by effect size). Each predictor was tested independently.}
  \Description{Descriptive Statistics and Kruskal–Wallis Tests For Low and High Trust Groups. This table shows the factors that had the largest individual differences between the high and low trust groups (top 15 by effect size). Each predictor was tested independently.}
  \label{kw_large}
\renewcommand{\arraystretch}{1.2}
  \vspace{-.5em}
    \begin{tabular}{p{0.25\linewidth} | p{0.05\linewidth} | p{0.05\linewidth} | p{0.05\linewidth} | p{0.05\linewidth} | p{0.06\linewidth} | p{0.06\linewidth} | p{0.08\linewidth} | p{0.1\linewidth}}
  \hline
 & \multicolumn{2}{c|}{\textbf{Low Trust}} & \multicolumn{2}{c|}{\textbf{High Trust}} & \multicolumn{4}{c}{\textbf{Comparison}}\\
 \hline
\textbf{Predictor} & \textbf{Mean} & \textbf{SD} & \textbf{Mean} & \textbf{SD} & \textbf{{\footnotesize H-stat}} & \textbf{{\footnotesize P-value}} & \textbf{{\footnotesize Cohen's D}} & \textbf{Effect} \\ 
  \hline
Overall Risk-Benefit & 3.23 & 0.87 & 4.63 & 0.81 & 595.25 & < 0.01 & 1.66 & large \\ 
  Ease of Use & 2.55 & 1.13 & 3.72 & 0.88 & 345.75 & < 0.01 & 1.14 & large \\ 
  Reduce Accidents & 2.79 & 1.07 & 3.89 & 0.86 & 338.97 & < 0.01 & 1.12 & large \\ 
  Trust Tech Companies & 2.19 & 0.99 & 3.09 & 0.99 & 241.61 & < 0.01 & 0.91 & large \\ 
 Increase Fun & 2.38 & 1.22 & 3.46 & 1.16 & 237.14 & < 0.01 & 0.91 & large \\ 
  AV Feasibility & 2.58 & 1.02 & 3.43 & 0.87 & 235.40 & < 0.01 & 0.88 & large \\ 
  Improve Efficiency & 2.46 & 1.06 & 3.36 & 1.03 & 214.80 & < 0.01 & 0.86 & large \\ 
 Reduce Traffic & 2.82 & 1.17 & 3.71 & 1.04 & 194.96 & < 0.01 & 0.80 & large \\
   \bottomrule
    \end{tabular}
\end{table*}

\textbf{Machine Learning Pipeline ---}
Our machine learning pipeline consisted of two components: (1) feature transformation and (2) classification modeling. For effective modeling, we reduce dimensions through Linear Discriminant Analysis (LDA)~\cite{tharwat2017linear} on the feature set to attain 10 transformed feature vectors. Since the number of “high” and “low” people in the dataset was not perfectly balanced, we over-sample the minority in the training data to match the number of majority samples. We used the Simple Minority Oversampling TEchnique (SMOTE)~\cite{chawla2002smote} to achieve this synthetic balanced dataset, a common method used in similar ML analyses.

Next, we applied several classical machine learning models on the derived feature set in order to find the best performing method. Given our task of binary classification (“low” vs. “high” trust), we used models similar to \color{black} \citet{bedmutha2024conversense} and \citet {ayoub2021modeling} \color{black} to achieve a balance of interpretability and performance. Specifically, we applied Logistic Regression (LR), Support Vector Machines (SVM) with linear and radial kernels, Decision Tree Classifiers (DTC), Random Forest (RF), Naive Bayes, and Gradient-Boosted Decision Trees. We used XGBoost implementation~\cite{chen2016xgboost} for the gradient-boosted trees.

\textbf{Training Experiments ---}
We conducted cross-validation experiments to quantify the performance of different models. The pipeline was evaluated across a five-fold split on the overall dataset. For each instance, the data was first normalized using Standard Scaling based on the training data and then fed into the machine learning pipeline. Each type of model was evaluated using balanced accuracy. We also tracked the precision, recall and macro-F1 scores for each iteration given their robustness to imbalance~\cite{plotz2021applying}. For each model, we experimented with the various parameters possible for that model type. Final model selection was conducted on the mean balanced accuracy performance across all cross-validation folds.

\subsection{Model Investigations}
\label{sec:model_investigations}
After training different possible models, our goal was to understand how features contribute to their overall performance. This explainability brings insight into the most important personal factors that predict a person's trust judgement. To achieve this, we used SHapley Additive exPlanations (SHAP)~\cite{lundberg2020local} to investigate dimensions indicative of trust. SHAP scores are game-theoretic measures of importance for how each feature contributes to the overall prediction, providing local interpretability of the pipeline. SHAP not only allowed us to understand \emph{which} features are important, but also \emph{how} important via a relative importance coefficient. In our training process, we tracked SHAP scores for each feature for every fold of all models. We created explainer plots for each of these sub-models for every fold to identify key descriptors influencing trust. While some features ranked relatively higher or lower across folds, to get a sense of the overall contribution, we report on the mean absolute contribution of each feature across the folds similar to the method used by \citet{nepal2024moodcapture}.

We present two types of SHAP visualizations. First, \emph{bar graphs} show the mean absolute SHAP scores for the top 20 features contributing to model predictions of high and low trust together (high/low trust predictions are mirrors of each other); these give us a breakdown of overall feature importance for predicting trust. Second, we present SHAP \emph{beeswarm plots} that show how individual feature values contribute to the model's prediction of trust. These are also organized based on feature importance, but provide additional information allowing us to see how high and low scores for a particular feature predict the high or low trust classification (e.g. Fig. \ref{fig:fullset_shap}). For example, these allow us to make claims like higher \texttt{Overall Risk-Benefit} assessment is associated with higher trust.

\subsection{Deeper Feature Experiments (Systematic Ablation Studies)}
\label{sec:feature_investigations}
We conducted a series of experiments to derive a deeper understanding of how key features impact our trust models. Given how the risk and benefit dimensions used in this study showed to be the top predictors of trust in our full model (Section \ref{full_model_SHAP_sec}), we conducted two follow-up ablation studies to test (1) the accuracy of predicting trust from risk and benefit features only, and (2) the impact of removing risk and benefit features from our full model (leaving only other types of features like psychosocial and driving). For each ablation (systematic elimination) study, we followed a pipeline and training process similar to Section~\ref{sec:model_development} and report the metrics averaged across folds. We compare them to the scores found in the larger full feature model; we consider the net decrease in performance to be an indicator of importance.

\section{Results}
We present results from our survey of 1457 young adults. First, we show descriptive statistics and Kruskal-Wallis tests comparing high and low trust individuals, generating initial dataset understanding. Next, we introduce the results of our ML analysis, with model explanations provided via SHAP. Finally, we deepen the ML analysis by performing two ablation (systematic elimination) studies, which compare results with and without risk-benefit contributions. 

\begin{table*}[b] 
\caption{Model performances using the full feature set, reporting the mean (standard deviation) validated across five folds. All measures are calculated as macro weighted scores between the two classes of trust, ensuring balance between the high and low trust groups. All models demonstrated generally good predictability -- Random Forest models showed the highest performance and were thus selected to be used for the remainder of the analysis.}
\Description{This table shows model performances using the full feature set, reporting the mean (standard deviation) validated across five folds. All measures are calculated as macro weighted scores between the two classes of trust, ensuring balance between the high and low trust groups. All models demonstrated generally good predictability -- Random Forest models showed the highest performance and were thus selected to be used for the remainder of the analysis.}
\vspace{-.5em}
\renewcommand{\arraystretch}{1.2}
\begin{tabular}{lccccc}
\toprule
\textbf{Model} & \textbf{Bal. Accuracy} & \textbf{Precision} & \textbf{Recall} & \textbf{F1-score} & \textbf{AUROC} \\
\midrule
Naive Bayes & 0.797 (0.03) & 0.799 (0.03) & 0.797 (0.03) & 0.797 (0.03) & 0.875 (0.03)\\
Logistic Regression & 0.840 (0.02) & 0.859 (0.02) & 0.857 (0.02) & 0.840 (0.02) & 0.922 (0.02) \\
Support Vector Machines (Linear) & 0.840 (0.02) & 0.842 (0.02) & 0.840 (0.02) & 0.840 (0.02) & 0.922 (0.03) \\
Support Vector Machines (Radial) & 0.838 (0.02) & 0.840 (0.02) & 0.838 (0.02) & 0.838 (0.02) & 0.888 (0.03) \\
Decision Tree Classifier & 0.800 (0.03) & 0.801 (0.03) & 0.800 (0.03) & 0.800 (0.03) & 0.800 (0.03) \\
\textbf{Random Forest} & \textbf{0.858 (0.02)} & \textbf{0.859 (0.02)} & \textbf{0.858 (0.02)} & \textbf{0.858 (0.02)} & \textbf{0.934 (0.01)} \\
Gradient Boosted Trees (XGBoost) & 0.825 (0.03) & 0.825 (0.03) & 0.825 (0.03) & 0.824 (0.03) & 0.906 (0.03) \\
\bottomrule
\end{tabular}
\label{tab:model_main}
\end{table*}

\subsection{Descriptive Analysis: Comparison of Factors Using Kruskal–Wallis}
\label{sec:descriptive}
Descriptive statistics and initial comparisons using Kruskal–Wallis tests for the high and low trust groups can be found in Table \ref{kw_large} (factors with large effect size only) and Appendix \ref{appendixA.trust} (all factors). Kruskal–Wallis results show if individual predictor variable scores differ between people of high trust and people of low trust. Predictably, given the large sample size, nearly all group differences were significant. To give a more functional comparison, we report effect size (Cohen's D) and provide an interpretation based on accepted standards~\cite{fritz2012effect}.

We find 8 variables to have large effect sizes: \texttt{Overall Risk-Benefit}, \texttt{Ease of Use}, \texttt{Reduce Accidents}, \texttt{Trust in Tech Companies}, \texttt{Increase Fun}, \texttt{AV Feasibility}, \texttt{Improve Efficiency}, and \texttt{Reduce Traffic}. Each of these may be important for predicting how much a person trusts AV, and indeed, each of these factors were in the top 20 most important factors for our full feature ML model. However, we note that in the Kruskal-Wallis case, each factor was tested individually, limiting our ability to understand complex patterns and non-linear relationships from the larger, combined set of predictors. Using the more sophisticated and predictive machine learning approaches described next, we get a more nuanced understanding of how each variable contributes to a person's trust assessment.


\subsection{Machine Learning Analysis Using The Full Feature Set}
\label{full_set_results}
\subsubsection{Model Comparisons and Best Performance}
We conducted a series of experiments to identify the best performing overall models using the full feature set, optimizing for different sets of parameters. We report the best achieved scores for each model averaged across all five cross-validation folds in Table~\ref{tab:model_main}. 

We find that all models performed well in general, strengthening support for the hypothesis that personal factors can predict trust dispositions for AVs. Overall, Random Forest produced our most predictive model for trust using the full feature set, resulting in 85.8\% balanced accuracy. We also note that precision and recall are higher than chance and have similar magnitudes. This implies that the model is intrinsically robust and consistent across both high and low trust groups. The area under ROC curve observed in the best performing model (0.934) shows that our model pipelines and features can create a strong class separation and are indifferent to label imbalance.

\subsubsection{Explaining the Full Feature Model using SHAP}
\label{full_model_SHAP_sec}
\begin{figure*}[t]
    \centering
        \centering
        \includegraphics[width=.9\linewidth]{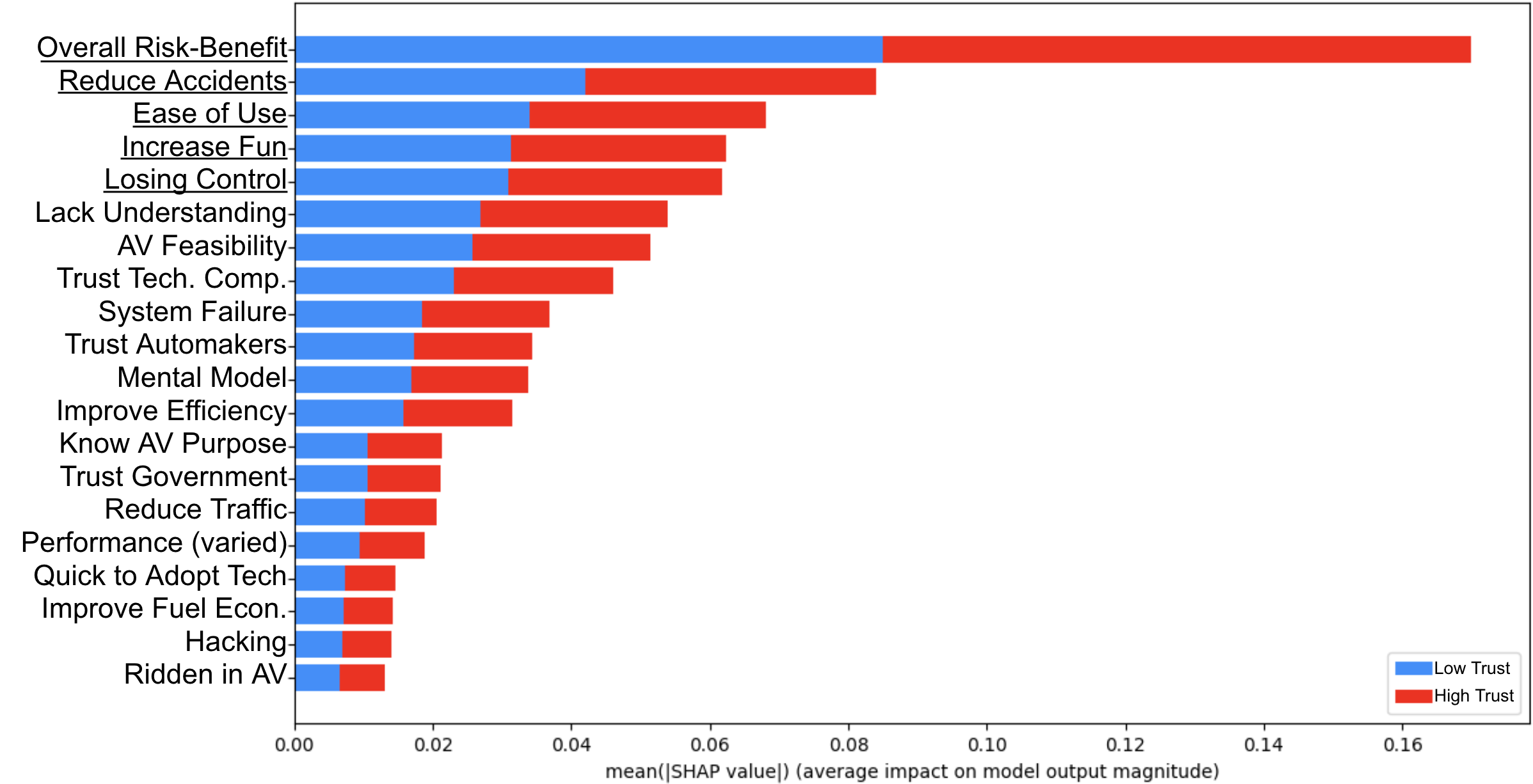}
        \vspace{-0.5em}
        \caption{Full Feature Model: mean absolute SHAP scores, averaged across 5 folds. This shows the most important features for predicting both high and low trust (shown in combination, as they are mirrors of each other). Risk and benefit factors were the most important.}
        \Description{This figure shows the results of the Full Feature Model: mean absolute SHAP scores, averaged across 5 folds. This shows the most important features for predicting both high and low trust (shown in combination, as they are mirrors of each other). Risk and benefit factors were the most important.}
        \label{fig:full_shap_mean}
\end{figure*}

\begin{figure*}[t]
    \centering
    \vspace{2em}
        \includegraphics[width=.45\linewidth]{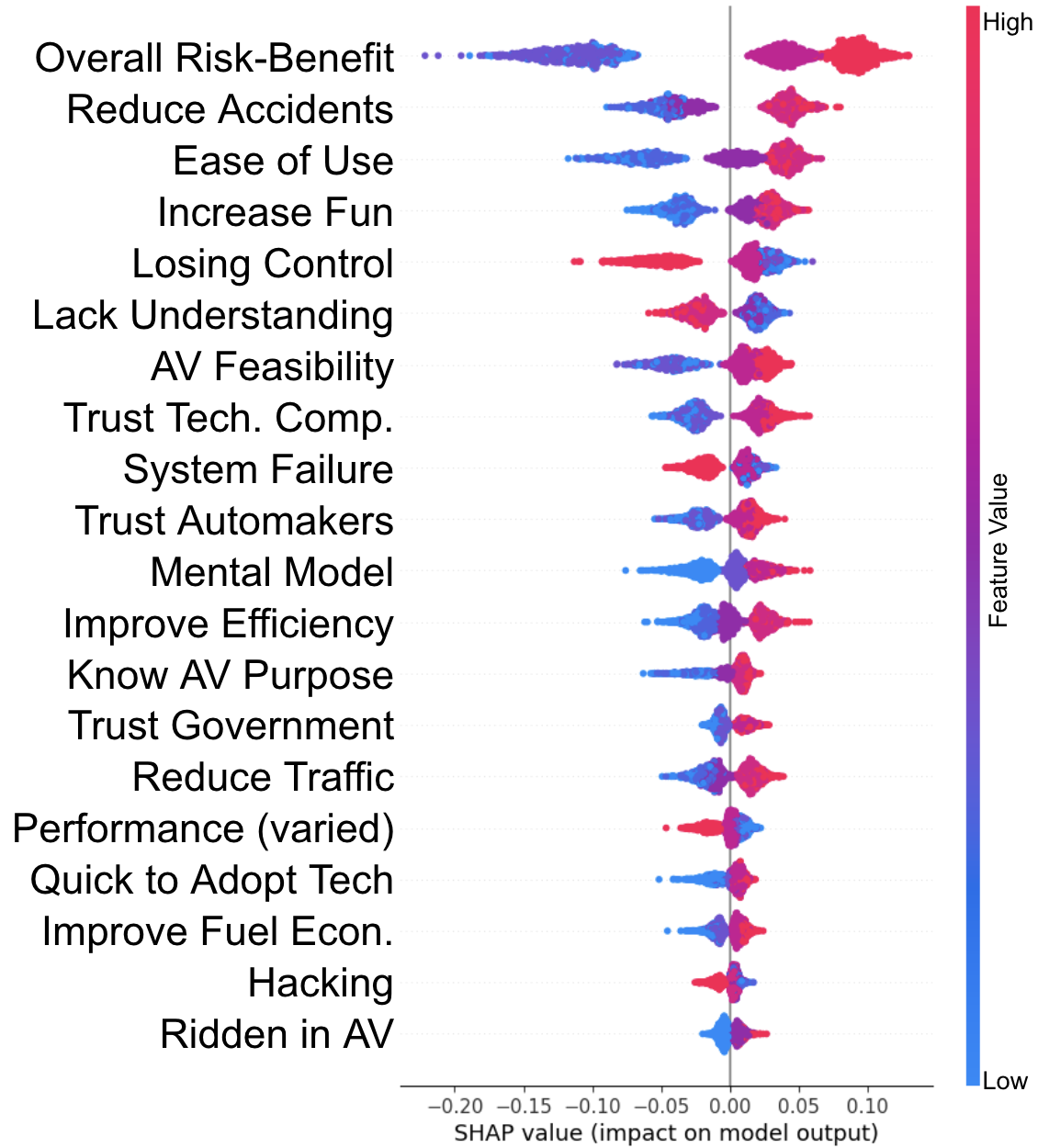}
        \vspace{-.5em}
        \caption{Full Feature Model SHAP summary plot showing feature value impact on trust. This plot shows how the \emph{value} of a feature impacts the model's output. The more extreme the SHAP value, the more indicative that value was of being high trust (positive SHAP values) or low trust (negative SHAP values). Features are organized by overall importance.}
        \Description{This figure shows the Full Feature Model SHAP summary plot, highlighting feature value impact on trust. This plot shows how the \emph{value} of a feature impacts the model's output. The more extreme the SHAP value, the more indicative that value was of being high trust (positive SHAP values) or low trust (negative SHAP values). Features are organized by overall importance.}
        \label{fig:fullset_shap}
\end{figure*}

Once the most predictive models were made, we identified the most important features via SHAP. We find that \texttt{Overall Risk-Benefit}, \texttt{Reduce Accidents}, \texttt{Ease of Use}, \texttt{Increase Fun} and \texttt{Losing Control} were the most important factors that influence trust in AVs (Fig.~\ref{fig:full_shap_mean}). Notably, 12 of the top 20 features were related to risk and benefit judgements. Fig.~\ref{fig:fullset_shap} shows how individual feature values contribute to the model's output. We observe that higher scores for \texttt{Overall Risk-Benefit}, \texttt{Reduce Accidents}, \texttt{Ease of Use}, and \texttt{Increase Fun} are positively associated with trust. \texttt{Losing Control} was negatively associated with trust.


\begin{table*}[t]
\caption{Summary of the three experiments on our dataset. The full feature set showed the best performance, followed closely by the feature subset (only risk and benefit features). The full feature set \emph{without} risks and benefits performed far worse.}
\Description{This table shows a summary of the three experiments on our dataset. The full feature set showed the best performance, followed closely by the feature subset (only risk and benefit features). The full feature set without risk and benefit features performed substantially worse.}
\renewcommand{\arraystretch}{1.3}
\begin{tabular}{lccc}
\toprule
\textbf{Models} & \textbf{Balanced Accuracy} & \textbf{F1 score} & \textbf{Net Accuracy Difference} \\
\midrule
Full Feature Set & 0.858 (0.02) & 0.858 (0.02) & -- \\
Risk + Benefits (\textit{feature subset})  & 0.846 (0.02) & 0.843 (0.02) & (-) 0.012 \\
Full Feature Set \textit{w/o} Risk + Benefits & 0.738 (0.02) & 0.738 (0.02) & (-) 0.120 \\
\bottomrule
\end{tabular}
\label{tab:ablation}
\end{table*}

\subsection{Ablation Studies: Systematic Isolation and Elimination of Risk-Benefit Factors}
Using our full feature model, risk and benefit-related factors were consistently among the most important for predicting a person's trust level. As such, we conducted two ablation (systematic elimination) studies to derive a deeper understanding of how risk and benefit factors specifically impact trust predictions. We followed a similar approach as we did while building and evaluating the full feature model (Section~\ref{full_set_results}). The first ablation study looks at only the risk and benefit factors from Table \ref{risk_benefit_factors} in isolation (henceforth referred to as the \emph{feature subset}). The second ablation study is this feature subset's complement, which instead \emph{eliminates} all risk-benefit factors from the full feature model. Table~\ref{tab:ablation} shows a summary of performance between these two ablation models compared to our full model.

We find that the full feature model outperforms both ablation models. The difference between the accuracy of the full feature model (85.8\%) and the full feature set without the risk and benefit factors (73.8\%) is quite large, dropping 12\%. We find much smaller differences between the full feature set and the isolated feature subset (84.6\%), losing only 1.2\% accuracy. These results imply that risk and benefit factors are very important for the prediction of trust in AVs.

\subsubsection{Feature Subset: Risk-Benefit Features Only}
The first ablation study, which we call the \emph{feature subset}, consisted only of risk and benefit measures. As discussed, the performance of this model was quite high. Investigating the top performing features in this subgroup using SHAP allows us to derive a better understanding of how different risks and benefits impacted our trust prediction.

\begin{figure*}[t]
        \centering
        \includegraphics[width=.9\linewidth]{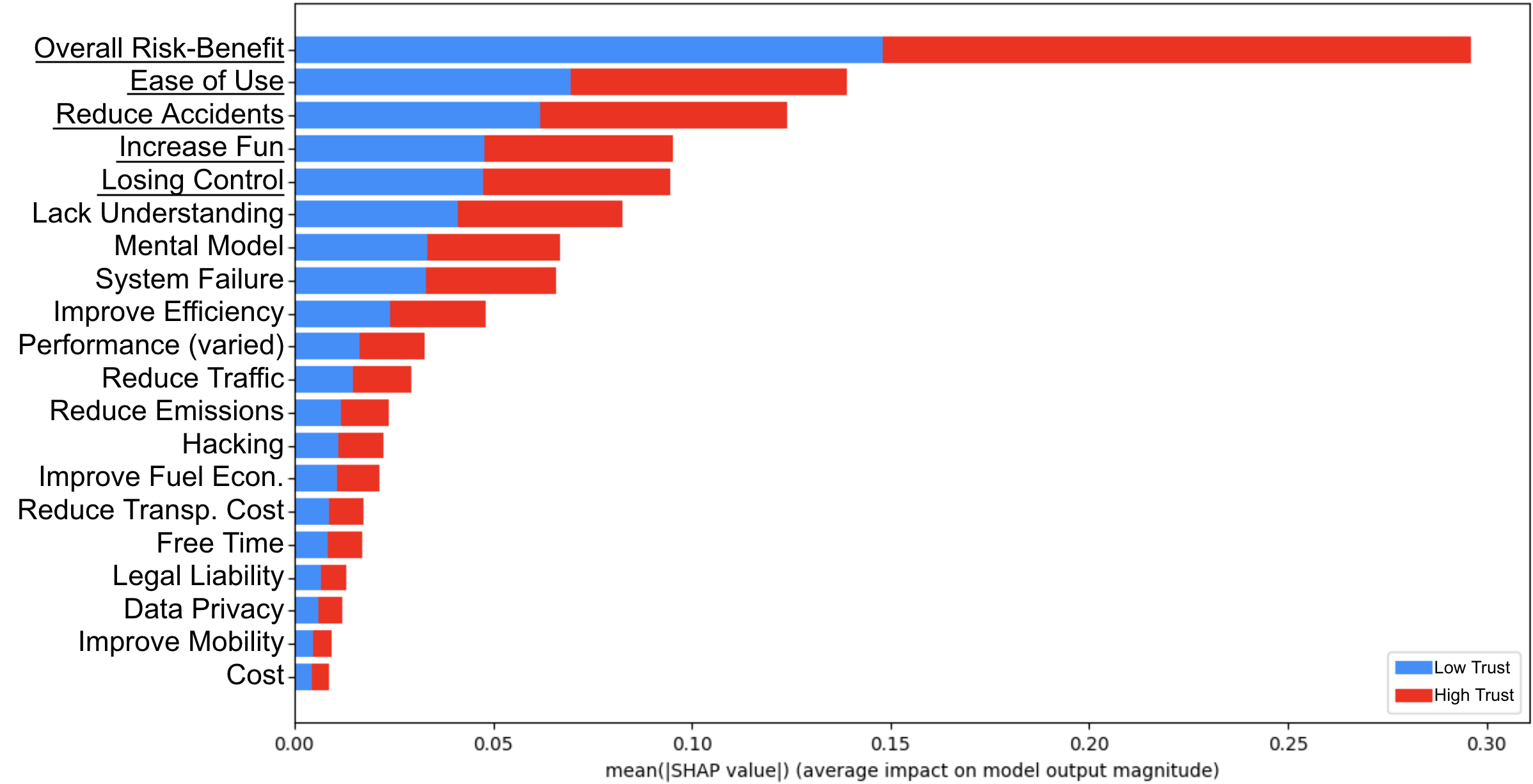}
        \vspace{-.5em}
        \caption{Feature Subset (risk and benefit only): mean absolute SHAP scores, averaged across 5 folds. This shows the most important features for predicting both high and low trust (shown in combination, as they are mirrors of each other). Results are similar to the full feature set model, supporting risk and benefit features as the most important predictors of trust.}
        \Description{This figure shows the results of the Feature Subset (risk and benefit only): mean absolute SHAP scores, averaged across 5 folds. This shows the most important features for predicting both high and low trust (shown in combination, as they are mirrors of each other). Results are similar to the full feature set model, supporting risk and benefit features as the most important predictors of trust.}
        \label{fig:ablation2_shap_mean}
\end{figure*}

\begin{figure*}[b]
        \centering
        \vspace{2em}
        \includegraphics[width=.45\linewidth]{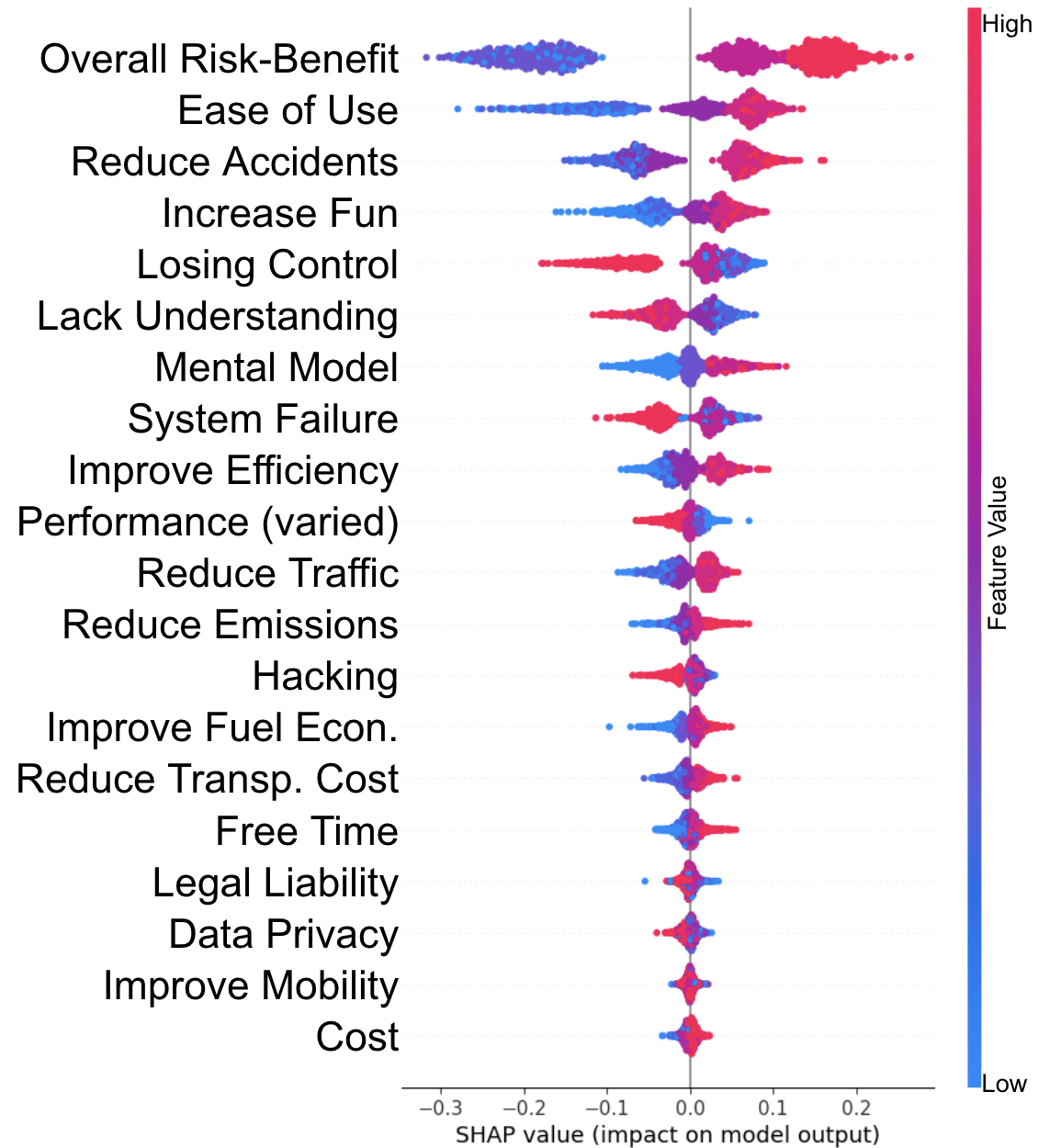}
        \vspace{-.5em}
        \caption{Feature Subset (risk and benefit only) SHAP summary plot showing feature value impact on trust. This plot shows how the \emph{value} of a feature impacts the model's output. The more extreme the SHAP value, the more indicative that value was of being high trust (positive SHAP values) or low trust (negative SHAP values). Features are organized by overall importance.}
        \Description{This figure shows the Feature Subset (risk and benefit only) SHAP summary plot, highlighting feature value impacts on trust. This plot shows how the value of a feature impacts the model's output. The more extreme the SHAP value, the more indicative that value was of being high trust (positive SHAP values) or low trust (negative SHAP values). Features are organized by overall importance.}
        \label{fig:ablation2_shap}
\end{figure*}

\begin{figure*}[t]
    \centering
        \centering    
        \includegraphics[width=.9\linewidth]{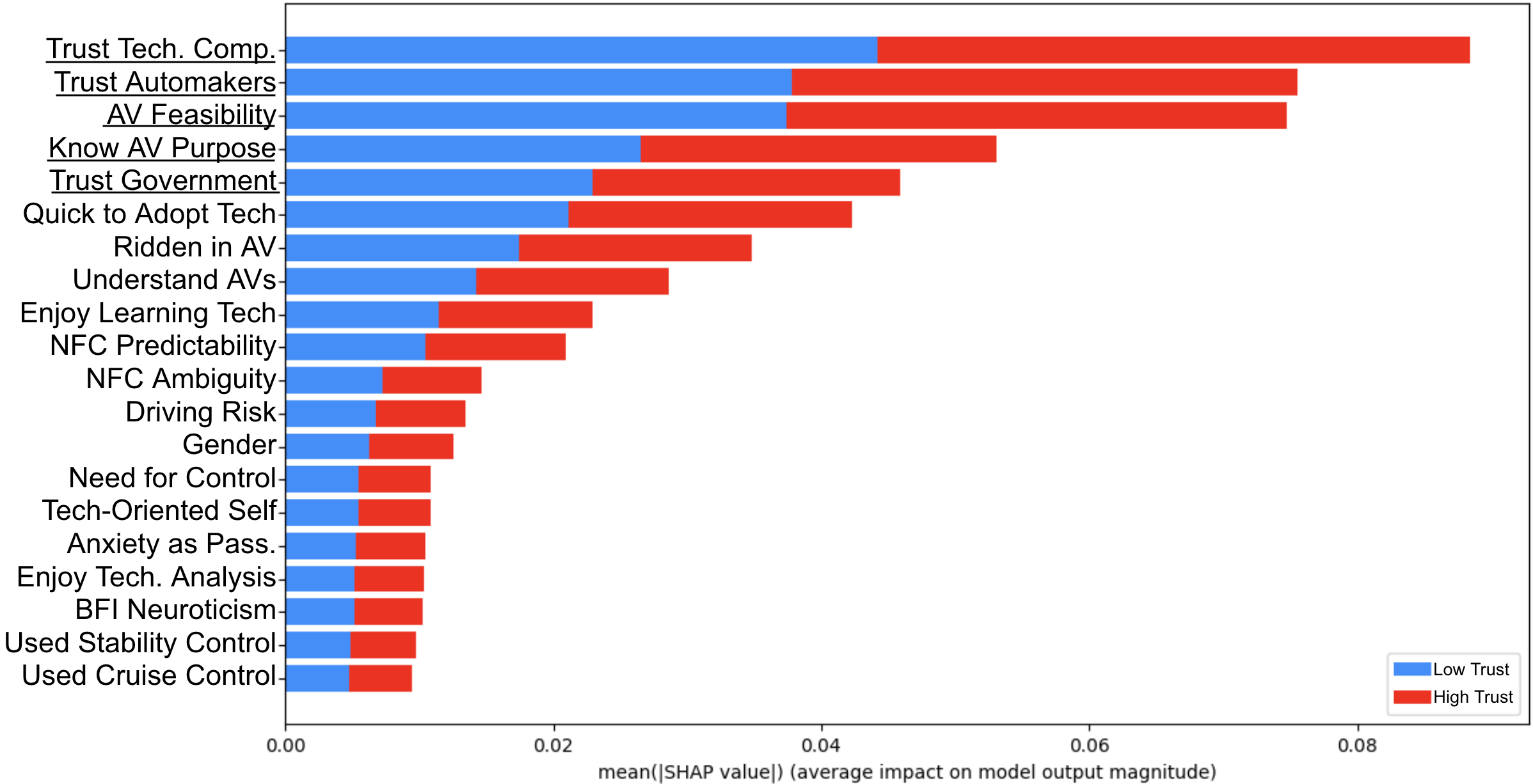}
        \caption{Full Feature Model \emph{without} risk and benefit factors: mean absolute SHAP scores, averaged across 5 folds. This shows the most important features for predicting both high and low trust (shown in combination, as they are mirrors of each other). Without risks and benefits, factors such as trust in institutions, AV feasibility, knowledge, and experience were the most important predictors of trust.}
        \Description{This figure shows the results of the Full Feature Model without risk and benefit factors: mean absolute SHAP scores, averaged across 5 folds. This shows the most important features for predicting both high and low trust (shown in combination, as they are mirrors of each other). Without risks and benefits, factors such as trust in institutions, AV feasibility, knowledge, and experience were the most important predictors of trust.}
        \label{fig:ablation1_shap_mean}
\end{figure*}

\begin{figure*}[b]
        \centering
        \includegraphics[width=.45\linewidth]{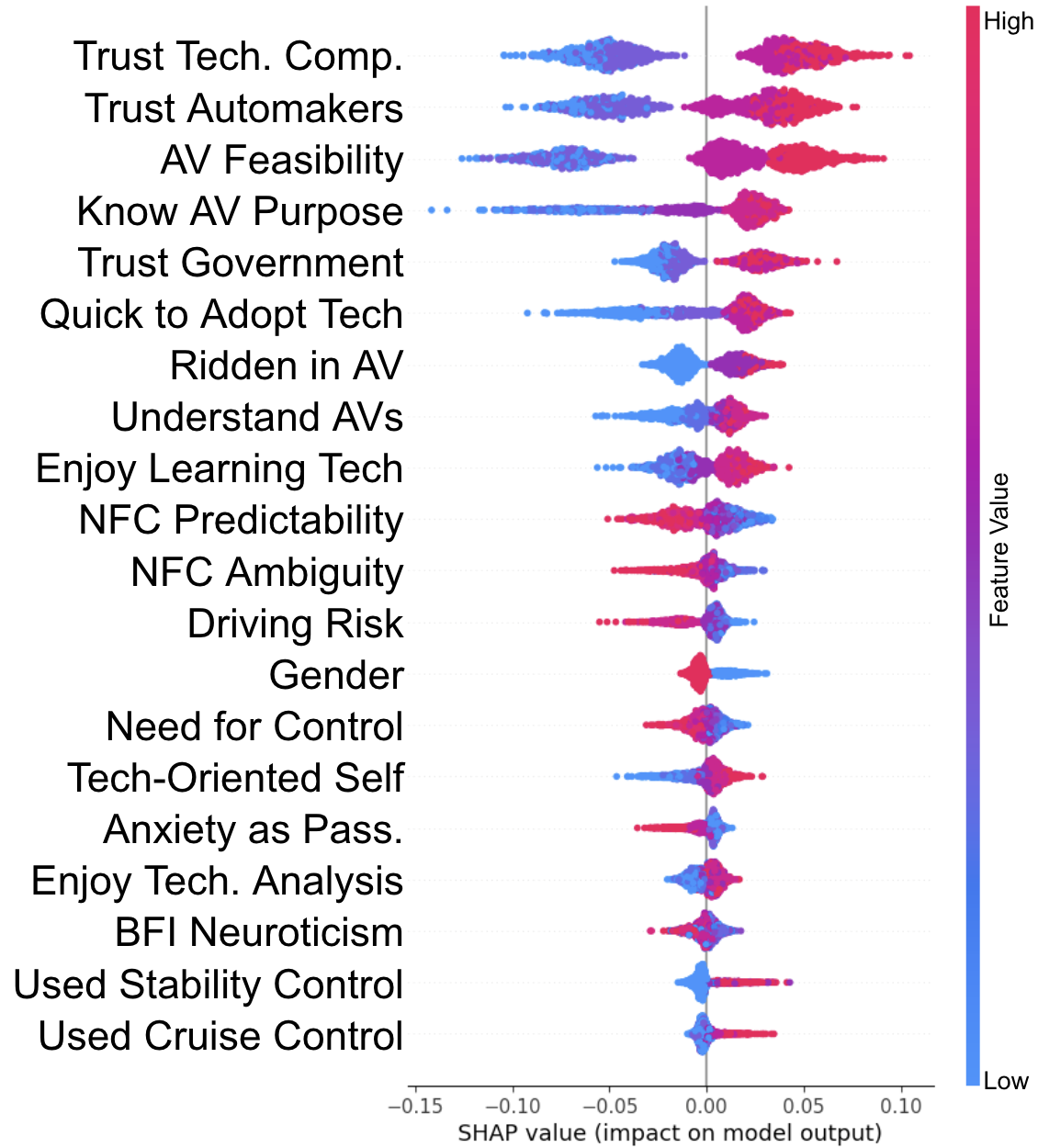}
        \caption{Full Feature Model \emph{without} risk/benefit factors: SHAP summary plot showing feature value impact on trust. This plot shows how the \emph{value} of a feature impacts the model's output. The more extreme the SHAP value, the more indicative that value was of being high trust (positive SHAP values) or low trust (negative SHAP values). Features are organized by importance.}
        \Description{This figure shows the Full Feature Model without risk/benefit factors: SHAP summary plot, highlighting feature value impact on trust. This plot shows how the value of a feature impacts the model's output. The more extreme the SHAP value, the more indicative that value was of being high trust (positive SHAP values) or low trust (negative SHAP values). Features are organized by overall importance.}
        \label{fig:ablation1_shap}
\end{figure*}

Fig.~\ref{fig:ablation2_shap_mean} shows the top contributing factors in terms of feature importance via mean absolute contribution, averaged across five folds. Fig.~\ref{fig:ablation2_shap} illustrates how different values of a feature predict that a person will be either high or low trust. We find \texttt{Overall Risk-Benefit}, \texttt{Ease of Use}, \texttt{Reduce Accidents}, \texttt{Increase Fun} and \texttt{Losing Control} are the most impactful features, with high scores in all showing a positive association with trust except \texttt{Losing Control}. This is near identical to the full feature model, with only slight differences in relative importance. The implication of this ablation study is that -- even when removing all demographic, psychosocial, driving, and tech-related factors -- factors specific to the risk-benefit evaluation of AVs can still predict trust quite well on their own.


\subsubsection{Full Model Without Risk-Benefit Features}
\label{ablation2section}
We also investigated the effect of eliminating the \emph{feature subset} from the full feature model. This second ablation experiment gives us insight into two additional areas of inquiry. First, if we do not have access to a person's risk-benefit assessment, can we still predict trust from other personal factors? Given the plethora of prior research on dispositional traits and attitudes discussed in Section \ref{prior_personal}, we hypothesized that this answer should be \emph{yes}. Second, \emph{if} we can predict trust from other personal factors, what factors are most important? 

To answer these questions, we ran experiments similar to the development of other models in this study, to find the best possible performance with this group of features. Unlike the full feature model and the feature subset (ablation experiment 1), we found that balanced model accuracy of this second ablation study was quite a bit lower, dropping to 73.8\%, a full 12\% dip compared to the full feature model. This further emphasizes the importance of risk-benefit features.

Though 73.8\% is not excellent predictive accuracy, from a theoretical lens it is still \emph{predictive enough} to draw initial insights on how non-risk-benefit features impact a person's judgement of high and low trust. We found that \texttt{Trust in Tech Companies}, \texttt{Trust in Automakers}, \texttt{AV Feasibility}, \texttt{Know AV Purpose}, and \texttt{Trust in Government Authorities} were the strongest predictors (Fig. \ref{fig:ablation1_shap_mean}); higher values of each were associated with higher trust (Fig. \ref{fig:ablation1_shap}).

 

\section{Discussion}
We provide evidence that young adult trust in autonomous vehicles (AVs) can be predicted via machine learning. We used a comprehensive set of personal factors with 130 distinct input features, derived from survey. We compared the predictive power of several machine learning models, all of which performed quite well. Random forest models performed best -- 85.8\% (Table \ref{tab:model_main}) -- at categorizing people into high and low trust groups. By applying the explainable AI technique SHAP to our best performing models, our analysis not only shows \emph{which} factors were most predictive of trust, but also \emph{how} each feature uniquely contributed to the model's conclusion. In the following sections, we contextualize our findings within the scope of human-AV interaction research and provide insights for how our analysis can be used to build trustworthy vehicles for diverse groups. Many of these insights may generalize to other AI-driven domains or for groups beyond young adults.

A general principle that resulted from our study is that \emph{perceptions of risks and benefits} are the most important factors for predicting trust. \color{black}The importance of risks and benefit evaluation for trust and acceptance has been highlighted in the past, especially for high-stakes contexts like AVs~\cite{ayoub2021modeling, hewitt2019assessing, zhang2023human}. Unlike prior studies, however, our analysis allowed us to isolate \emph{specific} types of AV-related risks and benefits, such as individual concerns over system failures and usability. As such, the present study is the first to articulate the \emph{relative importance} of these risk and benefit perceptions compared to other types of personal factors for predicting trust. \color{black} In our full feature model, 12 of the 20 most important features (based on mean SHAP value) were related to a person's assessment of the risks and benefits of using AVs. Using ablation studies, the importance of risk-benefit perceptions was confirmed: risk and benefit factors alone had similar predictive ability as our full model, whereas a knockout model \emph{without} risk and benefit factors performed far worse.  

We find a person's overall assessment of the risks and benefits of an AV was the most highly predictive feature for predicting trust level (importance rank 1) in both the full feature model and feature subset. This is unsurprising, as we would assume that the ratio of benefits to risks -- as discussed at length in cost-benefit analyses and (ir)rational decision-making~\cite{wang2019designing} -- would impact a person's decision to trust an AV. Indeed, \citet{ayoub2021modeling} also found that assessment of benefits and risks were the most important factors predicting trust in their model, though they do not specify which risks and benefits these may be. In the present study, we are able to articulate precisely \emph{which} risk-benefit factors are most important. From an AV design perspective, the importance of overall risk-benefit assessment on trust judgements implies that addressing high priority risks and benefits using strategic design elements can drastically impact a person's trust judgements. In the following sections, we will discuss risks and benefits, respectively.

\vspace{.5em}
\noindent
\textbf{Risks associated with system performance and usability failures were the most consequential ---}
Six of the top 20 most important features to our full model were related to risks. These included concerns with: Ease of Use (rank 3), Losing Control (5), Lacking Understanding of AV decision-making (6), System Failures (9), Performance of AVs across varied terrain (16), and Hacking (19). High ratings of these concerns were associated with lower trust. 

These concerns highlight areas of priority for human-AV interaction design and research. For instance, concerns over ease of use, giving up control, and lack of decision-understanding are fundamental to the development of explainable AI systems for AVs~\cite{koo2015did, miller2019explanation}, which aim to increase transparency and usability with complex AI-based systems. Other specific concerns with vehicle performance and security -- such as concern with equipment failures, performance, and hacking -- may also be addressed through a mix of UI enhancements and education campaigns (assuming that the systems are, in fact, performing well and secure). These may include visualizations reassuring riders that the vehicle is safe, private, and operating within the bounds of its ability. Importantly, elements should be designed to empower a rider by providing decision agency, not to mislead them by artificially minimizing risks.

\vspace{.5em}
\noindent
\textbf{Benefits need to add value beyond what can be achieved by human-controlled driving ---}
Five of the top 20 features were related to benefits. Perceiving that AVs can Reduce Accidents (rank 2), Increase the Fun of driving (4), Improve Efficiency (12), Reduce Traffic (15), and Improve Fuel Economy (18) were all associated with higher trust. In all cases, these are benefits that move beyond what is possible with current human-controlled driving and vehicle infrastructure. Our findings support prior work that trust and adoption of AVs are likely tied closely together \cite{hewitt2019assessing}, meaning that AVs will only be opted-for when their benefits far outpace what is possible by the current status quo of non-autonomously driven vehicles (as well as the risks of AVs discussed previously).

For future AV design and research, these findings suggest that communicating the unique benefits of AVs, particularly those that exceed the capabilities of human-controlled driving, is crucial for building trust. This emphasis should be coupled with strategies to mitigate perceived risks, ensuring that the value proposition of AVs is compelling enough to encourage widespread adoption. Design elements highlighting specific elements -- such as showing fuel savings or estimations of emission reduction -- can help build confidence in potential riders.

\vspace{.5em}
\noindent
\textbf{Attitudes towards AV feasibility, affinity for technology, and institutional trust helped to predict trust, but to a smaller degree than risk and benefit perceptions ---} Though as a whole less important than risk and benefit factors, we found that certain perceptions towards AVs, technology, and institutions as a whole helped predict trust. Unsurprisingly, we found that people who considered AVs more practically Feasible (rank 7), understood the Purpose of AVs (13), and who are generally Quick to Adopt New Technology (17) were more likely to trust AVs. These make sense intuitively; to trust AVs, a person would need to understand why AVs exist and believe that the technology is realistic. Given that widespread adoption has yet to be accomplished, it also makes sense that people who are quick to adopt new technologies would be more likely to trust AVs, as these people may be interested in new technologies -- like AVs -- and may have a higher tolerance for uncertainty regarding use of less established systems. This latter finding aligns with prior work on affinity for technology predicting AV trust~\cite{kraus2021s}.

Interestingly, all three trust in institutions measures were important predictors of trust. Trust in Tech Companies ranked 8 in importance, Trust in Automakers ranked 10, and Trust in Government Authorities ranked 14. We provide two potential explanations for these findings. First, many people likely recognize that AVs will function well \emph{if and only if} the groups that produce and regulate them function well. This strengthens the importance of brand reputation, regulation transparency, and thorough testing criteria. Effects may also be explained by differences between people who are generally more or less trusting as a dispositional characteristic: for some people, trust may simply be easier than for others~\cite{hancock2023and}. Special care should be taken by designers to reach individuals who do not trust easily. In these cases, we would expect heightened sensitivity to errors during use~\cite{luo2020trust} and a higher benefit to risk ratio necessary for adoption. We will discuss specific design implications -- including what these mean for future AV policy -- in Section \ref{design_implications}.

\vspace{.5em}
\noindent
\textbf{Psychosocial and many driving and technology-related factors were far less important than expected ---}
In Section \ref{prior_personal}, we discussed a plethora of prior work suggesting that psychosocial factors -- including personality, self-esteem, risk perception, and need for control -- will impact a person's trust judgement for AVs. \color{black} Our results generally contrast with these prior studies, as many of these factors did not appear to be important for our model's prediction of trust. \color{black} Further, exempting the few factors noted above, technology and driving factors -- including driving style, driving cognitions, and technology self-efficacy -- also showed little impact. 

There are a few explanations for this discrepancy. First, it is possible that the risk-benefit factors and others previously noted were fully sufficient to predict trust on their own, and thus less predictive (but still potentially relevant) factors were prioritized less by our ML models. If this were true, we would likely still see effects in our traditional high-low trust group comparison using Kruskal-Wallis tests. According to the descriptive analysis in \ref{sec:descriptive} and continued in Appendix \ref{appendixA.trust}, however, not a single one of these factors showed large or even medium effect size differences comparing the high and low trust groups. Some -- including certain prior experiences, perceptions of driving risk, Need For Closure (Predictability), and some technology self-efficacy measures -- showed small effects. The rest showed negligible effects. Even in our second ablation study (knocking out risk-benefit factors, Section \ref{ablation2section}), it is not until the tenth most important factor that we start to see psychosocial or demographic variables: Need for Closure Predictability and Ambiguity (importance ranks 10 and 11, respectively), driving risk assessment (12), gender (13), need for control (14), etc. Not a single driving style or driving cognitions factor appeared in the top 20, even with ablations.

As such, our results generally conflict with past work suggesting that these psychosocial, driving, and tech factors may be \emph{important} predictors of AV trust. This is not to say that they should be omitted from future AV trust research, nor that the relationships found in prior work are false. Instead, our results suggest that other, more important factors should be prioritized in future AV trust research and design for our specific subpopulation of interest. With less data available to a research team (such as without risks/benefits) or for more general populations of study, these factors may become more important. Further, psychosocial and driving-related factors may still inform other areas of AV design and research, such as cases where personalization based on personality and other traits have improved interactions~\cite{alves2020incorporating}.

\vspace{.5em}
\noindent
\textbf{Similarity between AV and human decision-making was positively associated with trust ---}
Interestingly, we found that people who viewed AV decision-making as more similar to human decision-making (Mental Model, ranked 10th in importance) were more likely to trust AVs. This suggests that a person's mental model of AV decision-making plays a key role in their trust. When AV decision-making aligns more closely with how they perceive human decision-making, trust tends to increase. From a design perspective, this implies that making AV interactions more human-like, with communication that is easy for people to understand, could help bolster trust. Our results support current efforts to make AI explanations more similar to human explanations, particularly those given by experts~\cite{miller2019explanation, pazzani2022expert, kaufman2022cognitive, koo2015did}.

\vspace{.5em}
\noindent
\textbf{Isolating only the subpopulation of young adults allowed us to uncover deeper predictors of trust, and can be used as a methodological template for the study of future subpopulations ---}
By accounting for specific characteristics of the subpopulation, such as age or education, we focused this study on young adults -- the next generation of AV adopters. While age and education have been used as surface-level predictors of trust in previous research, isolating this group allowed us to explore deeper, more meaningful factors that are particularly relevant to this subpopulation. Specifically, instead of replicating prior findings on the importance of age~\cite{hulse2018perceptions, abraham2017autonomous} or education~\cite{hudson2019people}, our models based their predictions on potentially more important characteristics, like their risk-benefit assessment and attitudes towards AVs. The one exception in our findings was gender: similar to prior work by \citet{hulse2018perceptions} and \citet{mosaferchi2023personality}, our findings also suggest that those who identify as women are less likely to trust AVs than those who do not identify as women. This was not an important factor for our full model, but did show moderate importance for the second ablation study. As such, our findings support the premise that focusing on subpopulations may be an effective (though more time intensive) strategy for deriving deeper insights about a particular group of interest. In the present case, our work provides a deeper insight into designing for young adults, and may serve as a methodological template for future research on other subpopulations.

\vspace{.5em}
\noindent
\textbf{Prior AV experience matters, but less than what prior work suggests ---} At the very last rank of our top 20, we find that people who have prior experience riding in an AV were more likely to trust AVs. This aligns with prior work by \citet{choi2015investigating} and \citet{mosaferchi2023personality}, who also found a positive association between prior experience and trust. Prior experience was ranked number 20 in terms of importance, however, meaning that there are many features that were better predictors of trust. This has important implications, as AV companies like Waymo~\cite{WaymoOne} often claim that people will trust AVs as soon as they gain experience with them. Though this may be true to a degree, our findings show that other factors should be prioritized for AV design and research focused on building trust and increasing adoption.

\subsection{Implications for Trustworthy Human-AV Interaction Design and Research}
\label{design_implications}
Our results have important implications for the design and study of autonomous vehicles. In this study, we successfully predicted AV trust from the personal factors of young adults. As expected, some features in our model were more important than others. We can use these results as a basis for personalized and adaptable designs based on specific factors of importance, as well as specific insights for trustworthy Explainable AI (XAI) design, communication and education campaign guidelines, and policy suggestions.

\vspace{.5em}
\noindent
\textbf{Explainable AI and Transparency ---}
The high importance of concerns around ease of use, giving up control, and lack of understanding around AV decision-making highlights the critical need for continued research and design efforts on AV explainability and explainable AI (XAI) more generally. \color{black} Our results echo prior work on XAI, emphasizing the necessity for explainability contexts where transparency can mitigate user concerns over safety and control (see Section \ref{prior_personal}). For example, these results support the design and study of clear, real-time explanations for AV decisions. \color{black} A special focus should be put on XAI explanations that are understandable by riders from diverse communities with a wide range of backgrounds and expertise levels. Failure to satisfy concerns over ease of use, vehicle performance, and safety and privacy may result in refusal to ride with an AV.

\vspace{.5em}
\noindent
\textbf{Prioritize Safety and Security Features ---}
One area of focus for XAI in AVs should be on effectively communicating driving performance metrics, particularly when they can reassure a rider that the AV is safe and protected from both physical harm and privacy risks. Explanations for \emph{what} action an AV is taking and {why}~\cite{koo2015did} may give a sense of control and agency back to a driver. These may be effectively paired with visualizations emphasizing that the AV is performing properly and trained well for the driving environment, mitigating concerns over performance adaptability and calibrating trust to an appropriate level. Visualizations and communications should be realistic and not misleading, as under-communicating performance errors could lead to over-reliance.

\vspace{.5em}
\noindent
\textbf{Communicate Unique Benefits Effectively ---}
Given the large number of potential benefits that AVs offer~\cite{fagnant2015preparing} and the importance of \emph{believing} in these benefits to trust judgements, our research suggests that educational campaigns and mid-drive interactions should strive to highlight benefits to increase trust and adoption. In particular, emphasizing the \emph{added value} AVs can bring \emph{beyond} human-controlled driving may be most effective, as the alternative to AV adoption are non-AVs (status quo) which still function quite well in many contexts. Specific benefits which should be prioritized include reducing crashes and traffic, increasing driving efficiency, and increasing fun.

\vspace{.5em}
\noindent\textbf{Anthropomorphizing AV communications may be beneficial, but should be done with caution ---}
We found that people who conceptualized AV decision-making as closer to human decision-making had higher AV trust. This alludes to the power that humanizing AV communications may have on increasing trust, suggesting similar approaches to AV explanations could be potentially useful. For instance, expert-informed explainable AI has been studied in the past with promising results~\cite{bayer2022role, pazzani2022expert, soltani2022user, kaufman2023explainable}. This approach should be taken with caution, however, as certain theoretical frameworks such as the Computers Are Social Actors (CASA) paradigm~\cite{nass1994computers} show that people often treat computerized systems the way they would treat other humans, meaning systems which are too humanized may be assumed to have human-like biases and flaws in judgement and decision-making.

\vspace{.5em}
\noindent
\textbf{Personalization and tailoring of AV communications and behavior may be a promising future direction ---}
Key traits found in our current study and in prior work emphasizes that \emph{people differ} in their underlying traits, attitudes, and experiences, and that these differences impact trust. In the present study, we highlight some of the most important traits that should be looked at for predicting a young adult's trust in AVs. This work can be used as a foundational step towards tailoring AV communications and behaviors to the specific individual who may be riding in the AV. Once the most important traits have been identified for a specific subpopulation of interest, design elements can be tailored and tested specifically for building trust for that group. By pairing a certain “trait profile” with certain design and explanation choices (e.g. different amounts, types, or modalities of information), an AV may better meet the needs of different types of individuals. For example, groups with certain concerns or backgrounds can be shown certain types of information in a display, while groups with \emph{other} concerns or backgrounds can be shown \emph{other} UI or explanation elements. If done correctly, tailoring can be done automatically based on who is interacting with the AV.

\vspace{.5em}
\noindent
\textbf{\color{black}Implications for Young Adults and Generalizability to Other Populations ---}
\color{black}Though this study focused on young adults, it is likely that many insights may generalize to broader populations of interest. Certain features -- such as ease of use, transparency, and institutional trust -- were likely predictive for young adults due to their high exposure to and reliance on technology in daily life. As a generation that has grown up with digital and AI-based tools, prior work has demonstrated that young adults are more accustomed to the presence of autonomous systems and may have greater baseline expectations around usability, control, and information transparency \cite{anderson2018future}. Moreover, young adults may rely on a combination of brand reputation, ethical standards, and transparency practices when evaluating AV trustworthiness \cite{twenge2014declines}, potentially explaining why institutional trust emerged as a strong predictor for this group. Focusing on young adults in isolation may also explain the lack of importance of psychosocial, driving, and technology-related factors compared to prior studies on more general populations. Future work segmenting and comparing population clusters could help clarify whether this reduced importance of psychosocial factors is specific to young adults, or if it reflects limitations in the generalizability of prior work over diverse user groups. Though these are potential explanations for why certain findings in this study were found \textit{specifically} for a young adult population, preference for usability, transparency, as well as many risk and benefit factors are likely important to a wide number of populations. Though future research is necessary to support this claim, we have no reason to suspect that concerns over safety or losing control, for example, would only be relevant to young adults. However, it is probable that different predictors (or at least, different relative importance of predictors) would emerge for other age groups or cultural contexts \cite{hancock2011meta}. \color{black}

\vspace{.5em}
\noindent
\textbf{Policy, Regulation, and \color{black}Ethical \color{black} Implications ---}
Given the importance of institutional trust (in technology companies, automakers, and government agencies), our results suggest that clear and consistent regulatory standards are crucial for building public trust in AVs. Policymakers should focus on establishing transparent safety protocols, rigorous testing standards, certification processes, and data sharing policies that are easy for the public to understand. Clear regulations can reassure the public that AVs meet high safety standards and are regularly audited for compliance. These are similar to the recommendations made by \citet{fagnant2015preparing}, though current realization of these recommendations leaves much to be desired in this regard. Our work highlights \emph{transparency} in regulation as a way to increase trust, particularly for cases when a person has low trust in institutions in general. Broader regulation focused on the most important areas of risk discussed previously -- vehicle performance, usability, and security -- should be prioritized.

\color{black}
The development of industry guidelines and regulatory oversight can be a practical way to promote the ethical application and use of trust insights for AV technology. Understanding the individual needs of potential AV users can allow equitable adoption of and access to AV technology. While our research aims to improve user-centered design by identifying factors that predict trust, insights on specific needs could be used for targeted marketing without parallel investments in AV safety and reliability. Such practices could result in users adopting AVs based on trust-building marketing tactics rather than well-informed, safety-centered choices. To prevent misuse, we recommend responsible application of trust-predictive insights, emphasizing transparency, user autonomy, and informed choice. Beyond institutional regulation, companies should ensure that trust-building efforts go hand-in-hand with clear and accessible communication about AV capabilities and limitations. For example, AV manufacturers could provide detailed information on independent safety evaluations, test outcomes, and known limitations of their systems, such as performance in inclement weather or in complex urban settings. This transparency helps users form realistic mental models of AVs, enabling them to trust based on accurate expectations rather than solely on positive marketing.\color{black} 

\subsection{Limitations and Future Work}
This study was not without limitations. First, our study leveraged survey methods, which -- though standard practice -- may be subject to bias as it is based on self-reported data. To mitigate this concern, we employed a large sample size, used validated questionnaires, and replicated variable selection from prior work whenever possible. Self-report concern may be particularly strong with measurements of trust, as (though quite common) self-reported trust may not be as effective as measuring trust via behavioral reliance~\cite{hoff2015trust}. To mitigate this risk, we used a composite trust measure to capture many aspects of a trust judgement. \color{black}Variability in participants' familiarity with AV technology may introduce latent confounds, as those with more exposure might respond differently to questions about AV trust, risks, or benefits. However, we believe this limitation is minimized by focusing on a clear, every day commuting context, which provided a shared baseline for interpretation. This range in familiarity also reflects the diversity of real-world users, enhancing the generalizability of our findings. Future work may segment participants by familiarity levels to help control for baseline knowledge differences. Likewise, we recognize that institutional trust measures like “trust in tech companies” may be interpreted as broad constructs, encompassing varying levels of trust that could be influenced by specific companies or historical practices. While these widely used variables capture overarching attitudes toward technology and institutions, which are valid for the predictive task in this study as they reflect general perceptions relevant to user trust in AVs, future work could benefit from a more nuanced approach to provide clearer insights. For example, exploring sub-features within each institutional construct could help differentiate the effects of brand or institutional identity, transparency practices, or innovation history. \color{black} Future work may seek to use behavioral trust measures. We focused on young adults, a majority of whom were enrolled at an undergraduate institution in the United States. As such, results presented in this study may not widely generalize. Future work should replicate on other populations if results in other specific settings are sought. Finally, testing the design implications through actual implementation of the guidelines outlined in Section \ref{design_implications} will be necessary to assure their validity.

\section{Conclusion}
In this study, we predicted young adult trust in autonomous vehicles from a comprehensive set of personal factors. Using machine learning and explainablity techniques, we uncovered \emph{which} factors were most predictive of trust, and \emph{how} each feature uniquely contributed to the model's conclusion. We found that perceptions of AV risks and benefits, attitudes toward feasibility and usability, institutional trust, and prior AV experience were the most important trust predictors. A person's mental model also played a role, where those who believed AV decision-making as closer to human decision-making were more trusting. Contrary to previous findings, psychosocial factors -- including personality, self-esteem, risk perception, and need for control -- as well as most technology and driving-specific factors, including driving style, driving cognitions, and technology self-efficacy -- were not strong predictors of trust. Our results build a new understanding on how personal factors can be used to predict trust in AVs. We conclude our discussion with a set of design, research, and policy implications that can be used to improve future trust and adoption of AVs that can meet the needs of diverse user bases.

\begin{acks}
We would like to thank Rohan Bhide, Saumitra Sapre, Chloe Lee, and Janzen Molina for their help disseminating the survey. A special thanks to David Danks for his advice on modeling methods and to the Discovery Way Foundation for their generous usage of lab space.
\end{acks}

\newpage
\bibliographystyle{ACM-Reference-Format}
\bibliography{bibliography}


\begin{thebibliography}{112}


\ifx \showCODEN    \undefined \def \showCODEN     #1{\unskip}     \fi
\ifx \showDOI      \undefined \def \showDOI       #1{#1}\fi
\ifx \showISBNx    \undefined \def \showISBNx     #1{\unskip}     \fi
\ifx \showISBNxiii \undefined \def \showISBNxiii  #1{\unskip}     \fi
\ifx \showISSN     \undefined \def \showISSN      #1{\unskip}     \fi
\ifx \showLCCN     \undefined \def \showLCCN      #1{\unskip}     \fi
\ifx \shownote     \undefined \def \shownote      #1{#1}          \fi
\ifx \showarticletitle \undefined \def \showarticletitle #1{#1}   \fi
\ifx \showURL      \undefined \def \showURL       {\relax}        \fi
\providecommand\bibfield[2]{#2}
\providecommand\bibinfo[2]{#2}
\providecommand\natexlab[1]{#1}
\providecommand\showeprint[2][]{arXiv:#2}

\bibitem[Abraham et~al\mbox{.}(2017a)]%
        {abraham2017autonomous}
\bibfield{author}{\bibinfo{person}{Hillary Abraham}, \bibinfo{person}{Chaiwoo Lee}, \bibinfo{person}{Samantha Brady}, \bibinfo{person}{Craig Fitzgerald}, \bibinfo{person}{Bruce Mehler}, \bibinfo{person}{Bryan Reimer}, {and} \bibinfo{person}{Joseph~F Coughlin}.} \bibinfo{year}{2017}\natexlab{a}.
\newblock \showarticletitle{Autonomous vehicles and alternatives to driving: trust, preferences, and effects of age}. In \bibinfo{booktitle}{\emph{Proceedings of the transportation research board 96th annual meeting}}. Transportation Research Board Washington, DC, \bibinfo{pages}{8--12}.
\newblock


\bibitem[Abraham et~al\mbox{.}(2017b)]%
        {abraham2017case}
\bibfield{author}{\bibinfo{person}{Hillary Abraham}, \bibinfo{person}{Hale McAnulty}, \bibinfo{person}{Bruce Mehler}, {and} \bibinfo{person}{Bryan Reimer}.} \bibinfo{year}{2017}\natexlab{b}.
\newblock \showarticletitle{Case study of today’s automotive dealerships: Introduction and delivery of advanced driver assistance systems}.
\newblock \bibinfo{journal}{\emph{Transportation research record}} \bibinfo{volume}{2660}, \bibinfo{number}{1} (\bibinfo{year}{2017}), \bibinfo{pages}{7--14}.
\newblock


\bibitem[Acheampong and Cugurullo(2019)]%
        {acheampong2019capturing}
\bibfield{author}{\bibinfo{person}{Ransford~A Acheampong} {and} \bibinfo{person}{Federico Cugurullo}.} \bibinfo{year}{2019}\natexlab{}.
\newblock \showarticletitle{Capturing the behavioural determinants behind the adoption of autonomous vehicles: Conceptual frameworks and measurement models to predict public transport, sharing and ownership trends of self-driving cars}.
\newblock \bibinfo{journal}{\emph{Transportation research part F: traffic psychology and behaviour}}  \bibinfo{volume}{62} (\bibinfo{year}{2019}), \bibinfo{pages}{349--375}.
\newblock


\bibitem[Ajenaghughrure et~al\mbox{.}(2020)]%
        {ajenaghughrure2020risk}
\bibfield{author}{\bibinfo{person}{Ighoyota~Ben Ajenaghughrure}, \bibinfo{person}{Sonia~Claudia da Costa~Sousa}, {and} \bibinfo{person}{David Lamas}.} \bibinfo{year}{2020}\natexlab{}.
\newblock \showarticletitle{Risk and Trust in artificial intelligence technologies: A case study of Autonomous Vehicles}. In \bibinfo{booktitle}{\emph{2020 13th international conference on human system interaction (HSI)}}. IEEE, \bibinfo{pages}{118--123}.
\newblock


\bibitem[Akash et~al\mbox{.}(2018)]%
        {akash2018classification}
\bibfield{author}{\bibinfo{person}{Kumar Akash}, \bibinfo{person}{Wan-Lin Hu}, \bibinfo{person}{Neera Jain}, {and} \bibinfo{person}{Tahira Reid}.} \bibinfo{year}{2018}\natexlab{}.
\newblock \showarticletitle{A classification model for sensing human trust in machines using EEG and GSR}.
\newblock \bibinfo{journal}{\emph{ACM Transactions on Interactive Intelligent Systems (TiiS)}} \bibinfo{volume}{8}, \bibinfo{number}{4} (\bibinfo{year}{2018}), \bibinfo{pages}{1--20}.
\newblock


\bibitem[Alves et~al\mbox{.}(2020)]%
        {alves2020incorporating}
\bibfield{author}{\bibinfo{person}{Tom{\'a}s Alves}, \bibinfo{person}{Joana Nat{\'a}lio}, \bibinfo{person}{Joana Henriques-Calado}, {and} \bibinfo{person}{Sandra Gama}.} \bibinfo{year}{2020}\natexlab{}.
\newblock \showarticletitle{Incorporating personality in user interface design: A review}.
\newblock \bibinfo{journal}{\emph{Personality and Individual Differences}}  \bibinfo{volume}{155} (\bibinfo{year}{2020}), \bibinfo{pages}{109709}.
\newblock


\bibitem[Anderson and Rainie(2018)]%
        {anderson2018future}
\bibfield{author}{\bibinfo{person}{Janna Anderson} {and} \bibinfo{person}{Lee Rainie}.} \bibinfo{year}{2018}\natexlab{}.
\newblock \showarticletitle{The future of well-being in a tech-saturated world}.
\newblock  (\bibinfo{year}{2018}).
\newblock


\bibitem[Araujo et~al\mbox{.}(2020)]%
        {araujo2020ai}
\bibfield{author}{\bibinfo{person}{Theo Araujo}, \bibinfo{person}{Natali Helberger}, \bibinfo{person}{Sanne Kruikemeier}, {and} \bibinfo{person}{Claes~H De~Vreese}.} \bibinfo{year}{2020}\natexlab{}.
\newblock \showarticletitle{In AI we trust? Perceptions about automated decision-making by artificial intelligence}.
\newblock \bibinfo{journal}{\emph{AI \& society}} \bibinfo{volume}{35}, \bibinfo{number}{3} (\bibinfo{year}{2020}), \bibinfo{pages}{611--623}.
\newblock


\bibitem[Ayoub et~al\mbox{.}(2023)]%
        {ayoub2023real}
\bibfield{author}{\bibinfo{person}{Jackie Ayoub}, \bibinfo{person}{Lilit Avetisian}, \bibinfo{person}{X~Jessie Yang}, {and} \bibinfo{person}{Feng Zhou}.} \bibinfo{year}{2023}\natexlab{}.
\newblock \showarticletitle{Real-time trust prediction in conditionally automated driving using physiological measures}.
\newblock \bibinfo{journal}{\emph{IEEE Transactions on Intelligent Transportation Systems}} (\bibinfo{year}{2023}).
\newblock


\bibitem[Ayoub et~al\mbox{.}(2021)]%
        {ayoub2021modeling}
\bibfield{author}{\bibinfo{person}{Jackie Ayoub}, \bibinfo{person}{X~Jessie Yang}, {and} \bibinfo{person}{Feng Zhou}.} \bibinfo{year}{2021}\natexlab{}.
\newblock \showarticletitle{Modeling dispositional and initial learned trust in automated vehicles with predictability and explainability}.
\newblock \bibinfo{journal}{\emph{Transportation research part F: traffic psychology and behaviour}}  \bibinfo{volume}{77} (\bibinfo{year}{2021}), \bibinfo{pages}{102--116}.
\newblock


\bibitem[Bayer et~al\mbox{.}(2022)]%
        {bayer2022role}
\bibfield{author}{\bibinfo{person}{Sarah Bayer}, \bibinfo{person}{Henner Gimpel}, {and} \bibinfo{person}{Moritz Markgraf}.} \bibinfo{year}{2022}\natexlab{}.
\newblock \showarticletitle{The role of domain expertise in trusting and following explainable AI decision support systems}.
\newblock \bibinfo{journal}{\emph{Journal of Decision Systems}} \bibinfo{volume}{32}, \bibinfo{number}{1} (\bibinfo{year}{2022}), \bibinfo{pages}{110--138}.
\newblock


\bibitem[Bedmutha et~al\mbox{.}(2024a)]%
        {bedmutha2024exploring}
\bibfield{author}{\bibinfo{person}{Manas~Satish Bedmutha}, \bibinfo{person}{Sahithi Karumudi}, \bibinfo{person}{Kevin Patrick}, \bibinfo{person}{Heidi Rataj}, {and} \bibinfo{person}{Nadir Weibel}.} \bibinfo{year}{2024}\natexlab{a}.
\newblock \showarticletitle{Exploring User Willingness towards Mobile Sensing and Intervention: A Case Study on Mental Health of Undergraduate College Students}. In \bibinfo{booktitle}{\emph{Companion of the 2024 on ACM International Joint Conference on Pervasive and Ubiquitous Computing}}. \bibinfo{pages}{721--728}.
\newblock


\bibitem[Bedmutha et~al\mbox{.}(2024b)]%
        {bedmutha2024conversense}
\bibfield{author}{\bibinfo{person}{Manas~Satish Bedmutha}, \bibinfo{person}{Anuujin Tsedenbal}, \bibinfo{person}{Kelly Tobar}, \bibinfo{person}{Sarah Borsotto}, \bibinfo{person}{Kimberly~R Sladek}, \bibinfo{person}{Deepansha Singh}, \bibinfo{person}{Reggie Casanova-Perez}, \bibinfo{person}{Emily Bascom}, \bibinfo{person}{Brian Wood}, \bibinfo{person}{Janice Sabin}, \bibinfo{person}{Wanda Pratt}, \bibinfo{person}{Andrea Hartzler}, {and} \bibinfo{person}{Nadir Weibel}.} \bibinfo{year}{2024}\natexlab{b}.
\newblock \showarticletitle{ConverSense: An Automated Approach to Assess Patient-Provider Interactions using Social Signals}. In \bibinfo{booktitle}{\emph{Proceedings of the CHI Conference on Human Factors in Computing Systems}} (Honolulu, HI, USA) \emph{(\bibinfo{series}{CHI '24})}. \bibinfo{publisher}{Association for Computing Machinery}, \bibinfo{address}{New York, NY, USA}, Article \bibinfo{articleno}{448}, \bibinfo{numpages}{22}~pages.
\newblock
\showISBNx{9798400703300}
\urldef\tempurl%
\url{https://doi.org/10.1145/3613904.3641998}
\showDOI{\tempurl}


\bibitem[Bedu{\'e} and Fritzsche(2022)]%
        {bedue2022can}
\bibfield{author}{\bibinfo{person}{Patrick Bedu{\'e}} {and} \bibinfo{person}{Albrecht Fritzsche}.} \bibinfo{year}{2022}\natexlab{}.
\newblock \showarticletitle{Can we trust AI? an empirical investigation of trust requirements and guide to successful AI adoption}.
\newblock \bibinfo{journal}{\emph{Journal of Enterprise Information Management}} \bibinfo{volume}{35}, \bibinfo{number}{2} (\bibinfo{year}{2022}), \bibinfo{pages}{530--549}.
\newblock


\bibitem[Bennett et~al\mbox{.}(2019)]%
        {bennett2019attitudes}
\bibfield{author}{\bibinfo{person}{Roger Bennett}, \bibinfo{person}{Rohini Vijaygopal}, {and} \bibinfo{person}{Rita Kottasz}.} \bibinfo{year}{2019}\natexlab{}.
\newblock \showarticletitle{Attitudes towards autonomous vehicles among people with physical disabilities}.
\newblock \bibinfo{journal}{\emph{Transportation research part A: policy and practice}}  \bibinfo{volume}{127} (\bibinfo{year}{2019}), \bibinfo{pages}{1--17}.
\newblock


\bibitem[Berliner et~al\mbox{.}(2019)]%
        {berliner2019uncovering}
\bibfield{author}{\bibinfo{person}{Rosaria~M Berliner}, \bibinfo{person}{Scott Hardman}, {and} \bibinfo{person}{Gil Tal}.} \bibinfo{year}{2019}\natexlab{}.
\newblock \showarticletitle{Uncovering early adopter’s perceptions and purchase intentions of automated vehicles: Insights from early adopters of electric vehicles in California}.
\newblock \bibinfo{journal}{\emph{Transportation research part F: traffic psychology and behaviour}}  \bibinfo{volume}{60} (\bibinfo{year}{2019}), \bibinfo{pages}{712--722}.
\newblock


\bibitem[B{\"o}ckle et~al\mbox{.}(2021)]%
        {bockle2021can}
\bibfield{author}{\bibinfo{person}{Martin B{\"o}ckle}, \bibinfo{person}{Kwaku Yeboah-Antwi}, {and} \bibinfo{person}{Iana Kouris}.} \bibinfo{year}{2021}\natexlab{}.
\newblock \showarticletitle{Can you trust the black box? The effect of personality traits on trust in AI-enabled user interfaces}. In \bibinfo{booktitle}{\emph{International Conference on Human-Computer Interaction}}. Springer, \bibinfo{pages}{3--20}.
\newblock


\bibitem[Breiman(2001)]%
        {breiman2001random}
\bibfield{author}{\bibinfo{person}{Leo Breiman}.} \bibinfo{year}{2001}\natexlab{}.
\newblock \showarticletitle{Random forests}.
\newblock \bibinfo{journal}{\emph{Machine learning}}  \bibinfo{volume}{45} (\bibinfo{year}{2001}), \bibinfo{pages}{5--32}.
\newblock


\bibitem[Burger and Cooper(1979)]%
        {burger1979desirability}
\bibfield{author}{\bibinfo{person}{Jerry~M Burger} {and} \bibinfo{person}{Harris~M Cooper}.} \bibinfo{year}{1979}\natexlab{}.
\newblock \showarticletitle{The desirability of control}.
\newblock \bibinfo{journal}{\emph{Motivation and emotion}}  \bibinfo{volume}{3} (\bibinfo{year}{1979}), \bibinfo{pages}{381--393}.
\newblock


\bibitem[Butler et~al\mbox{.}(1999)]%
        {butler1999post}
\bibfield{author}{\bibinfo{person}{Dennis~J Butler}, \bibinfo{person}{H~Steven Moffic}, {and} \bibinfo{person}{NICK~W TURKAL}.} \bibinfo{year}{1999}\natexlab{}.
\newblock \showarticletitle{Post-traumatic stress reactions following motor vehicle accidents}.
\newblock \bibinfo{journal}{\emph{American Family Physician}} \bibinfo{volume}{60}, \bibinfo{number}{2} (\bibinfo{year}{1999}), \bibinfo{pages}{524--530}.
\newblock


\bibitem[Chawla et~al\mbox{.}(2002)]%
        {chawla2002smote}
\bibfield{author}{\bibinfo{person}{Nitesh~V Chawla}, \bibinfo{person}{Kevin~W Bowyer}, \bibinfo{person}{Lawrence~O Hall}, {and} \bibinfo{person}{W~Philip Kegelmeyer}.} \bibinfo{year}{2002}\natexlab{}.
\newblock \showarticletitle{SMOTE: synthetic minority over-sampling technique}.
\newblock \bibinfo{journal}{\emph{Journal of artificial intelligence research}}  \bibinfo{volume}{16} (\bibinfo{year}{2002}), \bibinfo{pages}{321--357}.
\newblock


\bibitem[Chen and Chan(2011)]%
        {chen2011review}
\bibfield{author}{\bibinfo{person}{Ke Chen} {and} \bibinfo{person}{Alan~HS Chan}.} \bibinfo{year}{2011}\natexlab{}.
\newblock \showarticletitle{A review of technology acceptance by older adults.}
\newblock \bibinfo{journal}{\emph{Gerontechnology}} (\bibinfo{year}{2011}).
\newblock


\bibitem[Chen and Guestrin(2016)]%
        {chen2016xgboost}
\bibfield{author}{\bibinfo{person}{Tianqi Chen} {and} \bibinfo{person}{Carlos Guestrin}.} \bibinfo{year}{2016}\natexlab{}.
\newblock \showarticletitle{Xgboost: A scalable tree boosting system}. In \bibinfo{booktitle}{\emph{Proceedings of the 22nd acm sigkdd international conference on knowledge discovery and data mining}}. \bibinfo{pages}{785--794}.
\newblock


\bibitem[Chien et~al\mbox{.}(2018)]%
        {chien2018effect}
\bibfield{author}{\bibinfo{person}{Shih-Yi Chien}, \bibinfo{person}{Michael Lewis}, \bibinfo{person}{Katia Sycara}, \bibinfo{person}{Jyi-Shane Liu}, {and} \bibinfo{person}{Asiye Kumru}.} \bibinfo{year}{2018}\natexlab{}.
\newblock \showarticletitle{The effect of culture on trust in automation: reliability and workload}.
\newblock \bibinfo{journal}{\emph{ACM Transactions on Interactive Intelligent Systems (TiiS)}} \bibinfo{volume}{8}, \bibinfo{number}{4} (\bibinfo{year}{2018}), \bibinfo{pages}{1--31}.
\newblock


\bibitem[Chien et~al\mbox{.}(2016)]%
        {chien2016relation}
\bibfield{author}{\bibinfo{person}{Shih-Yi Chien}, \bibinfo{person}{Katia Sycara}, \bibinfo{person}{Jyi-Shane Liu}, {and} \bibinfo{person}{Asiye Kumru}.} \bibinfo{year}{2016}\natexlab{}.
\newblock \showarticletitle{Relation between trust attitudes toward automation, Hofstede’s cultural dimensions, and big five personality traits}. In \bibinfo{booktitle}{\emph{Proceedings of the human factors and ergonomics society annual meeting}}, Vol.~\bibinfo{volume}{60}. SAGE Publications Sage CA: Los Angeles, CA, \bibinfo{pages}{841--845}.
\newblock


\bibitem[Choi and Ji(2015)]%
        {choi2015investigating}
\bibfield{author}{\bibinfo{person}{Jong~Kyu Choi} {and} \bibinfo{person}{Yong~Gu Ji}.} \bibinfo{year}{2015}\natexlab{}.
\newblock \showarticletitle{Investigating the importance of trust on adopting an autonomous vehicle}.
\newblock \bibinfo{journal}{\emph{International Journal of Human-Computer Interaction}} \bibinfo{volume}{31}, \bibinfo{number}{10} (\bibinfo{year}{2015}), \bibinfo{pages}{692--702}.
\newblock


\bibitem[Choung et~al\mbox{.}(2023)]%
        {choung2023trust}
\bibfield{author}{\bibinfo{person}{Hyesun Choung}, \bibinfo{person}{Prabu David}, {and} \bibinfo{person}{Arun Ross}.} \bibinfo{year}{2023}\natexlab{}.
\newblock \showarticletitle{Trust and ethics in AI}.
\newblock \bibinfo{journal}{\emph{Ai \& Society}} \bibinfo{volume}{38}, \bibinfo{number}{2} (\bibinfo{year}{2023}), \bibinfo{pages}{733--745}.
\newblock


\bibitem[Dohmen et~al\mbox{.}(2005)]%
        {dohmen2005individual}
\bibfield{author}{\bibinfo{person}{Thomas~J Dohmen}, \bibinfo{person}{Armin Falk}, \bibinfo{person}{David Huffman}, \bibinfo{person}{Uwe Sunde}, \bibinfo{person}{J{\"u}rgen Schupp}, {and} \bibinfo{person}{Gert~G Wagner}.} \bibinfo{year}{2005}\natexlab{}.
\newblock \showarticletitle{Individual risk attitudes: New evidence from a large, representative, experimentally-validated survey}.
\newblock  (\bibinfo{year}{2005}).
\newblock


\bibitem[Ehlers et~al\mbox{.}(2007)]%
        {ehlers2007driving}
\bibfield{author}{\bibinfo{person}{Anke Ehlers}, \bibinfo{person}{Joanne~E Taylor}, \bibinfo{person}{Thomas Ehring}, \bibinfo{person}{Stefan~G Hofmann}, \bibinfo{person}{Frank~P Deane}, \bibinfo{person}{Walton~T Roth}, {and} \bibinfo{person}{John~V Podd}.} \bibinfo{year}{2007}\natexlab{}.
\newblock \showarticletitle{The driving cognitions questionnaire: Development and preliminary psychometric properties}.
\newblock \bibinfo{journal}{\emph{Journal of anxiety disorders}} \bibinfo{volume}{21}, \bibinfo{number}{4} (\bibinfo{year}{2007}), \bibinfo{pages}{493--509}.
\newblock


\bibitem[Fagnant and Kockelman(2015)]%
        {fagnant2015preparing}
\bibfield{author}{\bibinfo{person}{Daniel~J Fagnant} {and} \bibinfo{person}{Kara Kockelman}.} \bibinfo{year}{2015}\natexlab{}.
\newblock \showarticletitle{Preparing a nation for autonomous vehicles: opportunities, barriers and policy recommendations}.
\newblock \bibinfo{journal}{\emph{Transportation Research Part A: Policy and Practice}}  \bibinfo{volume}{77} (\bibinfo{year}{2015}), \bibinfo{pages}{167--181}.
\newblock


\bibitem[Ferronato and Bashir(2020)]%
        {ferronato2020examination}
\bibfield{author}{\bibinfo{person}{Priscilla Ferronato} {and} \bibinfo{person}{Masooda Bashir}.} \bibinfo{year}{2020}\natexlab{}.
\newblock \showarticletitle{An examination of dispositional trust in human and autonomous system interactions}. In \bibinfo{booktitle}{\emph{Human-Computer Interaction. Human Values and Quality of Life: Thematic Area, HCI 2020, Held as Part of the 22nd International Conference, HCII 2020, Copenhagen, Denmark, July 19--24, 2020, Proceedings, Part III 22}}. Springer, \bibinfo{pages}{420--435}.
\newblock


\bibitem[Fraedrich et~al\mbox{.}(2015)]%
        {fraedrich2015transition}
\bibfield{author}{\bibinfo{person}{Eva Fraedrich}, \bibinfo{person}{Sven Beiker}, {and} \bibinfo{person}{Barbara Lenz}.} \bibinfo{year}{2015}\natexlab{}.
\newblock \showarticletitle{Transition pathways to fully automated driving and its implications for the sociotechnical system of automobility}.
\newblock \bibinfo{journal}{\emph{European Journal of Futures Research}}  \bibinfo{volume}{3} (\bibinfo{year}{2015}), \bibinfo{pages}{1--11}.
\newblock


\bibitem[Franke et~al\mbox{.}(2019)]%
        {franke2019personal}
\bibfield{author}{\bibinfo{person}{Thomas Franke}, \bibinfo{person}{Christiane Attig}, {and} \bibinfo{person}{Daniel Wessel}.} \bibinfo{year}{2019}\natexlab{}.
\newblock \showarticletitle{A personal resource for technology interaction: development and validation of the affinity for technology interaction (ATI) scale}.
\newblock \bibinfo{journal}{\emph{International Journal of Human--Computer Interaction}} \bibinfo{volume}{35}, \bibinfo{number}{6} (\bibinfo{year}{2019}), \bibinfo{pages}{456--467}.
\newblock


\bibitem[Fritz et~al\mbox{.}(2012)]%
        {fritz2012effect}
\bibfield{author}{\bibinfo{person}{Catherine~O Fritz}, \bibinfo{person}{Peter~E Morris}, {and} \bibinfo{person}{Jennifer~J Richler}.} \bibinfo{year}{2012}\natexlab{}.
\newblock \showarticletitle{Effect size estimates: current use, calculations, and interpretation.}
\newblock \bibinfo{journal}{\emph{Journal of experimental psychology: General}} \bibinfo{volume}{141}, \bibinfo{number}{1} (\bibinfo{year}{2012}), \bibinfo{pages}{2}.
\newblock


\bibitem[Funk et~al\mbox{.}(2019)]%
        {funk2019trust}
\bibfield{author}{\bibinfo{person}{Cary Funk}, \bibinfo{person}{Meg Hefferon}, \bibinfo{person}{Brian Kennedy}, {and} \bibinfo{person}{Courtney Johnson}.} \bibinfo{year}{2019}\natexlab{}.
\newblock \showarticletitle{Trust and mistrust in Americans’ views of scientific experts}.
\newblock \bibinfo{journal}{\emph{Pew Research Center}}  \bibinfo{volume}{2} (\bibinfo{year}{2019}), \bibinfo{pages}{1--96}.
\newblock


\bibitem[Gunning and Aha(2019)]%
        {gunning2019darpa}
\bibfield{author}{\bibinfo{person}{David Gunning} {and} \bibinfo{person}{David Aha}.} \bibinfo{year}{2019}\natexlab{}.
\newblock \showarticletitle{DARPA’s explainable artificial intelligence (XAI) program}.
\newblock \bibinfo{journal}{\emph{AI magazine}} \bibinfo{volume}{40}, \bibinfo{number}{2} (\bibinfo{year}{2019}), \bibinfo{pages}{44--58}.
\newblock


\bibitem[Hamburger et~al\mbox{.}(2022)]%
        {hamburger2022personality}
\bibfield{author}{\bibinfo{person}{Yair~Amichai Hamburger}, \bibinfo{person}{Yaron Sela}, \bibinfo{person}{Sharon Kaufman}, \bibinfo{person}{Tamar Wellingstein}, \bibinfo{person}{Noy Stein}, {and} \bibinfo{person}{Joel Sivan}.} \bibinfo{year}{2022}\natexlab{}.
\newblock \showarticletitle{Personality and the autonomous vehicle: Overcoming psychological barriers to the driverless car}.
\newblock \bibinfo{journal}{\emph{Technology in Society}}  \bibinfo{volume}{69} (\bibinfo{year}{2022}), \bibinfo{pages}{101971}.
\newblock


\bibitem[Hancock et~al\mbox{.}(2023)]%
        {hancock2023and}
\bibfield{author}{\bibinfo{person}{PA Hancock}, \bibinfo{person}{Theresa~T Kessler}, \bibinfo{person}{Alexandra~D Kaplan}, \bibinfo{person}{Kimberly Stowers}, \bibinfo{person}{J~Christopher Brill}, \bibinfo{person}{Deborah~R Billings}, \bibinfo{person}{Kristin~E Schaefer}, {and} \bibinfo{person}{James~L Szalma}.} \bibinfo{year}{2023}\natexlab{}.
\newblock \showarticletitle{How and why humans trust: A meta-analysis and elaborated model}.
\newblock \bibinfo{journal}{\emph{Frontiers in psychology}}  \bibinfo{volume}{14} (\bibinfo{year}{2023}).
\newblock


\bibitem[Hancock et~al\mbox{.}(2011)]%
        {hancock2011meta}
\bibfield{author}{\bibinfo{person}{Peter~A Hancock}, \bibinfo{person}{Deborah~R Billings}, \bibinfo{person}{Kristin~E Schaefer}, \bibinfo{person}{Jessie~YC Chen}, \bibinfo{person}{Ewart~J De~Visser}, {and} \bibinfo{person}{Raja Parasuraman}.} \bibinfo{year}{2011}\natexlab{}.
\newblock \showarticletitle{A meta-analysis of factors affecting trust in human-robot interaction}.
\newblock \bibinfo{journal}{\emph{Human factors}} \bibinfo{volume}{53}, \bibinfo{number}{5} (\bibinfo{year}{2011}), \bibinfo{pages}{517--527}.
\newblock


\bibitem[Hassija et~al\mbox{.}(2024)]%
        {hassija2024interpreting}
\bibfield{author}{\bibinfo{person}{Vikas Hassija}, \bibinfo{person}{Vinay Chamola}, \bibinfo{person}{Atmesh Mahapatra}, \bibinfo{person}{Abhinandan Singal}, \bibinfo{person}{Divyansh Goel}, \bibinfo{person}{Kaizhu Huang}, \bibinfo{person}{Simone Scardapane}, \bibinfo{person}{Indro Spinelli}, \bibinfo{person}{Mufti Mahmud}, {and} \bibinfo{person}{Amir Hussain}.} \bibinfo{year}{2024}\natexlab{}.
\newblock \showarticletitle{Interpreting black-box models: a review on explainable artificial intelligence}.
\newblock \bibinfo{journal}{\emph{Cognitive Computation}} \bibinfo{volume}{16}, \bibinfo{number}{1} (\bibinfo{year}{2024}), \bibinfo{pages}{45--74}.
\newblock


\bibitem[He et~al\mbox{.}(2022)]%
        {he2022modelling}
\bibfield{author}{\bibinfo{person}{Xiaolin He}, \bibinfo{person}{Jork Stapel}, \bibinfo{person}{Meng Wang}, {and} \bibinfo{person}{Riender Happee}.} \bibinfo{year}{2022}\natexlab{}.
\newblock \showarticletitle{Modelling perceived risk and trust in driving automation reacting to merging and braking vehicles}.
\newblock \bibinfo{journal}{\emph{Transportation research part F: traffic psychology and behaviour}}  \bibinfo{volume}{86} (\bibinfo{year}{2022}), \bibinfo{pages}{178--195}.
\newblock


\bibitem[Hegner et~al\mbox{.}(2019)]%
        {hegner2019automatic}
\bibfield{author}{\bibinfo{person}{Sabrina~M Hegner}, \bibinfo{person}{Ardion~D Beldad}, {and} \bibinfo{person}{Gary~J Brunswick}.} \bibinfo{year}{2019}\natexlab{}.
\newblock \showarticletitle{In automatic we trust: investigating the impact of trust, control, personality characteristics, and extrinsic and intrinsic motivations on the acceptance of autonomous vehicles}.
\newblock \bibinfo{journal}{\emph{International Journal of Human--Computer Interaction}} \bibinfo{volume}{35}, \bibinfo{number}{19} (\bibinfo{year}{2019}), \bibinfo{pages}{1769--1780}.
\newblock


\bibitem[Hewitt et~al\mbox{.}(2019)]%
        {hewitt2019assessing}
\bibfield{author}{\bibinfo{person}{Charlie Hewitt}, \bibinfo{person}{Ioannis Politis}, \bibinfo{person}{Theocharis Amanatidis}, {and} \bibinfo{person}{Advait Sarkar}.} \bibinfo{year}{2019}\natexlab{}.
\newblock \showarticletitle{Assessing public perception of self-driving cars: The autonomous vehicle acceptance model}. In \bibinfo{booktitle}{\emph{Proceedings of the 24th international conference on intelligent user interfaces}}. \bibinfo{pages}{518--527}.
\newblock


\bibitem[Hoff and Bashir(2015)]%
        {hoff2015trust}
\bibfield{author}{\bibinfo{person}{Kevin~Anthony Hoff} {and} \bibinfo{person}{Masooda Bashir}.} \bibinfo{year}{2015}\natexlab{}.
\newblock \showarticletitle{Trust in automation: Integrating empirical evidence on factors that influence trust}.
\newblock \bibinfo{journal}{\emph{Human factors}} \bibinfo{volume}{57}, \bibinfo{number}{3} (\bibinfo{year}{2015}), \bibinfo{pages}{407--434}.
\newblock


\bibitem[Hohenberger et~al\mbox{.}(2016)]%
        {hohenberger2016and}
\bibfield{author}{\bibinfo{person}{Christoph Hohenberger}, \bibinfo{person}{Matthias Sp{\"o}rrle}, {and} \bibinfo{person}{Isabell~M Welpe}.} \bibinfo{year}{2016}\natexlab{}.
\newblock \showarticletitle{How and why do men and women differ in their willingness to use automated cars? The influence of emotions across different age groups}.
\newblock \bibinfo{journal}{\emph{Transportation Research Part A: Policy and Practice}}  \bibinfo{volume}{94} (\bibinfo{year}{2016}), \bibinfo{pages}{374--385}.
\newblock


\bibitem[Howard and Dai(2014)]%
        {howard2014public}
\bibfield{author}{\bibinfo{person}{Daniel Howard} {and} \bibinfo{person}{Danielle Dai}.} \bibinfo{year}{2014}\natexlab{}.
\newblock \showarticletitle{Public perceptions of self-driving cars: The case of Berkeley, California}. In \bibinfo{booktitle}{\emph{Transportation research board 93rd annual meeting}}, Vol.~\bibinfo{volume}{14}. The National Academies of Sciences, Engineering, and Medicine Washington, DC~…, \bibinfo{pages}{1--16}.
\newblock


\bibitem[Huang et~al\mbox{.}(2024)]%
        {huangexploring}
\bibfield{author}{\bibinfo{person}{Chunxi Huang}, \bibinfo{person}{Jiyao Wang}, \bibinfo{person}{Song Yan}, {and} \bibinfo{person}{Dengbo He}.} \bibinfo{year}{2024}\natexlab{}.
\newblock \showarticletitle{Exploring Factors Related to Drivers’ Mental Model of and Trust in Advanced Driver Assistance Systems Using an ABN-Based Mixed Approach}.
\newblock \bibinfo{journal}{\emph{IEEE Transactions on Human-Machine Systems}} (\bibinfo{year}{2024}).
\newblock


\bibitem[Hudson et~al\mbox{.}(2019)]%
        {hudson2019people}
\bibfield{author}{\bibinfo{person}{John Hudson}, \bibinfo{person}{Marta Orviska}, {and} \bibinfo{person}{Jan Hunady}.} \bibinfo{year}{2019}\natexlab{}.
\newblock \showarticletitle{People’s attitudes to autonomous vehicles}.
\newblock \bibinfo{journal}{\emph{Transportation research part A: policy and practice}}  \bibinfo{volume}{121} (\bibinfo{year}{2019}), \bibinfo{pages}{164--176}.
\newblock


\bibitem[Hulse et~al\mbox{.}(2018)]%
        {hulse2018perceptions}
\bibfield{author}{\bibinfo{person}{Lynn~M Hulse}, \bibinfo{person}{Hui Xie}, {and} \bibinfo{person}{Edwin~R Galea}.} \bibinfo{year}{2018}\natexlab{}.
\newblock \showarticletitle{Perceptions of autonomous vehicles: Relationships with road users, risk, gender and age}.
\newblock \bibinfo{journal}{\emph{Safety science}}  \bibinfo{volume}{102} (\bibinfo{year}{2018}), \bibinfo{pages}{1--13}.
\newblock


\bibitem[Jian et~al\mbox{.}(2000)]%
        {jian2000foundations}
\bibfield{author}{\bibinfo{person}{Jiun-Yin Jian}, \bibinfo{person}{Ann~M Bisantz}, {and} \bibinfo{person}{Colin~G Drury}.} \bibinfo{year}{2000}\natexlab{}.
\newblock \showarticletitle{Foundations for an empirically determined scale of trust in automated systems}.
\newblock \bibinfo{journal}{\emph{International journal of cognitive ergonomics}} \bibinfo{volume}{4}, \bibinfo{number}{1} (\bibinfo{year}{2000}), \bibinfo{pages}{53--71}.
\newblock


\bibitem[Jobin et~al\mbox{.}(2019)]%
        {jobin2019global}
\bibfield{author}{\bibinfo{person}{Anna Jobin}, \bibinfo{person}{Marcello Ienca}, {and} \bibinfo{person}{Effy Vayena}.} \bibinfo{year}{2019}\natexlab{}.
\newblock \showarticletitle{The global landscape of AI ethics guidelines}.
\newblock \bibinfo{journal}{\emph{Nature machine intelligence}} \bibinfo{volume}{1}, \bibinfo{number}{9} (\bibinfo{year}{2019}), \bibinfo{pages}{389--399}.
\newblock


\bibitem[Kaufman et~al\mbox{.}(2024a)]%
        {kaufman2024warning}
\bibfield{author}{\bibinfo{person}{Robert Kaufman}, \bibinfo{person}{Aaron Broukhim}, {and} \bibinfo{person}{Michael Haupt}.} \bibinfo{year}{2024}\natexlab{a}.
\newblock \showarticletitle{WARNING This Contains Misinformation: The Effect of Cognitive Factors, Beliefs, and Personality on Misinformation Warning Tag Attitudes}.
\newblock \bibinfo{journal}{\emph{arXiv preprint arXiv:2407.02710}} (\bibinfo{year}{2024}).
\newblock


\bibitem[Kaufman et~al\mbox{.}(2024b)]%
        {kaufman2024did}
\bibfield{author}{\bibinfo{person}{Robert Kaufman}, \bibinfo{person}{Aaron Broukhim}, \bibinfo{person}{David Kirsh}, {and} \bibinfo{person}{Nadir Weibel}.} \bibinfo{year}{2024}\natexlab{b}.
\newblock \showarticletitle{What Did My Car Say? Impact of Autonomous Vehicle Explanation Errors and Driving Context On Comfort, Reliance, Satisfaction, and Driving Confidence}.
\newblock \bibinfo{journal}{\emph{arXiv preprint arXiv:2409.05731}} (\bibinfo{year}{2024}).
\newblock


\bibitem[Kaufman et~al\mbox{.}(2024c)]%
        {kaufman2024effects}
\bibfield{author}{\bibinfo{person}{Robert Kaufman}, \bibinfo{person}{Jean Costa}, {and} \bibinfo{person}{Everlyne Kimani}.} \bibinfo{year}{2024}\natexlab{c}.
\newblock \showarticletitle{Effects of multimodal explanations for autonomous driving on driving performance, cognitive load, expertise, confidence, and trust}.
\newblock \bibinfo{journal}{\emph{Scientific reports}} \bibinfo{volume}{14}, \bibinfo{number}{1} (\bibinfo{year}{2024}), \bibinfo{pages}{13061}.
\newblock


\bibitem[Kaufman and Kirsh(2023)]%
        {kaufman2023explainable}
\bibfield{author}{\bibinfo{person}{Robert Kaufman} {and} \bibinfo{person}{David Kirsh}.} \bibinfo{year}{2023}\natexlab{}.
\newblock \showarticletitle{Explainable AI And Visual Reasoning: Insights From Radiology}.
\newblock \bibinfo{journal}{\emph{arXiv preprint arXiv:2304.03318}} (\bibinfo{year}{2023}).
\newblock


\bibitem[Kaufman et~al\mbox{.}(2024d)]%
        {kaufman2024developing}
\bibfield{author}{\bibinfo{person}{Robert Kaufman}, \bibinfo{person}{David Kirsh}, {and} \bibinfo{person}{Nadir Weibel}.} \bibinfo{year}{2024}\natexlab{d}.
\newblock \showarticletitle{Developing Situational Awareness for Joint Action with Autonomous Vehicles}.
\newblock \bibinfo{journal}{\emph{arXiv preprint arXiv:2404.11800}} (\bibinfo{year}{2024}).
\newblock


\bibitem[Kaufman et~al\mbox{.}(2022)]%
        {kaufman2022s}
\bibfield{author}{\bibinfo{person}{Robert~A Kaufman}, \bibinfo{person}{Michael~Robert Haupt}, {and} \bibinfo{person}{Steven~P Dow}.} \bibinfo{year}{2022}\natexlab{}.
\newblock \showarticletitle{Who's in the Crowd Matters: Cognitive Factors and Beliefs Predict Misinformation Assessment Accuracy}.
\newblock \bibinfo{journal}{\emph{Proceedings of the ACM on Human-Computer Interaction}} \bibinfo{volume}{6}, \bibinfo{number}{CSCW2} (\bibinfo{year}{2022}), \bibinfo{pages}{1--18}.
\newblock


\bibitem[Kaufman and Kirsh(2022)]%
        {kaufman2022cognitive}
\bibfield{author}{\bibinfo{person}{Robert~A Kaufman} {and} \bibinfo{person}{David Kirsh}.} \bibinfo{year}{2022}\natexlab{}.
\newblock \showarticletitle{Cognitive differences in human and AI explanation}. In \bibinfo{booktitle}{\emph{Proceedings of the Annual Meeting of the Cognitive Science Society}}, Vol.~\bibinfo{volume}{44}.
\newblock


\bibitem[Kenesei et~al\mbox{.}(2022)]%
        {kenesei2022trust}
\bibfield{author}{\bibinfo{person}{Zs{\'o}fia Kenesei}, \bibinfo{person}{Katalin {\'A}sv{\'a}nyi}, \bibinfo{person}{L{\'a}szl{\'o} K{\"o}k{\'e}ny}, \bibinfo{person}{Melinda J{\'a}szber{\'e}nyi}, \bibinfo{person}{M{\'a}rk Miskolczi}, \bibinfo{person}{Tam{\'a}s Gyulav{\'a}ri}, {and} \bibinfo{person}{Jhanghiz Syahrivar}.} \bibinfo{year}{2022}\natexlab{}.
\newblock \showarticletitle{Trust and perceived risk: How different manifestations affect the adoption of autonomous vehicles}.
\newblock \bibinfo{journal}{\emph{Transportation research part A: policy and practice}}  \bibinfo{volume}{164} (\bibinfo{year}{2022}), \bibinfo{pages}{379--393}.
\newblock


\bibitem[Khastgir et~al\mbox{.}(2018)]%
        {khastgir2018calibrating}
\bibfield{author}{\bibinfo{person}{Siddartha Khastgir}, \bibinfo{person}{Stewart Birrell}, \bibinfo{person}{Gunwant Dhadyalla}, {and} \bibinfo{person}{Paul Jennings}.} \bibinfo{year}{2018}\natexlab{}.
\newblock \showarticletitle{Calibrating trust through knowledge: Introducing the concept of informed safety for automation in vehicles}.
\newblock \bibinfo{journal}{\emph{Transportation research part C: emerging technologies}}  \bibinfo{volume}{96} (\bibinfo{year}{2018}), \bibinfo{pages}{290--303}.
\newblock


\bibitem[Koo et~al\mbox{.}(2015)]%
        {koo2015did}
\bibfield{author}{\bibinfo{person}{Jeamin Koo}, \bibinfo{person}{Jungsuk Kwac}, \bibinfo{person}{Wendy Ju}, \bibinfo{person}{Martin Steinert}, \bibinfo{person}{Larry Leifer}, {and} \bibinfo{person}{Clifford Nass}.} \bibinfo{year}{2015}\natexlab{}.
\newblock \showarticletitle{Why did my car just do that? Explaining semi-autonomous driving actions to improve driver understanding, trust, and performance}.
\newblock \bibinfo{journal}{\emph{International Journal on Interactive Design and Manufacturing (IJIDeM)}}  \bibinfo{volume}{9} (\bibinfo{year}{2015}), \bibinfo{pages}{269--275}.
\newblock


\bibitem[Kraus et~al\mbox{.}(2021)]%
        {kraus2021s}
\bibfield{author}{\bibinfo{person}{Johannes Kraus}, \bibinfo{person}{David Scholz}, {and} \bibinfo{person}{Martin Baumann}.} \bibinfo{year}{2021}\natexlab{}.
\newblock \showarticletitle{What’s driving me? Exploration and validation of a hierarchical personality model for trust in automated driving}.
\newblock \bibinfo{journal}{\emph{Human factors}} \bibinfo{volume}{63}, \bibinfo{number}{6} (\bibinfo{year}{2021}), \bibinfo{pages}{1076--1105}.
\newblock


\bibitem[Kraus et~al\mbox{.}(2020)]%
        {kraus2020scared}
\bibfield{author}{\bibinfo{person}{Johannes Kraus}, \bibinfo{person}{David Scholz}, \bibinfo{person}{Eva-Maria Messner}, \bibinfo{person}{Matthias Messner}, {and} \bibinfo{person}{Martin Baumann}.} \bibinfo{year}{2020}\natexlab{}.
\newblock \showarticletitle{Scared to trust?--predicting trust in highly automated driving by depressiveness, negative self-evaluations and state anxiety}.
\newblock \bibinfo{journal}{\emph{Frontiers in Psychology}}  \bibinfo{volume}{10} (\bibinfo{year}{2020}), \bibinfo{pages}{2917}.
\newblock


\bibitem[Lee and See(2004)]%
        {lee2004trust}
\bibfield{author}{\bibinfo{person}{John~D Lee} {and} \bibinfo{person}{Katrina~A See}.} \bibinfo{year}{2004}\natexlab{}.
\newblock \showarticletitle{Trust in automation: Designing for appropriate reliance}.
\newblock \bibinfo{journal}{\emph{Human factors}} \bibinfo{volume}{46}, \bibinfo{number}{1} (\bibinfo{year}{2004}), \bibinfo{pages}{50--80}.
\newblock


\bibitem[Lee and Rich(2021)]%
        {lee2021included}
\bibfield{author}{\bibinfo{person}{Min~Kyung Lee} {and} \bibinfo{person}{Katherine Rich}.} \bibinfo{year}{2021}\natexlab{}.
\newblock \showarticletitle{Who is included in human perceptions of AI?: Trust and perceived fairness around healthcare AI and cultural mistrust}. In \bibinfo{booktitle}{\emph{Proceedings of the 2021 CHI conference on human factors in computing systems}}. \bibinfo{pages}{1--14}.
\newblock


\bibitem[Levinson et~al\mbox{.}(2011)]%
        {levinson2011towards}
\bibfield{author}{\bibinfo{person}{Jesse Levinson}, \bibinfo{person}{Jake Askeland}, \bibinfo{person}{Jan Becker}, \bibinfo{person}{Jennifer Dolson}, \bibinfo{person}{David Held}, \bibinfo{person}{Soeren Kammel}, \bibinfo{person}{J~Zico Kolter}, \bibinfo{person}{Dirk Langer}, \bibinfo{person}{Oliver Pink}, \bibinfo{person}{Vaughan Pratt}, {et~al\mbox{.}}} \bibinfo{year}{2011}\natexlab{}.
\newblock \showarticletitle{Towards fully autonomous driving: Systems and algorithms}. In \bibinfo{booktitle}{\emph{2011 IEEE intelligent vehicles symposium (IV)}}. IEEE, \bibinfo{pages}{163--168}.
\newblock


\bibitem[Li et~al\mbox{.}(2020)]%
        {li2020personality}
\bibfield{author}{\bibinfo{person}{Wenmin Li}, \bibinfo{person}{Nailang Yao}, \bibinfo{person}{Yanwei Shi}, \bibinfo{person}{Weiran Nie}, \bibinfo{person}{Yuhai Zhang}, \bibinfo{person}{Xiangrong Li}, \bibinfo{person}{Jiawen Liang}, \bibinfo{person}{Fang Chen}, {and} \bibinfo{person}{Zaifeng Gao}.} \bibinfo{year}{2020}\natexlab{}.
\newblock \showarticletitle{Personality openness predicts driver trust in automated driving}.
\newblock \bibinfo{journal}{\emph{Automotive Innovation}}  \bibinfo{volume}{3} (\bibinfo{year}{2020}), \bibinfo{pages}{3--13}.
\newblock


\bibitem[Liao and Sundar(2022)]%
        {liao2022designing}
\bibfield{author}{\bibinfo{person}{Q~Vera Liao} {and} \bibinfo{person}{S~Shyam Sundar}.} \bibinfo{year}{2022}\natexlab{}.
\newblock \showarticletitle{Designing for responsible trust in AI systems: A communication perspective}. In \bibinfo{booktitle}{\emph{Proceedings of the 2022 ACM Conference on Fairness, Accountability, and Transparency}}. \bibinfo{pages}{1257--1268}.
\newblock


\bibitem[Liao and Varshney(2021)]%
        {liao2021human}
\bibfield{author}{\bibinfo{person}{Q~Vera Liao} {and} \bibinfo{person}{Kush~R Varshney}.} \bibinfo{year}{2021}\natexlab{}.
\newblock \showarticletitle{Human-centered explainable ai (xai): From algorithms to user experiences}.
\newblock \bibinfo{journal}{\emph{arXiv preprint arXiv:2110.10790}} (\bibinfo{year}{2021}).
\newblock


\bibitem[Liao et~al\mbox{.}(2022)]%
        {liao2022driver}
\bibfield{author}{\bibinfo{person}{Xishun Liao}, \bibinfo{person}{Shashank Mehrotra}, \bibinfo{person}{Samson Ho}, \bibinfo{person}{Yuki Gorospe}, \bibinfo{person}{Xingwei Wu}, {and} \bibinfo{person}{Teruhisa Mistu}.} \bibinfo{year}{2022}\natexlab{}.
\newblock \showarticletitle{Driver profile modeling based on driving style, personality traits, and mood states}. In \bibinfo{booktitle}{\emph{2022 IEEE 25th international conference on intelligent transportation systems (ITSC)}}. IEEE, \bibinfo{pages}{709--716}.
\newblock


\bibitem[Lindqvist et~al\mbox{.}(2021)]%
        {lindqvist2021gender}
\bibfield{author}{\bibinfo{person}{Anna Lindqvist}, \bibinfo{person}{Marie~Gustafsson Send{\'e}n}, {and} \bibinfo{person}{Emma~A Renstr{\"o}m}.} \bibinfo{year}{2021}\natexlab{}.
\newblock \showarticletitle{What is gender, anyway: a review of the options for operationalising gender}.
\newblock \bibinfo{journal}{\emph{Psychology \& sexuality}} \bibinfo{volume}{12}, \bibinfo{number}{4} (\bibinfo{year}{2021}), \bibinfo{pages}{332--344}.
\newblock


\bibitem[Liu et~al\mbox{.}(2019)]%
        {liu2019public}
\bibfield{author}{\bibinfo{person}{Peng Liu}, \bibinfo{person}{Run Yang}, {and} \bibinfo{person}{Zhigang Xu}.} \bibinfo{year}{2019}\natexlab{}.
\newblock \showarticletitle{Public acceptance of fully automated driving: Effects of social trust and risk/benefit perceptions}.
\newblock \bibinfo{journal}{\emph{Risk Analysis}} \bibinfo{volume}{39}, \bibinfo{number}{2} (\bibinfo{year}{2019}), \bibinfo{pages}{326--341}.
\newblock


\bibitem[Liu et~al\mbox{.}(2014)]%
        {liu2014generic}
\bibfield{author}{\bibinfo{person}{Xin Liu}, \bibinfo{person}{Gilles Tredan}, {and} \bibinfo{person}{Anwitaman Datta}.} \bibinfo{year}{2014}\natexlab{}.
\newblock \showarticletitle{A Generic Trust Framework for Large-scale Open Systems Using Machine Learning}.
\newblock \bibinfo{journal}{\emph{Computational Intelligence}} \bibinfo{volume}{30}, \bibinfo{number}{4} (\bibinfo{year}{2014}), \bibinfo{pages}{700--721}.
\newblock


\bibitem[Lohaus et~al\mbox{.}(2024)]%
        {lohaus2024automated}
\bibfield{author}{\bibinfo{person}{Leonie Lohaus}, \bibinfo{person}{Marcel Woide}, \bibinfo{person}{Nicole Damm}, \bibinfo{person}{Zeynep Demiral}, \bibinfo{person}{Hannah Friedrich}, \bibinfo{person}{Anna Pet{\'a}kov{\'a}}, {and} \bibinfo{person}{Francesco Walker}.} \bibinfo{year}{2024}\natexlab{}.
\newblock \showarticletitle{Automated or Human: Which Driver Wins the Race for the Passengers’ Trust? Examining Passenger Trust in Human-Driven and Automated Vehicles Following a Dangerous Situation}.
\newblock \bibinfo{journal}{\emph{Computers in Human Behavior}} (\bibinfo{year}{2024}), \bibinfo{pages}{108387}.
\newblock


\bibitem[L{\'o}pez and Maag(2015)]%
        {lopez2015towards}
\bibfield{author}{\bibinfo{person}{Jorge L{\'o}pez} {and} \bibinfo{person}{Stephane Maag}.} \bibinfo{year}{2015}\natexlab{}.
\newblock \showarticletitle{Towards a generic trust management framework using a machine-learning-based trust model}. In \bibinfo{booktitle}{\emph{2015 IEEE Trustcom/BigDataSE/ISPA}}, Vol.~\bibinfo{volume}{1}. IEEE, \bibinfo{pages}{1343--1348}.
\newblock


\bibitem[Lundberg et~al\mbox{.}(2020)]%
        {lundberg2020local}
\bibfield{author}{\bibinfo{person}{Scott~M Lundberg}, \bibinfo{person}{Gabriel Erion}, \bibinfo{person}{Hugh Chen}, \bibinfo{person}{Alex DeGrave}, \bibinfo{person}{Jordan~M Prutkin}, \bibinfo{person}{Bala Nair}, \bibinfo{person}{Ronit Katz}, \bibinfo{person}{Jonathan Himmelfarb}, \bibinfo{person}{Nisha Bansal}, {and} \bibinfo{person}{Su-In Lee}.} \bibinfo{year}{2020}\natexlab{}.
\newblock \showarticletitle{From local explanations to global understanding with explainable AI for trees}.
\newblock \bibinfo{journal}{\emph{Nature machine intelligence}} \bibinfo{volume}{2}, \bibinfo{number}{1} (\bibinfo{year}{2020}), \bibinfo{pages}{56--67}.
\newblock


\bibitem[Luo et~al\mbox{.}(2020)]%
        {luo2020trust}
\bibfield{author}{\bibinfo{person}{Ruikun Luo}, \bibinfo{person}{Jian Chu}, {and} \bibinfo{person}{X~Jessie Yang}.} \bibinfo{year}{2020}\natexlab{}.
\newblock \showarticletitle{Trust dynamics in human-AV (automated vehicle) interaction}. In \bibinfo{booktitle}{\emph{Extended abstracts of the 2020 CHI conference on human factors in computing systems}}. \bibinfo{pages}{1--7}.
\newblock


\bibitem[Ma and Feng(2023)]%
        {ma2023analysing}
\bibfield{author}{\bibinfo{person}{Jun Ma} {and} \bibinfo{person}{Xuejing Feng}.} \bibinfo{year}{2023}\natexlab{}.
\newblock \showarticletitle{Analysing the Effects of Scenario-Based Explanations on Automated Vehicle HMIs from Objective and Subjective Perspectives}.
\newblock \bibinfo{journal}{\emph{Sustainability}} \bibinfo{volume}{16}, \bibinfo{number}{1} (\bibinfo{year}{2023}), \bibinfo{pages}{63}.
\newblock


\bibitem[Ma et~al\mbox{.}(2024)]%
        {ma2024understanding}
\bibfield{author}{\bibinfo{person}{Xingjian Ma}, \bibinfo{person}{Xizi Xiao}, \bibinfo{person}{Ranjana Mehta}, {and} \bibinfo{person}{Anthony~D McDonald}.} \bibinfo{year}{2024}\natexlab{}.
\newblock \showarticletitle{Understanding Reliance Decisions in Automated Vehicles Using Random Forest Analysis}. In \bibinfo{booktitle}{\emph{Proceedings of the Human Factors and Ergonomics Society Annual Meeting}}. SAGE Publications Sage CA: Los Angeles, CA, \bibinfo{pages}{10711813241262982}.
\newblock


\bibitem[Mack et~al\mbox{.}(2021)]%
        {mack2021politics}
\bibfield{author}{\bibinfo{person}{Elizabeth~A Mack}, \bibinfo{person}{Steven~R Miller}, \bibinfo{person}{Chu-Hsiang Chang}, \bibinfo{person}{Jenna~A Van~Fossen}, \bibinfo{person}{Shelia~R Cotten}, \bibinfo{person}{Peter~T Savolainen}, {and} \bibinfo{person}{John Mann}.} \bibinfo{year}{2021}\natexlab{}.
\newblock \showarticletitle{The politics of new driving technologies: Political ideology and autonomous vehicle adoption}.
\newblock \bibinfo{journal}{\emph{Telematics and Informatics}}  \bibinfo{volume}{61} (\bibinfo{year}{2021}), \bibinfo{pages}{101604}.
\newblock


\bibitem[Miller(2019)]%
        {miller2019explanation}
\bibfield{author}{\bibinfo{person}{Tim Miller}.} \bibinfo{year}{2019}\natexlab{}.
\newblock \showarticletitle{Explanation in artificial intelligence: Insights from the social sciences}.
\newblock \bibinfo{journal}{\emph{Artificial intelligence}}  \bibinfo{volume}{267} (\bibinfo{year}{2019}), \bibinfo{pages}{1--38}.
\newblock


\bibitem[Monteiro et~al\mbox{.}(2022)]%
        {monteiro2022efficient}
\bibfield{author}{\bibinfo{person}{Renan~P Monteiro}, \bibinfo{person}{Gabriel Lins de~Holanda Coelho}, \bibinfo{person}{Paul~HP Hanel}, \bibinfo{person}{Emerson~Di{\'o}genes de Medeiros}, {and} \bibinfo{person}{Phillip Dyamond~Gomes da Silva}.} \bibinfo{year}{2022}\natexlab{}.
\newblock \showarticletitle{The efficient assessment of self-esteem: proposing the brief rosenberg self-esteem scale}.
\newblock \bibinfo{journal}{\emph{Applied Research in Quality of Life}} \bibinfo{volume}{17}, \bibinfo{number}{2} (\bibinfo{year}{2022}), \bibinfo{pages}{931--947}.
\newblock


\bibitem[Mosaferchi et~al\mbox{.}(2023)]%
        {mosaferchi2023personality}
\bibfield{author}{\bibinfo{person}{Saeedeh Mosaferchi}, \bibinfo{person}{Rosaria Califano}, {and} \bibinfo{person}{Alessandro Naddeo}.} \bibinfo{year}{2023}\natexlab{}.
\newblock \showarticletitle{How Personality, Demographics, and Technology Affinity Affect Trust in Autonomous Vehicles: A Case Study}.
\newblock \bibinfo{journal}{\emph{Human Factors in Transportation}} \bibinfo{volume}{95}, \bibinfo{number}{95} (\bibinfo{year}{2023}).
\newblock


\bibitem[Muir(1994)]%
        {muir1994trust}
\bibfield{author}{\bibinfo{person}{Bonnie~M Muir}.} \bibinfo{year}{1994}\natexlab{}.
\newblock \showarticletitle{Trust in automation: Part I. Theoretical issues in the study of trust and human intervention in automated systems}.
\newblock \bibinfo{journal}{\emph{Ergonomics}} \bibinfo{volume}{37}, \bibinfo{number}{11} (\bibinfo{year}{1994}), \bibinfo{pages}{1905--1922}.
\newblock


\bibitem[Nass et~al\mbox{.}(1994)]%
        {nass1994computers}
\bibfield{author}{\bibinfo{person}{Clifford Nass}, \bibinfo{person}{Jonathan Steuer}, {and} \bibinfo{person}{Ellen~R Tauber}.} \bibinfo{year}{1994}\natexlab{}.
\newblock \showarticletitle{Computers are social actors}. In \bibinfo{booktitle}{\emph{Proceedings of the SIGCHI conference on Human factors in computing systems}}. \bibinfo{pages}{72--78}.
\newblock


\bibitem[Nepal et~al\mbox{.}(2024)]%
        {nepal2024moodcapture}
\bibfield{author}{\bibinfo{person}{Subigya Nepal}, \bibinfo{person}{Arvind Pillai}, \bibinfo{person}{Weichen Wang}, \bibinfo{person}{Tess Griffin}, \bibinfo{person}{Amanda~C Collins}, \bibinfo{person}{Michael Heinz}, \bibinfo{person}{Damien Lekkas}, \bibinfo{person}{Shayan Mirjafari}, \bibinfo{person}{Matthew Nemesure}, \bibinfo{person}{George Price}, {et~al\mbox{.}}} \bibinfo{year}{2024}\natexlab{}.
\newblock \showarticletitle{MoodCapture: Depression Detection Using In-the-Wild Smartphone Images}. In \bibinfo{booktitle}{\emph{Proceedings of the CHI Conference on Human Factors in Computing Systems}}. \bibinfo{pages}{1--18}.
\newblock


\bibitem[Pataranutaporn et~al\mbox{.}(2023)]%
        {pataranutaporn2023influencing}
\bibfield{author}{\bibinfo{person}{Pat Pataranutaporn}, \bibinfo{person}{Ruby Liu}, \bibinfo{person}{Ed Finn}, {and} \bibinfo{person}{Pattie Maes}.} \bibinfo{year}{2023}\natexlab{}.
\newblock \showarticletitle{Influencing human--AI interaction by priming beliefs about AI can increase perceived trustworthiness, empathy and effectiveness}.
\newblock \bibinfo{journal}{\emph{Nature Machine Intelligence}} \bibinfo{volume}{5}, \bibinfo{number}{10} (\bibinfo{year}{2023}), \bibinfo{pages}{1076--1086}.
\newblock


\bibitem[Pazzani et~al\mbox{.}(2022)]%
        {pazzani2022expert}
\bibfield{author}{\bibinfo{person}{Michael Pazzani}, \bibinfo{person}{Severine Soltani}, \bibinfo{person}{Robert Kaufman}, \bibinfo{person}{Samson Qian}, {and} \bibinfo{person}{Albert Hsiao}.} \bibinfo{year}{2022}\natexlab{}.
\newblock \showarticletitle{Expert-informed, user-centric explanations for machine learning}. In \bibinfo{booktitle}{\emph{Proceedings of the AAAI Conference on Artificial Intelligence}}, Vol.~\bibinfo{volume}{36}. \bibinfo{pages}{12280--12286}.
\newblock


\bibitem[Pl{\"O}tz(2021)]%
        {plotz2021applying}
\bibfield{author}{\bibinfo{person}{Thomas Pl{\"O}tz}.} \bibinfo{year}{2021}\natexlab{}.
\newblock \showarticletitle{Applying machine learning for sensor data analysis in interactive systems: Common pitfalls of pragmatic use and ways to avoid them}.
\newblock \bibinfo{journal}{\emph{ACM Computing Surveys (CSUR)}} \bibinfo{volume}{54}, \bibinfo{number}{6} (\bibinfo{year}{2021}), \bibinfo{pages}{1--25}.
\newblock


\bibitem[Qualtrics(2024)]%
        {Qualtrics2024}
\bibfield{author}{\bibinfo{person}{Qualtrics}.} \bibinfo{year}{2024}\natexlab{}.
\newblock \bibinfo{title}{Qualtrics: Online Survey Software}.
\newblock
\newblock
\newblock
\shownote{Provo, UT, USA. \url{https://www.qualtrics.com}}.


\bibitem[Rammstedt and John(2007)]%
        {rammstedt2007measuring}
\bibfield{author}{\bibinfo{person}{Beatrice Rammstedt} {and} \bibinfo{person}{Oliver~P John}.} \bibinfo{year}{2007}\natexlab{}.
\newblock \showarticletitle{Measuring personality in one minute or less: A 10-item short version of the Big Five Inventory in English and German}.
\newblock \bibinfo{journal}{\emph{Journal of research in Personality}} \bibinfo{volume}{41}, \bibinfo{number}{1} (\bibinfo{year}{2007}), \bibinfo{pages}{203--212}.
\newblock


\bibitem[Roets and Van~Hiel(2011)]%
        {roets2011item}
\bibfield{author}{\bibinfo{person}{Arne Roets} {and} \bibinfo{person}{Alain Van~Hiel}.} \bibinfo{year}{2011}\natexlab{}.
\newblock \showarticletitle{Item selection and validation of a brief, 15-item version of the Need for Closure Scale}.
\newblock \bibinfo{journal}{\emph{Personality and individual differences}} \bibinfo{volume}{50}, \bibinfo{number}{1} (\bibinfo{year}{2011}), \bibinfo{pages}{90--94}.
\newblock


\bibitem[Saha et~al\mbox{.}(2021)]%
        {saha2021person}
\bibfield{author}{\bibinfo{person}{Koustuv Saha}, \bibinfo{person}{Ted Grover}, \bibinfo{person}{Stephen~M Mattingly}, \bibinfo{person}{Vedant~Das Swain}, \bibinfo{person}{Pranshu Gupta}, \bibinfo{person}{Gonzalo~J Martinez}, \bibinfo{person}{Pablo Robles-Granda}, \bibinfo{person}{Gloria Mark}, \bibinfo{person}{Aaron Striegel}, {and} \bibinfo{person}{Munmun De~Choudhury}.} \bibinfo{year}{2021}\natexlab{}.
\newblock \showarticletitle{Person-centered predictions of psychological constructs with social media contextualized by multimodal sensing}.
\newblock \bibinfo{journal}{\emph{Proceedings of the ACM on Interactive, Mobile, Wearable and Ubiquitous Technologies}} \bibinfo{volume}{5}, \bibinfo{number}{1} (\bibinfo{year}{2021}), \bibinfo{pages}{1--32}.
\newblock


\bibitem[Schneider and Handali(2019)]%
        {schneider2019personalized}
\bibfield{author}{\bibinfo{person}{Johanes Schneider} {and} \bibinfo{person}{Joshua Handali}.} \bibinfo{year}{2019}\natexlab{}.
\newblock \showarticletitle{Personalized explanation in machine learning: A conceptualization}.
\newblock \bibinfo{journal}{\emph{arXiv preprint arXiv:1901.00770}} (\bibinfo{year}{2019}).
\newblock


\bibitem[Scott and Bruce(1995)]%
        {scott1995decision}
\bibfield{author}{\bibinfo{person}{Susanne~G Scott} {and} \bibinfo{person}{Reginald~A Bruce}.} \bibinfo{year}{1995}\natexlab{}.
\newblock \showarticletitle{Decision-making style: The development and assessment of a new measure}.
\newblock \bibinfo{journal}{\emph{Educational and psychological measurement}} \bibinfo{volume}{55}, \bibinfo{number}{5} (\bibinfo{year}{1995}), \bibinfo{pages}{818--831}.
\newblock


\bibitem[Soltani et~al\mbox{.}(2022)]%
        {soltani2022user}
\bibfield{author}{\bibinfo{person}{Severine Soltani}, \bibinfo{person}{Robert~A Kaufman}, {and} \bibinfo{person}{Michael~J Pazzani}.} \bibinfo{year}{2022}\natexlab{}.
\newblock \showarticletitle{User-centric enhancements to explainable AI algorithms for image classification}. In \bibinfo{booktitle}{\emph{Proceedings of the Annual Meeting of the Cognitive Science Society}}, Vol.~\bibinfo{volume}{44}.
\newblock


\bibitem[Stamos et~al\mbox{.}(2019)]%
        {stamos2019investigating}
\bibfield{author}{\bibinfo{person}{Angelos Stamos}, \bibinfo{person}{Efthymios Altsitsiadis}, {and} \bibinfo{person}{Siegfried Dewitte}.} \bibinfo{year}{2019}\natexlab{}.
\newblock \showarticletitle{Investigating the effect of childhood socioeconomic background on interpersonal trust: Lower childhood socioeconomic status predicts lower levels of trust}.
\newblock \bibinfo{journal}{\emph{Personality and Individual Differences}}  \bibinfo{volume}{145} (\bibinfo{year}{2019}), \bibinfo{pages}{19--25}.
\newblock


\bibitem[Steinke et~al\mbox{.}(2012)]%
        {steinke2012trust}
\bibfield{author}{\bibinfo{person}{Frederick Steinke}, \bibinfo{person}{Tobias Fritsch}, {and} \bibinfo{person}{Lina Silbermann}.} \bibinfo{year}{2012}\natexlab{}.
\newblock \showarticletitle{Trust in ambient assisted living (AAL)-a systematic review of trust in automation and assistance systems}.
\newblock \bibinfo{journal}{\emph{International Journal on Advances in Life Sciences}} \bibinfo{volume}{4}, \bibinfo{number}{3-4} (\bibinfo{year}{2012}).
\newblock


\bibitem[Taubman-Ben-Ari et~al\mbox{.}(2004)]%
        {taubman2004multidimensional}
\bibfield{author}{\bibinfo{person}{Orit Taubman-Ben-Ari}, \bibinfo{person}{Mario Mikulincer}, {and} \bibinfo{person}{Omri Gillath}.} \bibinfo{year}{2004}\natexlab{}.
\newblock \showarticletitle{The multidimensional driving style inventory—scale construct and validation}.
\newblock \bibinfo{journal}{\emph{Accident Analysis \& Prevention}} \bibinfo{volume}{36}, \bibinfo{number}{3} (\bibinfo{year}{2004}), \bibinfo{pages}{323--332}.
\newblock


\bibitem[Teo and Van~Schaik(2012)]%
        {teo2012understanding}
\bibfield{author}{\bibinfo{person}{Timothy Teo} {and} \bibinfo{person}{Paul Van~Schaik}.} \bibinfo{year}{2012}\natexlab{}.
\newblock \showarticletitle{Understanding the intention to use technology by preservice teachers: An empirical test of competing theoretical models}.
\newblock \bibinfo{journal}{\emph{International Journal of Human-Computer Interaction}} \bibinfo{volume}{28}, \bibinfo{number}{3} (\bibinfo{year}{2012}), \bibinfo{pages}{178--188}.
\newblock


\bibitem[Tharwat et~al\mbox{.}(2017)]%
        {tharwat2017linear}
\bibfield{author}{\bibinfo{person}{Alaa Tharwat}, \bibinfo{person}{Tarek Gaber}, \bibinfo{person}{Abdelhameed Ibrahim}, {and} \bibinfo{person}{Aboul~Ella Hassanien}.} \bibinfo{year}{2017}\natexlab{}.
\newblock \showarticletitle{Linear discriminant analysis: A detailed tutorial}.
\newblock \bibinfo{journal}{\emph{AI communications}} \bibinfo{volume}{30}, \bibinfo{number}{2} (\bibinfo{year}{2017}), \bibinfo{pages}{169--190}.
\newblock


\bibitem[Tibshirani(1996)]%
        {tibshirani1996regression}
\bibfield{author}{\bibinfo{person}{Robert Tibshirani}.} \bibinfo{year}{1996}\natexlab{}.
\newblock \showarticletitle{Regression shrinkage and selection via the lasso}.
\newblock \bibinfo{journal}{\emph{Journal of the Royal Statistical Society Series B: Statistical Methodology}} \bibinfo{volume}{58}, \bibinfo{number}{1} (\bibinfo{year}{1996}), \bibinfo{pages}{267--288}.
\newblock


\bibitem[Twenge et~al\mbox{.}(2014)]%
        {twenge2014declines}
\bibfield{author}{\bibinfo{person}{Jean~M Twenge}, \bibinfo{person}{W~Keith Campbell}, {and} \bibinfo{person}{Nathan~T Carter}.} \bibinfo{year}{2014}\natexlab{}.
\newblock \showarticletitle{Declines in trust in others and confidence in institutions among American adults and late adolescents, 1972--2012}.
\newblock \bibinfo{journal}{\emph{Psychological science}} \bibinfo{volume}{25}, \bibinfo{number}{10} (\bibinfo{year}{2014}), \bibinfo{pages}{1914--1923}.
\newblock


\bibitem[Wang et~al\mbox{.}(2019)]%
        {wang2019designing}
\bibfield{author}{\bibinfo{person}{Danding Wang}, \bibinfo{person}{Qian Yang}, \bibinfo{person}{Ashraf Abdul}, {and} \bibinfo{person}{Brian~Y Lim}.} \bibinfo{year}{2019}\natexlab{}.
\newblock \showarticletitle{Designing theory-driven user-centric explainable AI}. In \bibinfo{booktitle}{\emph{Proceedings of the 2019 CHI conference on human factors in computing systems}}. \bibinfo{pages}{1--15}.
\newblock


\bibitem[Wang et~al\mbox{.}(2020)]%
        {wang2020survey}
\bibfield{author}{\bibinfo{person}{Jingwen Wang}, \bibinfo{person}{Xuyang Jing}, \bibinfo{person}{Zheng Yan}, \bibinfo{person}{Yulong Fu}, \bibinfo{person}{Witold Pedrycz}, {and} \bibinfo{person}{Laurence~T Yang}.} \bibinfo{year}{2020}\natexlab{}.
\newblock \showarticletitle{A survey on trust evaluation based on machine learning}.
\newblock \bibinfo{journal}{\emph{ACM Computing Surveys (CSUR)}} \bibinfo{volume}{53}, \bibinfo{number}{5} (\bibinfo{year}{2020}), \bibinfo{pages}{1--36}.
\newblock


\bibitem[{Waymo LLC}(2024)]%
        {WaymoOne}
\bibfield{author}{\bibinfo{person}{{Waymo LLC}}.} \bibinfo{year}{2024}\natexlab{}.
\newblock \bibinfo{title}{Waymo One}.
\newblock
\newblock
\urldef\tempurl%
\url{https://waymo.com/waymo-one/}
\showURL{%
\tempurl}
\newblock
\shownote{Accessed: 2024-07-22}.


\bibitem[Yarkoni and Westfall(2017)]%
        {yarkoni2017choosing}
\bibfield{author}{\bibinfo{person}{Tal Yarkoni} {and} \bibinfo{person}{Jacob Westfall}.} \bibinfo{year}{2017}\natexlab{}.
\newblock \showarticletitle{Choosing prediction over explanation in psychology: Lessons from machine learning}.
\newblock \bibinfo{journal}{\emph{Perspectives on Psychological Science}} \bibinfo{volume}{12}, \bibinfo{number}{6} (\bibinfo{year}{2017}), \bibinfo{pages}{1100--1122}.
\newblock


\bibitem[Yi et~al\mbox{.}(2023)]%
        {yi2023measurement}
\bibfield{author}{\bibinfo{person}{Binlin Yi}, \bibinfo{person}{Haotian Cao}, \bibinfo{person}{Xiaolin Song}, \bibinfo{person}{Jianqiang Wang}, \bibinfo{person}{Wenfeng Guo}, {and} \bibinfo{person}{Zhi Huang}.} \bibinfo{year}{2023}\natexlab{}.
\newblock \showarticletitle{Measurement and real-time recognition of driver trust in conditionally automated vehicles: Using multimodal feature fusions network}.
\newblock \bibinfo{journal}{\emph{Transportation research record}} \bibinfo{volume}{2677}, \bibinfo{number}{8} (\bibinfo{year}{2023}), \bibinfo{pages}{311--330}.
\newblock


\bibitem[Yoo et~al\mbox{.}(2011)]%
        {yoo2011measuring}
\bibfield{author}{\bibinfo{person}{Boonghee Yoo}, \bibinfo{person}{Naveen Donthu}, {and} \bibinfo{person}{Tomasz Lenartowicz}.} \bibinfo{year}{2011}\natexlab{}.
\newblock \showarticletitle{Measuring Hofstede's five dimensions of cultural values at the individual level: Development and validation of CVSCALE}.
\newblock \bibinfo{journal}{\emph{Journal of international consumer marketing}} \bibinfo{volume}{23}, \bibinfo{number}{3-4} (\bibinfo{year}{2011}), \bibinfo{pages}{193--210}.
\newblock


\bibitem[Zhang et~al\mbox{.}(2023)]%
        {zhang2023human}
\bibfield{author}{\bibinfo{person}{Qidi Zhang}, \bibinfo{person}{Tingru Zhang}, {and} \bibinfo{person}{Liang Ma}.} \bibinfo{year}{2023}\natexlab{}.
\newblock \showarticletitle{Human acceptance of autonomous vehicles: Research status and prospects}.
\newblock \bibinfo{journal}{\emph{International journal of industrial ergonomics}}  \bibinfo{volume}{95} (\bibinfo{year}{2023}), \bibinfo{pages}{103458}.
\newblock


\bibitem[Zhang et~al\mbox{.}(2024)]%
        {zhang2024effects}
\bibfield{author}{\bibinfo{person}{Yaping Zhang}, \bibinfo{person}{Qianli Ma}, \bibinfo{person}{Jianhong Qu}, {and} \bibinfo{person}{Ronggang Zhou}.} \bibinfo{year}{2024}\natexlab{}.
\newblock \showarticletitle{Effects of driving style on takeover performance during automated driving: Under the influence of warning system factors}.
\newblock \bibinfo{journal}{\emph{Applied Ergonomics}}  \bibinfo{volume}{117} (\bibinfo{year}{2024}), \bibinfo{pages}{104229}.
\newblock


\bibitem[Zolfaghar and Aghaie(2012)]%
        {zolfaghar2012syntactical}
\bibfield{author}{\bibinfo{person}{Kiyana Zolfaghar} {and} \bibinfo{person}{Abdollah Aghaie}.} \bibinfo{year}{2012}\natexlab{}.
\newblock \showarticletitle{A syntactical approach for interpersonal trust prediction in social web applications: Combining contextual and structural data}.
\newblock \bibinfo{journal}{\emph{Knowledge-Based Systems}}  \bibinfo{volume}{26} (\bibinfo{year}{2012}), \bibinfo{pages}{93--102}.
\newblock


\end{thebibliography}

\newpage
 \begin{appendices}
 \onecolumn
\section{Comparison of High and Low Trust Groups For Each Predictor}

\setcounter{table}{0}
\renewcommand{\thetable}{A\arabic{table}}

\label{appendixA.trust}
Table~\ref{tab:appendix} contains the list of all variable names used as predictor inputs. Participants are categorized as Low Trust and High Trust based on if they are on the bottom half of possible values or the top half of possible values for the trust composite score scale. \textbf{This allows us to compare predictor variable scores between people high and low trust.} Comparisons use the non-parametric Kruskal-Wallis test. Variable descriptions are included in Tables \ref{driving_factors}, \ref{demo_psych_factors}, and \ref{risk_benefit_factors}. The top 15 large and medium effect sizes are included in Table \ref{kw_large}.
\newline



\begin{longtable}{p{0.3\linewidth} | p{0.05\linewidth} | p{0.05\linewidth} | p{0.05\linewidth} | p{0.05\linewidth} | p{0.06\linewidth} | p{0.07\linewidth} | p{0.08\linewidth} | p{0.07\linewidth}}
\caption{List of all variable names used as predictor inputs to compare high and low trust groups.}
\label{tab:appendix}
  \\ \hline
 & \multicolumn{2}{c|}{\textbf{Low Trust}} & \multicolumn{2}{c|}{\textbf{High Trust}} & \multicolumn{4}{c}{\textbf{Comparison}}\\
 \hline
\textbf{Predictor} & \textbf{Mean} & \textbf{SD} & \textbf{Mean} & \textbf{SD} & \textbf{{\footnotesize H-stat}} & \textbf{{\footnotesize P-value}} & \textbf{{\footnotesize Cohen's D}} & \textbf{Effect} \\ 
  \hline
\endfirsthead
\multicolumn{9}{r}{\textit{Table \thetable\ continued from previous page\newline}}
\\ \hline
& \multicolumn{2}{c|}{\textbf{Low Trust}} & \multicolumn{2}{c|}{\textbf{High Trust}} & \multicolumn{4}{c}{\textbf{Comparison}}\\
 \hline
\textbf{Predictor} & \textbf{Mean} & \textbf{SD} & \textbf{Mean} & \textbf{SD} & \textbf{{\footnotesize H-stat}} & \textbf{{\footnotesize P-value}} & \textbf{{\footnotesize Cohen's D}} & \textbf{Effect} \\ 
  \hline
\endhead
Overall Risk-Benefit & 3.23 & 0.87 & 4.63 & 0.81 & 595.25 & < 0.01 & 1.66 & large \\ 
Ease of Use & 2.55 & 1.13 & 3.72 & 0.88 & 345.75 & < 0.01 & 1.14 & large \\ 
Reduce Accidents & 2.79 & 1.07 & 3.89 & 0.86 & 338.97 & < 0.01 & 1.12 & large \\ 
Trust Tech Companies & 2.19 & 0.99 & 3.09 & 0.99 & 241.61 & < 0.01 & 0.91 & large \\ 
Increase Fun & 2.38 & 1.22 & 3.46 & 1.16 & 237.14 & < 0.01 & 0.91 & large \\ 
AV Feasibility & 2.58 & 1.02 & 3.43 & 0.87 & 235.40 & < 0.01 & 0.88 & large \\ 
Improve Efficiency & 2.46 & 1.06 & 3.36 & 1.03 & 214.80 & < 0.01 & 0.86 & large \\ 
Reduce Traffic & 2.82 & 1.17 & 3.71 & 1.04 & 194.96 & < 0.01 & 0.80 & large \\ 
\midrule
Trust Automakers & 2.47 & 1.07 & 3.28 & 0.96 & 191.45 & < 0.01 & 0.79 & medium \\ 
Mental Model & 1.70 & 0.88 & 2.42 & 1.04 & 191.01 & < 0.01 & 0.76 & medium \\ 
Losing Control & 4.18 & 1.06 & 3.37 & 1.13 & 208.25 & < 0.01 & -0.74 & medium \\ 
Reduce Emissions & 3.02 & 1.15 & 3.76 & 0.99 & 142.02 & < 0.01 & 0.69 & medium \\ 
Know AV Purpose & 3.16 & 1.22 & 3.91 & 0.89 & 137.61 & < 0.01 & 0.69 & medium \\ 
Improve Fuel Econ. & 3.26 & 1.07 & 3.94 & 0.89 & 145.74 & < 0.01 & 0.68 & medium \\ 
Reduce Transport Cost & 2.80 & 1.11 & 3.52 & 1.07 & 133.91 & < 0.01 & 0.66 & medium \\
Quick to Adopt Tech & 3.04 & 1.11 & 3.71 & 0.93 & 131.84 & < 0.01 & 0.65 & medium \\ 
Lack Understanding & 3.55 & 1.20 & 2.86 & 1.10 & 127.35 & < 0.01 & -0.59 & medium \\ 
Trust Government & 2.02 & 0.99 & 2.63 & 1.07 & 115.61 & < 0.01 & 0.59 & medium \\ 
System Failure & 4.40 & 0.92 & 3.87 & 0.99 & 144.07 & < 0.01 & -0.56 & medium \\ 
Enjoy Learning Tech & 2.74 & 1.21 & 3.39 & 1.16 & 96.65 & < 0.01 & 0.55 & medium \\ 
Free Time & 2.63 & 1.30 & 3.29 & 1.30 & 83.72 & < 0.01 & 0.51 & medium \\ 
Understand How AVs Work & 2.32 & 1.19 & 2.92 & 1.21 & 83.46 & < 0.01 & 0.50 & medium \\ 
Performance (varied) & 3.93 & 1.06 & 3.40 & 1.09 & 92.62 & < 0.01 & -0.50 & medium \\ 
\midrule
Ridden in AV & 1.39 & 0.57 & 1.69 & 0.68 & 79.10 & < 0.01 & 0.49 & small \\ 
Tech-Oriented Self & 2.99 & 1.24 & 3.53 & 1.13 & 66.30 & < 0.01 & 0.46 & small \\ 
Enjoy Tech Analysis & 2.72 & 1.28 & 3.13 & 1.23 & 37.00 & < 0.01 & 0.33 & small \\ 
Hacking & 3.87 & 1.19 & 3.51 & 1.20 & 40.61 & < 0.01 & -0.31 & small \\ 
Improve Mobility & 4.08 & 0.98 & 4.35 & 0.81 & 29.67 & < 0.01 & 0.30 & small \\ 
Use Automatic Stability Control & 1.25 & 0.65 & 1.48 & 0.90 & 29.31 & < 0.01 & 0.30 & small \\ 
Legal Liability & 4.04 & 1.03 & 3.74 & 1.04 & 36.57 & < 0.01 & -0.29 & small \\ 
Understand Computer Vision & 1.81 & 1.14 & 2.15 & 1.29 & 24.92 & < 0.01 & 0.28 & small \\ 
Understand AV Algorithms & 1.60 & 1.03 & 1.90 & 1.17 & 30.56 & < 0.01 & 0.27 & small \\ 
Data Privacy & 3.48 & 1.30 & 3.14 & 1.29 & 25.06 & < 0.01 & -0.26 & small \\ 
Driving Risk & 2.94 & 1.05 & 2.68 & 0.97 & 21.13 & < 0.01 & -0.26 & small \\ 
NFC Predictability & 11.87 & 2.57 & 11.21 & 2.57 & 26.55 & < 0.01 & -0.26 & small \\ 
Have Built AI Systems & 1.19 & 0.66 & 1.39 & 0.95 & 22.70 & < 0.01 & 0.25 & small \\ 
Used Adaptive Cruise Control & 1.71 & 0.91 & 1.96 & 1.08 & 14.47 & < 0.01 & 0.25 & small \\ 
Gender (Female vs. Not) & 0.79 & 0.41 & 0.69 & 0.46 & 19.99 & < 0.01 & -0.24 & small \\ 
Good With Technology & 3.54 & 1.10 & 3.79 & 1.02 & 18.34 & < 0.01 & 0.23 & small \\ 
Use Collision Avoidance System & 1.30 & 0.75 & 1.50 & 0.95 & 16.16 & < 0.01 & 0.23 & small \\ 
Use ML For Work & 1.40 & 0.93 & 1.63 & 1.11 & 21.45 & < 0.01 & 0.23 & small \\ 
Lots of Tech Experience & 2.84 & 1.26 & 3.12 & 1.22 & 17.03 & < 0.01 & 0.22 & small \\ 
Experience Training/Evaluating ML & 1.33 & 0.87 & 1.54 & 1.09 & 16.96 & < 0.01 & 0.22 & small \\ 
Use Parallel Parking Assist & 1.18 & 0.63 & 1.34 & 0.81 & 20.72 & < 0.01 & 0.22 & small \\ 
Use Vehicle Guidance System & 1.20 & 0.65 & 1.36 & 0.85 & 17.03 & < 0.01 & 0.21 & small \\ 
Use Blindspot Detection & 1.95 & 1.28 & 2.22 & 1.36 & 14.45 & < 0.01 & 0.21 & small \\ 
\midrule
Confidence Learning Tech & 2.99 & 1.24 & 3.23 & 1.21 & 13.38 & < 0.01 & 0.20 & neg. \\ 
CVSCALE Power Distance & 7.51 & 3.06 & 8.13 & 3.42 & 13.06 & < 0.01 & 0.19 & neg. \\ 
Injury Concerns (DC) & 2.02 & 1.13 & 1.82 & 0.95 & 6.43 & 0.01 & -0.19 & neg. \\ 
I help solve tech probs. & 2.90 & 1.31 & 3.14 & 1.28 & 11.97 & < 0.01 & 0.19 & neg. \\ 
I am knowledgeable about tech. & 3.37 & 1.13 & 3.57 & 1.09 & 11.29 & < 0.01 & 0.18 & neg. \\ 
Familiar with AVs & 3.08 & 1.53 & 3.34 & 1.43 & 8.98 & < 0.01 & 0.18 & neg. \\ 
I am AV Expert & 1.20 & 0.56 & 1.31 & 0.72 & 8.62 & < 0.01 & 0.18 & neg. \\ 
I work with AVs & 1.05 & 0.28 & 1.11 & 0.49 & 9.30 & < 0.01 & 0.17 & neg. \\ 
Anxiety as Passenger & 2.36 & 1.07 & 2.18 & 0.97 & 8.58 & < 0.01 & -0.17 & neg. \\ 
NFC Ambiguity & 11.96 & 2.40 & 11.56 & 2.26 & 13.89 & < 0.01 & -0.17 & neg. \\ 
BFI Neuroticism & 0.90 & 2.00 & 0.56 & 2.01 & 10.71 & < 0.01 & -0.17 & neg. \\ 
Thrill from breaking law (MDSI) & 1.40 & 0.79 & 1.55 & 0.98 & 5.36 & 0.02 & 0.17 & neg. \\ 
Risk Willingness & 5.61 & 1.79 & 5.90 & 1.86 & 8.89 & < 0.01 & 0.16 & neg. \\ 
Die in accident (DC) & 2.05 & 1.15 & 1.88 & 1.02 & 5.20 & 0.02 & -0.16 & neg. \\ 
Plan routes badly (MDSI) & 1.65 & 0.90 & 1.79 & 0.98 & 6.60 & 0.01 & 0.15 & neg. \\ 
Enjoy dangerous driving (MDSI) & 1.59 & 1.00 & 1.75 & 1.13 & 7.07 & 0.01 & 0.15 & neg. \\ 
BFI Conscientiousness & 1.09 & 1.69 & 0.84 & 1.58 & 8.60 & < 0.01 & -0.15 & neg. \\ 
BFI Extroversion & -0.26 & 1.99 & 0.04 & 1.96 & 7.10 & 0.01 & 0.15 & neg. \\ 
Self Esteem & 4.29 & 3.20 & 4.75 & 2.95 & 6.20 & 0.01 & 0.15 & neg. \\ 
Relax activities mid-drive (MDSI) & 2.57 & 1.23 & 2.75 & 1.23 & 6.55 & 0.01 & 0.15 & neg. \\ 
Use Lane Control & 1.68 & 1.05 & 1.84 & 1.15 & 5.82 & 0.02 & 0.15 & neg. \\ 
Use Auto E-Brake & 1.61 & 0.94 & 1.76 & 1.06 & 4.99 & 0.03 & 0.14 & neg. \\ 
SES & 4.55 & 5.43 & 5.32 & 5.69 & 7.23 & 0.01 & 0.14 & neg. \\ 
Honk horn at others (MDSI) & 2.21 & 1.26 & 2.05 & 1.16 & 3.44 & 0.06 & -0.13 & neg. \\ 
Need for Control & 20.31 & 3.34 & 19.87 & 3.35 & 6.24 & 0.01 & -0.13 & neg. \\ 
Feel frustrated while driving (MDSI) & 2.61 & 1.28 & 2.45 & 1.24 & 4.56 & 0.03 & -0.13 & neg. \\ 
Meditate while driving (MDSI) & 1.63 & 0.98 & 1.76 & 1.00 & 7.28 & 0.01 & 0.13 & neg. \\ 
Distracted driver (MDSI) & 2.58 & 1.24 & 2.73 & 1.26 & 4.19 & 0.04 & 0.12 & neg. \\ 
Forget high bean lights on (MDSI) & 1.53 & 0.91 & 1.64 & 0.96 & 6.10 & 0.01 & 0.12 & neg. \\ 
Thrill of flirting with death (MDSI) & 1.30 & 0.78 & 1.40 & 0.87 & 4.67 & 0.03 & 0.11 & neg. \\ 
CVSCALE Collectivism & 18.18 & 4.56 & 18.71 & 4.92 & 5.54 & 0.02 & 0.11 & neg. \\ 
Drive through red lights (MDSI) & 1.63 & 0.98 & 1.74 & 1.05 & 4.50 & 0.03 & 0.11 & neg. \\ 
Nervous while driving (MDSI) & 2.69 & 1.39 & 2.54 & 1.35 & 3.37 & 0.07 & -0.11 & neg. \\ 
Passengers will be hurt (DC) & 1.96 & 1.17 & 1.84 & 1.04 & 1.47 & 0.23 & -0.11 & neg. \\ 
I have control over driving (MDSI) & 4.06 & 0.94 & 4.16 & 0.91 & 3.77 & 0.05 & 0.11 & neg. \\ 
Age & 20.58 & 2.22 & 20.83 & 2.54 & 2.12 & 0.15 & 0.11 & neg. \\ 
Worry of bad weather (MDSI) & 3.78 & 1.16 & 3.66 & 1.17 & 4.13 & 0.04 & -0.10 & neg. \\ 
Swear at other drivers (MDSI) & 2.57 & 1.40 & 2.43 & 1.39 & 3.30 & 0.07 & -0.10 & neg. \\ 
Attempt to drive from park (MDSI) & 1.62 & 0.98 & 1.72 & 1.07 & 2.95 & 0.09 & 0.10 & neg. \\ 
Comfortable while driving (MDSI) & 3.86 & 1.12 & 3.97 & 1.01 & 1.58 & 0.21 & 0.10 & neg. \\ 
Mix wiper and light switch (MDSI) & 1.71 & 1.07 & 1.82 & 1.12 & 2.68 & 0.10 & 0.10 & neg. \\ 
High beam when annoyed (MDSI) & 1.52 & 0.99 & 1.61 & 1.06 & 3.63 & 0.06 & 0.09 & neg. \\ 
Enjoy taking driving risks (MDSI) & 1.64 & 0.91 & 1.73 & 1.02 & 1.41 & 0.24 & 0.09 & neg. \\ 
Daydream while driving (MDSI) & 2.59 & 1.32 & 2.71 & 1.27 & 3.07 & 0.08 & 0.09 & neg. \\ 
Lack control over accidents (DC) & 2.93 & 1.26 & 2.82 & 1.26 & 2.37 & 0.12 & -0.09 & neg. \\ 
Muscle relaxation mid-drive (MDSI) & 1.93 & 1.15 & 2.03 & 1.18 & 2.50 & 0.11 & 0.09 & neg. \\ 
Fix appearance mid-drive (MDSI) & 2.39 & 1.26 & 2.50 & 1.31 & 2.06 & 0.15 & 0.09 & neg. \\ 
Plan long trips in advance (MDSI) & 4.01 & 1.10 & 3.91 & 1.15 & 2.26 & 0.13 & -0.09 & neg. \\ 
Others will notice I'm anxious (DC) & 2.08 & 1.26 & 1.98 & 1.18 & 1.78 & 0.18 & -0.09 & neg. \\ 
BFI Agreeableness & 5.23 & 2.26 & 5.42 & 2.15 & 2.49 & 0.11 & 0.09 & neg. \\ 
People will think bad driver (DC) & 2.48 & 1.26 & 2.37 & 1.21 & 1.82 & 0.18 & -0.09 & neg. \\ 
I will not react quickly enough (DC) & 2.44 & 1.18 & 2.35 & 1.11 & 1.40 & 0.24 & -0.09 & neg. \\ 
BFI Openness & 1.03 & 1.82 & 0.90 & 1.74 & 2.46 & 0.12 & -0.08 & neg. \\ 
People will laugh at me (DC) & 1.84 & 1.13 & 1.76 & 1.02 & 0.38 & 0.54 & -0.07 & neg. \\ 
I will injure someone (DC) & 1.83 & 1.09 & 1.76 & 1.02 & 1.31 & 0.25 & -0.07 & neg. \\ 
Get lost in thoughts (MDSI) & 1.92 & 1.08 & 2.00 & 1.07 & 2.53 & 0.11 & 0.07 & neg. \\ 
Drive cautiously (MDSI) & 4.15 & 0.79 & 4.09 & 0.92 & 0.12 & 0.72 & -0.07 & neg. \\ 
I will cause accident (DC) & 1.92 & 1.07 & 1.85 & 0.98 & 0.40 & 0.53 & -0.06 & neg. \\ 
Drive below speed limit (MDSI) & 2.09 & 1.19 & 2.16 & 1.25 & 0.69 & 0.41 & 0.06 & neg. \\ 
Impatient in rush hour (MDSI) & 3.38 & 1.24 & 3.32 & 1.28 & 0.62 & 0.43 & -0.06 & neg. \\ 
Misjudge space mid-drive (MDSI) & 2.40 & 1.41 & 2.47 & 1.40 & 1.13 & 0.29 & 0.05 & neg. \\ 
“Better safe than sorry” (MDSI) & 4.03 & 0.96 & 3.98 & 0.99 & 0.71 & 0.40 & -0.05 & neg. \\ 
Strategize get through traffic (MDSI) & 3.20 & 1.32 & 3.27 & 1.29 & 0.73 & 0.39 & 0.05 & neg. \\ 
Distressed while driving (MDSI) & 2.44 & 1.25 & 2.38 & 1.24 & 0.65 & 0.42 & -0.05 & neg. \\ 
Enjoy driving on limit (MDSI) & 2.66 & 1.13 & 2.71 & 1.23 & 0.56 & 0.46 & 0.04 & neg. \\ 
Decision Style & 2.04 & 2.17 & 1.95 & 2.12 & 1.01 & 0.32 & -0.04 & neg. \\ 
Patiently wait at green lights (MDSI) & 3.62 & 1.17 & 3.67 & 1.14 & 0.39 & 0.53 & 0.04 & neg. \\ 
Always ready to react (MDSI) & 3.85 & 1.00 & 3.81 & 0.97 & 1.05 & 0.30 & -0.04 & neg. \\ 
Impatient at green lights (MDSI) & 2.38 & 1.22 & 2.42 & 1.23 & 0.47 & 0.49 & 0.04 & neg. \\ 
Switch b/n lanes in traffic (MDSI) & 2.71 & 1.24 & 2.76 & 1.23 & 0.49 & 0.48 & 0.04 & neg. \\ 
Driving Frequency & 43.50 & 38.96 & 44.95 & 39.82 & 0.31 & 0.58 & 0.04 & neg. \\ 
Misjudge speed w/ passing (MDSI) & 2.03 & 1.01 & 2.07 & 1.03 & 0.32 & 0.57 & 0.04 & neg. \\ 
Politics & 3.72 & 0.86 & 3.69 & 0.85 & 0.16 & 0.69 & -0.03 & neg. \\ 
Fear of losing self-control (DC) & 1.57 & 0.95 & 1.60 & 0.96 & 0.21 & 0.65 & 0.03 & neg. \\ 
Try to relax when driving (MDSI) & 3.54 & 1.07 & 3.57 & 1.07 & 0.20 & 0.66 & 0.02 & neg. \\ 
Number of Collisions & 0.35 & 0.63 & 0.34 & 0.60 & 0.04 & 0.83 & -0.02 & neg. \\ 
Patience yielding (MDSI) & 4.27 & 0.93 & 4.25 & 0.89 & 0.59 & 0.44 & -0.02 & neg. \\ 
Education & 1.75 & 1.29 & 1.76 & 1.42 & 0.04 & 0.84 & 0.01 & neg. \\ 
Blow horn when frustrated (MDSI) & 1.89 & 1.24 & 1.90 & 1.22 & 0.11 & 0.74 & 0.01 & neg. \\ 
Drive assertively (MDSI) & 2.60 & 1.29 & 2.59 & 1.29 & 0.02 & 0.90 & -0.01 & neg. \\ 
Fear of criticism (DC) & 2.27 & 1.26 & 2.26 & 1.24 & < 0.01 & 0.97 & -0.01 & neg. \\ 
CVSCALE Uncertainty Avoidance & 21.39 & 3.06 & 21.37 & 3.06 & 0.02 & 0.88 & -0.01 & neg. \\ 
Cost & 4.15 & 1.05 & 4.15 & 0.97 & 0.33 & 0.57 & -0.00 & neg. \\ 
I will cause traffic/anger (DC) & 2.34 & 1.31 & 2.34 & 1.28 & 0.03 & 0.86 & -0.00 & neg. \\ 
   \hline
 \multicolumn{9}{l}{\textit{MDSI = Multidimensional Driving Style Inventory~\cite{taubman2004multidimensional}; DC = Driving Cognitions Questionnaire~\cite{ehlers2007driving}}}   
\end{longtable}

\end{appendices}

\end{document}